\documentclass[manuscript,screen]{acmart}

\usepackage{array} 
\usepackage{multirow} 
\usepackage{tabularx}
\usepackage{graphicx}
\usepackage{footnote}
\usepackage{makecell}
\usepackage{subcaption}

\AtBeginDocument{%
  }

\setcopyright{acmlicensed}
\copyrightyear{2024}
\acmYear{2024}
\acmDOI{XXXXXXX.XXXXXXX}

\acmISBN{978-1-4503-XXXX-X/18/06}

\begin{document}

\title{Survey: Transformer-based Models in Data Modality Conversion}

\author{Elyas Rashno}
\email{elyas.rashno@queensu.ca}
\orcid{0000-0003-0813-3417}
\author{Amir Eskandari}
\authornotemark[1]
\email{amir.eskandari@queensu.ca}
\author{Aman Anand}
\authornote{Amir Eskandari and Aman Anand contributed equally to this research.}
\email{aman.anand@queensu.ca}
\author{Farhana Zulkernine}
\email{farhana.zulkernine@queensu.ca}
\affiliation{%
  \institution{School of Computing, Queen's University}
  \city{Kingston}
  \state{ON}
  \country{Canada}
}








\renewcommand{\shortauthors}{Trovato et al.}

\begin{abstract}
Transformers have made significant strides across various artificial intelligence domains, including natural language processing, computer vision, and audio processing. This success has naturally garnered considerable interest from both academic and industry researchers. Consequently, numerous Transformer variants (often referred to as X-formers) have been developed for these fields. However, a thorough and systematic review of these modality-specific conversions remains lacking. Modality Conversion involves the transformation of data from one form of representation to another, mimicking the way humans integrate and interpret sensory information. This paper provides a comprehensive review of transformer-based models applied to the primary modalities of text, vision, and speech, discussing their architectures, conversion methodologies, and applications. By synthesizing the literature on modality conversion, this survey aims to underline the versatility and scalability of transformers in advancing AI-driven content generation and understanding.
\end{abstract}



\maketitle

\section{Introduction}
Artificial Intelligence (AI) is inspired by human perceptions, such as vision, hearing, and reading, and seeks to replicate these abilities \cite{intro_1_hu2024transformer}. Typically, a modality is linked to a particular sensor that creates a distinct communication channel, such as sight, speech, and written language. Humans possess a fundamental process in sensory perception that allows them to efficiently engage with the world in dynamic and unconstrained situations by integrating data from several sensory modalities. Each modality functions as a separate source of information that is distinguished by its own specific statistical features. A photograph depicting "elephants playing in the water" delivers visual information through numerous pixels, whereas a similar verbal description conveys this sight using distinct words. Similarly, voice can communicate the same occurrence using spectrograms or speech characteristics. A data conversion AI system must receive input from a specific modality, process, understand, and reproduce its content in a different modality, imitating human-like perception. Modality Conversion (MC) is a broad methodology for constructing artificial intelligence models that can extract and transform information from one modality of representation to another \cite{intro_1_hu2024transformer}.

Transformer-based (TB) techniques have significantly transformed the process of converting data from one modality to another by utilizing their advanced attention processes to accurately represent and translate various forms of input. These models demonstrate exceptional performance in tasks such as converting text to speech, speech to text, speech to image, images to text, and even performing cross-modal translation, such as generating images from text \cite{intro_5_han2023survey}. Transformers facilitate a smooth and highly precise conversion by capturing complex interdependencies and contextual interactions across various data modalities. Due to their adaptability and scalability, they play a crucial role in expanding applications in natural language processing, computer vision, and multi-modal data integration. This drives advancements in AI-driven content production and understanding \cite{intro_5_han2023survey}.

\begin{figure}[h]
\centering
\includegraphics[width=\linewidth]{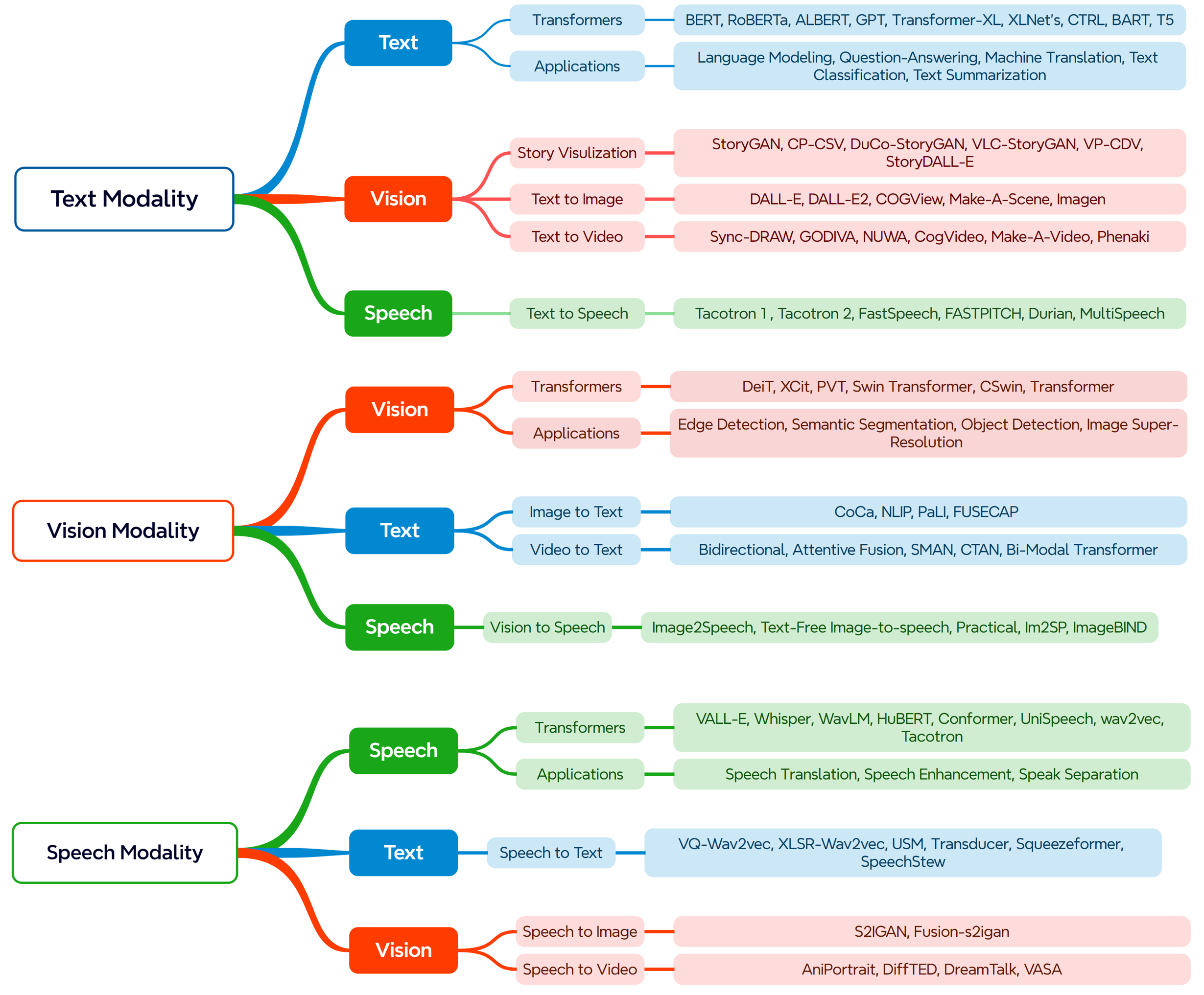}
\caption{Overview of the paper's structure, featuring three modalities: text (blue), vision (red), and speech (green). Each section introduces one modality, identified by its color and name. It covers well-known TBs and their primary applications. The right-hand boxes for each modality illustrate conversion processes and notable applications.}
\label{fig_intro}
\end{figure}


\begin{itemize}

    \item \textbf{Related Surveys:} Numerous surveys have explored TB models across text processing \cite{back_1_ammus2021survey}, computer vision \cite{back_6_johnson2021vision}, and speech processing \cite{back_17_lee2021comparative} domains. Each of these surveys typically reviews papers focusing on a single modality, processing the input to generate the desired output based on specific applications. There are also surveys on data fusion that aim to integrate data from different modalities. These papers generally review various types of fusion models and the types of inputs, such as text, vision, and speech. For instance, the survey by Davis et al. \cite{back_25_davis2022multimodal} on multimodal learning with transformers examines the synergistic use of multiple modalities, demonstrating substantial improvements in tasks requiring comprehensive understanding from diverse data sources. In conclusion, there is no existing survey paper that comprehensively reviews the literature on data conversion across different modalities (text, vision, and speech). 
    
    \item \textbf{Paper Contribution:} In this paper, we present a comprehensive review of TB models used for data modality conversion. We focus on three primary modalities: text, vision, and speech. The input can be in any of these modalities for each transformer model, and the output can be in a different or the same modality. For instance, given a text input, the output could be a translated text (machine translation), an image (story visualization), or speech. Similarly, the outputs can be converted to any of the other three modalities for vision and speech inputs. We have systematically reviewed all the relevant literature on modality conversion using transformer-based models ( Fig. \ref{fig_intro}).

    \item \textbf{Scope:} Our review is restricted to papers published from 2017 to 2024, given that transformers, introduced by Vaswani et al. in 2017 \cite{vaswani2021scaling}, are a relatively new technology. Focusing on this period allows us to include the most recent and pertinent advancements in transformers related to modality representation and conversion. The citation analysis reveals a total of 95 methods from 2017 to 2024, with peak interest between 2020 and 2024. This survey aims to serve researchers and practitioners by consolidating the state-of-the-art transformer models across these domains.


\end{itemize}

\begin{figure}[t]
\centering
\includegraphics[width=\linewidth]{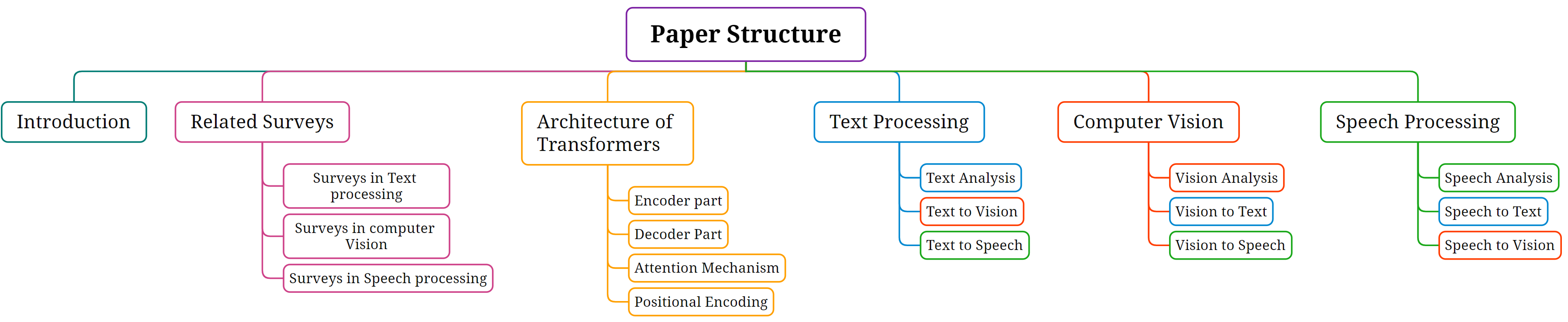}
\DeclareGraphicsExtensions.
\caption{The overall structure of the paper is as follows: Related surveys will be introduced in the second section. The basic transformer model (Vanilla) will be detailed in the third section. The last three sections will cover methods for text, vision, and speech processing, respectively.}
\label{paper_structure}
\end{figure}

The rest of this survey is structured as follows: Section 2 gathers all relevant surveys on TB models. Section 3 provides an introduction to the architecture and key components of the Transformer. Sections 4, 5, and 6 review TB models where the input is text, vision, and speech, respectively, and the output can be any of these three modalities. Section 7 discusses additional aspects of the Transformer that may be of interest to researchers and summarizes the paper \ref{paper_structure}.

\section{Related surveys}

\begin{figure}[t]
\centering
\includegraphics[width=\linewidth]{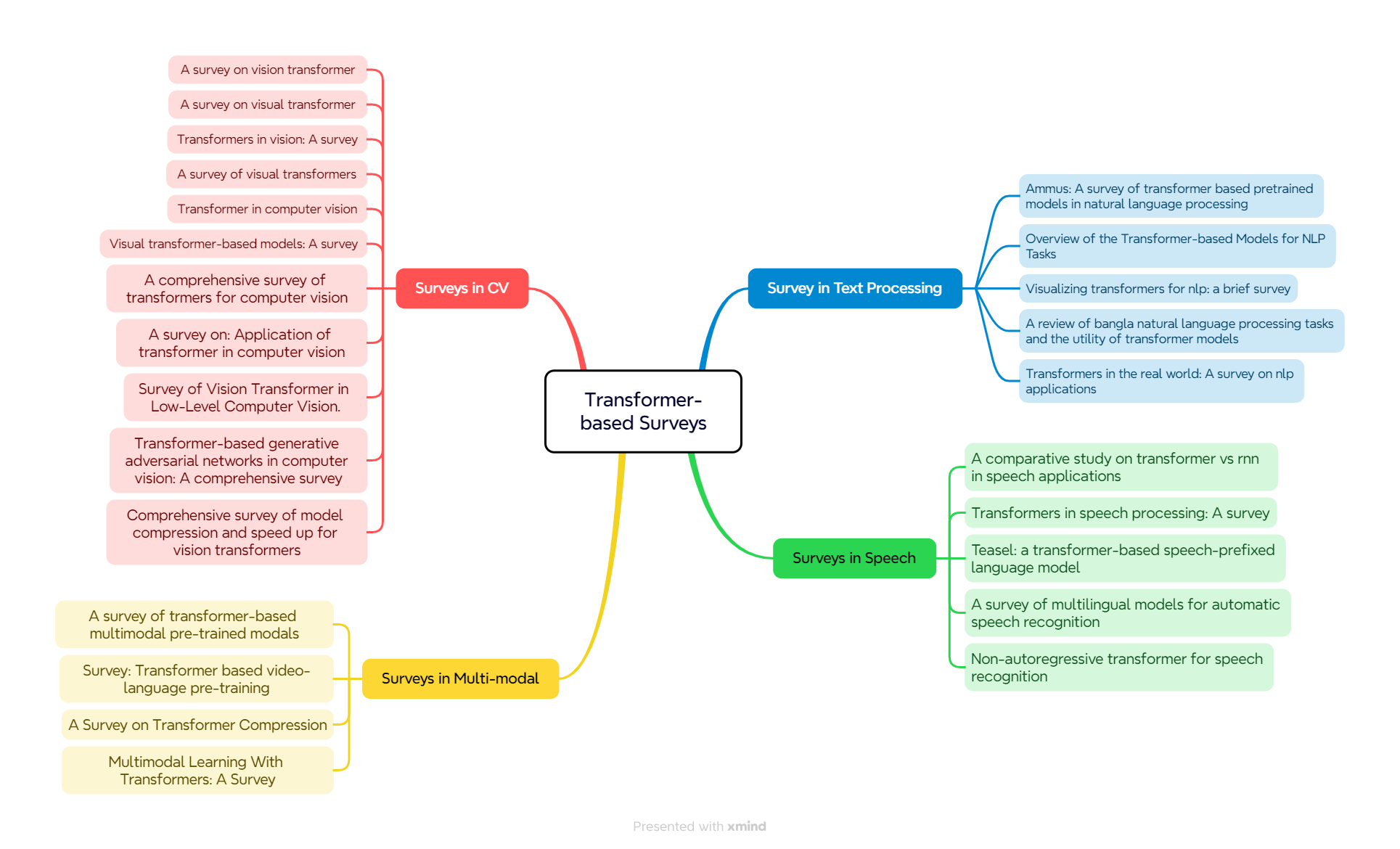}
\caption{Recent surveys related to text, vision, speech, and multi-modality.}
\label{fig_surveys}
\end{figure}

Numerous studies have investigated TB models in the areas of text\cite{Back_2_zhang2022overview}, vision \cite{back_7_lee2021visual}, speech processing \cite{back_18_johnson2021survey}, and multi-modality \cite{back_23_brown2022videolanguage} (Fig. \ref{fig_surveys}). This section reviews these survey papers by categorizing them according to their respective modalities.

\textbf{Surveys in NLP:} In recent years, TB pre-trained models have revolutionized the field of natural language processing (NLP), enabling significant advancements in various applications. Comprehensive surveys such as those by Ammus et al. \cite{back_1_ammus2021survey} (2021) and Zhang et al. \cite{Back_2_zhang2022overview} have meticulously cataloged the evolution and efficacy of these models across numerous NLP tasks, highlighting their versatility and superior performance. Detailed visualizations by Lee et al. \cite{Back_3_lee2021visualizing} provide intuitive insights into the inner workings and interpretability of transformers, facilitating a deeper understanding of their mechanisms. Furthermore, the applicability of these models in less-resourced languages, as discussed by Rahman et al. \cite{back_4_rahman2021review}, underscores the transformative impact of TBs in global linguistic contexts. Practical deployments of these models in real-world scenarios, examined by Smith et al. \cite{back_5_smith2022transformers}, demonstrate their utility and effectiveness in diverse, real-life applications, reinforcing the importance of continuous innovation in transformer architectures for advancing NLP.

\textbf{Surveys in Computer Vision:} The application of TB models to computer vision has been extensively explored, with numerous surveys providing comprehensive overviews of their evolution, applications, and effectiveness. Surveys such as those by Johnson et al. \cite{back_6_johnson2021vision} and Lee et al. \cite{back_7_lee2021visual} detail the foundational aspects and developments in vision transformers, while Kim et al. \cite{back_8_kim2021vision} and Park et al. \cite{back_9_park2021visual} offer insights into the specific architectural advancements and model variations. Further, Liu et al. \cite{back_10_liu2021vision} and Wang et al. \cite{back_11_wang2021models} examine the utility of transformers in low-level computer vision tasks, and Smith et al. \cite{back_12_smith2021comprehensive} focus on the integration of TB models with generative adversarial networks (GANs). Practical applications and the impact of these models in real-world scenarios are highlighted by Brown et al. \cite{back_13_brown2021application} and Chen et al. \cite{back_14_chen2021lowlevel}, while Davis et al. \cite{back_15_davis2021generative} and Evans et al. \cite{back_16_evans2021compression} delve into the strategies for model compression and efficiency improvements, ensuring their scalability and performance optimization. Collectively, these surveys underscore the transformative impact of vision transformers across various subfields and their potential for future advancements in computer vision.

\textbf{Surveys in Speech Processing:} Recent studies have highlighted the transformative impact of TB models in speech processing. Comparisons with RNNs reveal transformers' superior performance and efficiency (Lee et al., \cite{back_17_lee2021comparative}). Extensive surveys detail their advantages in speech applications (Johnson et al., \cite{back_18_johnson2021survey}) and innovations like the TEASEL model for integrating speech and language processing (Kim et al., \cite{back_19_kim2021teasel}). Additionally, the multilingual capabilities of transformers in automatic speech recognition are emphasized (Chen et al., \cite{back_20_chen2021multilingual}), and non-autoregressive transformers show promise for faster, accurate speech-to-text conversion (Park et al., \cite{back_21_park2021nonautoregressive}). These studies collectively underscore transformers' advancements in model performance, efficiency, and multilingual support.


\textbf{Surveys in multi-modal processing:} TB models have shown significant advancements in multimodal learning, as highlighted by recent surveys. The comprehensive review by Smith et al. \cite{back_22_smith2022multimodal} on TB multimodal pre-trained models emphasizes the integration of different data modalities, enhancing model versatility and performance. Brown et al. \cite{back_23_brown2022videolanguage} provides an in-depth survey on video-language pre-training with transformers, demonstrating their capability to learn rich representations from synchronized video and textual data. Lee et al. \cite{back_24_lee2022compression} examine various techniques for transformer compression, addressing the need for efficient and scalable models in resource-constrained environments. Furthermore, the survey by Davis et al. \cite{back_25_davis2022multimodal} on multimodal learning with transformers explores the synergistic use of multiple modalities, showing substantial improvements in tasks that require comprehensive understanding from diverse data sources. These studies collectively illustrate the transformative potential of TB models in multimodal and pre-training contexts.

\section{Architecture of TB models}

\begin{figure}[!t]
\centering
\includegraphics[width=2in]{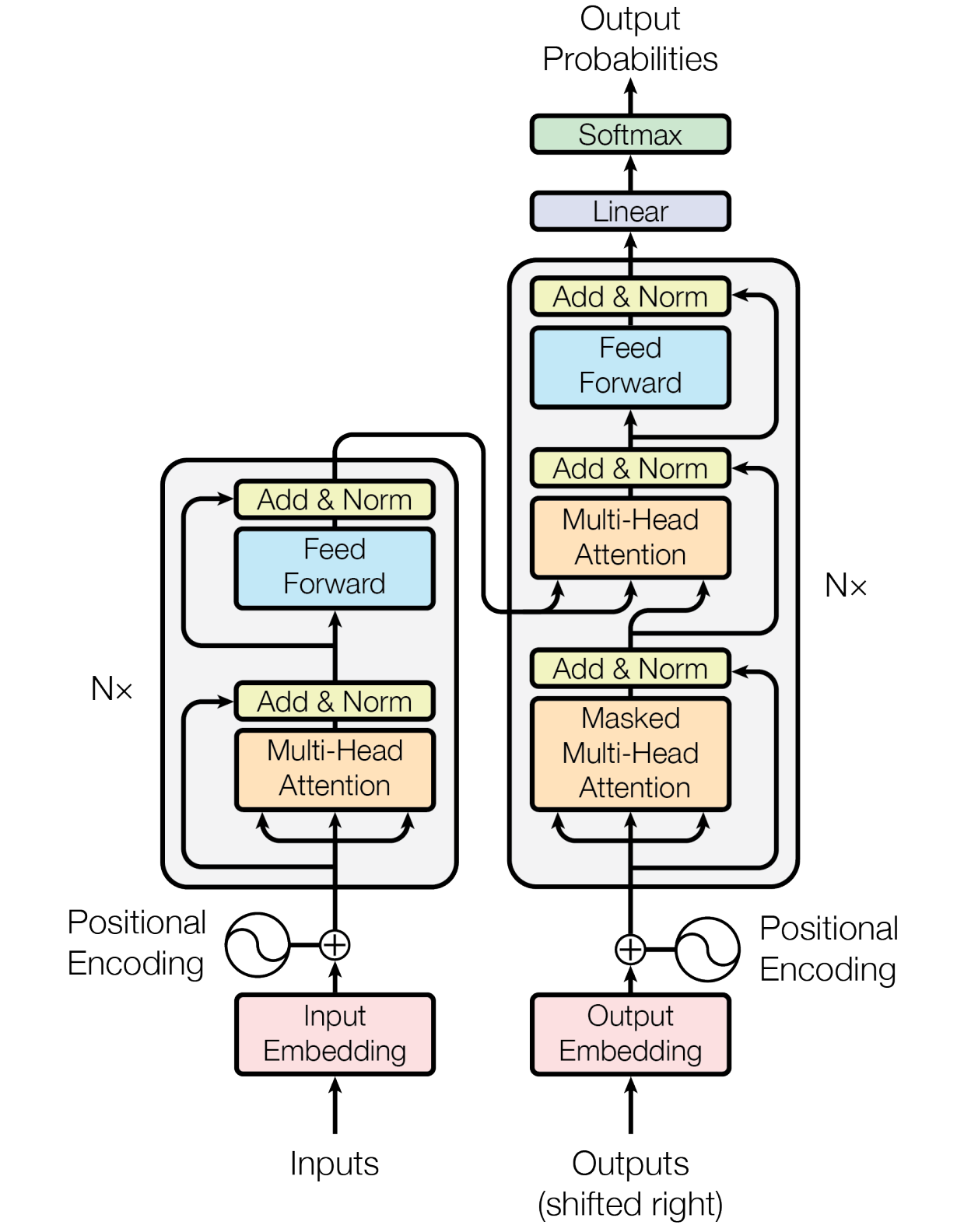}
\DeclareGraphicsExtensions.
\caption{Structure of the Vanilla Transformer \cite{intro_1vaswani2017attention}}
\label{fig_vanilla}
\end{figure}

The transformer is a type of architecture that has revolutionized the field of natural language processing (NLP) and has subsequently made significant impacts across various domains of artificial intelligence (AI). Introduced by Vaswani et al. in 2017 \cite{intro_1vaswani2017attention}, the TB model deviates from earlier sequence-to-sequence architectures by eschewing recurrent (RNN) or convolutional layers. Instead, it relies entirely on a mechanism known as self-attention to weigh the significance of different parts of the input data. The vanilla transformer model consists of two main components: an encoder and a decoder, each composed of a stack of identical layers (Fig. \ref{fig_vanilla}).

\textbf{Self-attention mechanism:} A key breakthrough in the Transformer model is the self-attention mechanism. This feature enables the model to assess the significance of various segments of the input sequence while handling each token. The mechanism functions concurrently across all positions within the sequence.

\textbf{Model architecture:} The Transformer's structure includes an encoder and a decoder, each comprising several layers of multi-head attention and feed-forward neural networks. The self-attention mechanism enables the model to focus on diverse parts of the input sequence, efficiently capturing long-range dependencies. Fig. \ref{fig_vanilla} depicts the basic architecture of the vanilla transformer.

\begin{itemize}
    \item Encoder-decoder: The Transformer model is commonly employed in sequence-to-sequence tasks, where the encoder processes the input sequence and the decoder produces the output sequence.
    \item Layer stacking: Both the encoder and decoder are composed of multiple identical layers stacked vertically. Each of these layers includes self-attention mechanisms and feed-forward neural networks.
\end{itemize}

\textbf{Positional encoding:} Unlike RNNs or CNNs, the Transformer lacks built-in positional information. To address this, it adds positional encodings to the input embeddings, providing data about the position of each token in the sequence \cite{intro_1vaswani2017attention}.

\textbf{Multi-head attention:} Rather than using a single attention mechanism, the Transformer utilizes multi-head attention. This technique computes attention multiple times in parallel, each with distinct learned linear projections, allowing the model to focus on various parts of the input sequence for different tasks.

\textbf{Position-wise feed-forward networks:} Following the multi-head attention mechanism, a position-wise feed-forward network is applied independently to each position within the sequence.

\textbf{Residual connections and layer normalization:} To combat the vanishing gradient problem, residual connections are employed. Additionally, layer normalization is applied after each sub-layer within each layer.

\section{Natural Language Processing}

In the rapid development of NLP, Pretrained Language Models (PLMs) \cite{NLP_1_hu2023survey} have established new benchmarks in performance across a range of linguistic tasks. In this section, we will first illustrate the architecture of NLP transformers and highlight prominent models. Subsequently, we will study various downstream tasks in NLP. Finally, we will show the application of TB models in converting textual data to visual or speech modalities.

\subsection{TB Architecture in NLP}
In NLP, three types of TB models namely Encoder-only, Decoder-only, and Encoder-Decoder, which will be explained.
\subsubsection{Encoder-only}
Encoder-only architectures within the domain of PLMs endeavor to encapsulate the entirety of semantic and contextual data present within a text corpus, subsequently transforming this information into a condensed feature vector representation. The most popular encoder-only architecture is BERT, proposed by Devlin et al. \cite{NLP_2_devlin2018bert}. It represents a paradigm shift as a pre-trained model that leverages the TB's architecture and is subsequently fine-tuned to excel in various NLP applications, including sentiment analysis, entity recognition, and question-answering. BERT's pre-training involves predicting masked tokens in a sentence, allowing the model to learn deep bidirectional context. This methodology enables BERT to capture nuanced word relationships, substantially improving performance across diverse NLP tasks.
RoBERTa (A Robustly Optimized BERT Pretraining Approach) \cite{NLP_4_liu2019roberta} developed by Facebook AI and builds upon the foundation laid by BERT, employing dynamic masking patterns and eliminating the next sentence prediction objective. It further refines the pre-training process through larger mini-batches and more data, resulting in improved performance on downstream tasks. A smaller, faster, cheaper, and lighter version of BERT, DistilBERT \cite{NLP_7_sanh2019distilbert} is trained by distilling BERT’s knowledge into a smaller model that retains most of its predecessor's performance but with significantly reduced size and complexity.  ALBERT (A Lite BERT) \cite{NLP_8_lan2019albert} is another variant of BERT that aims to reduce the model size drastically while maintaining performance. Another PLM is Electra \cite{NLP_9_clark2020electra} which uses a masked language model for pretraining and uses a setup where it replaces some of the tokens in an input sequence with incorrect ones and trains the model to distinguish between the "real" and "fake" tokens. 

\subsubsection{Decoder only}

Decoder-only PLMs are designed primarily for generating text, capable of producing coherent and contextually relevant sequences of text based on a given prompt. The introduction of GPT (Generative Pretrained Transformer) by Radford et al. \cite{NLP_3_radford2018improving} marked another significant stride in transformer evolution. As a generative model pre-trained on extensive textual data, GPT's objective is to forecast subsequent tokens based on the preceding context. Exhibiting proficiency in text synthesis, language modeling, and question-answering, GPT differs from BERT in its generative training objective, which endows it with a broader, more holistic grasp of linguistic patterns. Another extension of the original TB model, Transformer-XL \cite{LM_3_dai2019transformer} introduces a mechanism to handle long-term dependencies, enabling the model to remember information from much earlier in the text than standard TB models. XLNet's \cite{QA_5_yang2019xlnet} architecture allows it to function effectively in a generative capacity as well. It combines the best of both worlds: the bidirectional context modeling of BERT and the generative capabilities of models like GPT. 
CTRL (Conditional Transformer Language Model) \cite{NLP_10_keskar2019ctrl} is developed by Salesforce, CTRL is a decoder-only model that generates text conditioned on control codes that specify domain, style, topic, dates, and other attributes. Moreover, the Reformer model \cite{QA_2_kitaev2020reformer} introduces efficiency improvements that enable the processing of very long documents, significantly reducing memory usage and computation time without sacrificing the quality of text generation.

\subsubsection{Encoder–Decoder}

Encoder-decoder PLMs are designed to handle a wide array of complex NLP tasks that involve both understanding input text (encoding) and generating new text based on that understanding (decoding). These models have been pivotal in advancing the capabilities of NLP applications, from machine translation to summarization and question-answering. Facebook AI developed BART (Bidirectional and Auto-Regressive Transformers) \cite{NLP_11_lewis2019bart}. BART combines bidirectional encoding (similar to BERT) with autoregressive decoding (similar to GPT), making it particularly effective for text generation tasks that require a deep understanding of context, such as summarization and translation. mBART (multilingual BART) \cite{NLP_6_chipman2022mbart}, an extension into multilingual contexts, is a sequence-to-sequence model pre-trained on large-scale monolingual corpora across multiple languages. This pre-training gives mBART the deep understanding of linguistic subtleties it needs to do translation work, even in languages with few resources. This makes the benefits of advanced NLP models available to everyone, regardless of language. Google introduced T5 or Text-to-Text Transformer \cite{NLP_5_ni2021sentence}. It redefines the paradigm by framing all NLP tasks as a text-to-text problem. The model handles every task, from translation to summarization, by converting one type of text into another using a consistent approach. This innovative perspective has simplified the application of transfer learning in NLP. In addition, BigBird \cite{NLP_11_lewis2019bart} is an encoder-decoder model that proposes a sparse attention mechanism, which reduces complexity and time consumption for tasks such as question answering and summarization.


\begin{table*}[h!]
\centering
\caption{Summary of NLP downstream tasks: The first column shows the task, and the second mentions the popular methods related to the task. The rest of the columns classify each method based on Attention Mechanism Types, TB Architectures, Base Models, and Datasets.}
\label{text_app}
\resizebox{\textwidth}{!}{%
\begin{tabular}{cccccc}
\hline
\textbf{NLP Task} & \textbf{Method} & \textbf{Attention Mechanism Type} & \textbf{Architecture type} & \textbf{Base Model} & \textbf{Dataset} \\ \hline

\multirow{3}{*}{Language Modeling} & Autoprompt \cite{LM_2_shin2020autoprompt} & Masked attention & Decoder-only & GPT-3 & OpenAI GPT-3 Dataset \\ 
& Transformer-XL \cite{LM_3_dai2019transformer} & Self-attention & Decoder-only & Transformer-XL & WikiText-103 \\ 
& Dynamic Evaluation \cite{LM_4_krause2019dynamic} & Self-attention & Decoder-only & RNNs & Penn Treebank \\ \hline

\multirow{7}{*}{Question Answering} & Reformer \cite{QA_2_kitaev2020reformer} & Multi-head Attention & Encoder-Decoder & Reformer & SQuAD \\ 
& SDNet \cite{QA_3_zhu2018sdnet}  & Self-attention & Encoder-only & BERT & SQuAD \\ 
& TANDA \cite{QA_4_garg2020tanda} & Masked multi-head attention & Encoder-only & RoBERTa & GLUE \\ 
& XLNet \cite{QA_5_yang2019xlnet} & Query-stream self-attention & Encoder-only & Transformer-XL & SQuAD \\ 
& TOD-BERT \cite{QA_7_wu2020tod} & Multi-head attention & Encoder-only & BERT & MultiWOZ \\ 
& DIALOGPT \cite{QA_8_zhang2019dialogpt} & Multi-head attention & Decoder-only & GPT-2 & Reddit Conversations \\ 
& SOLOIST \cite{QA_9_peng2020soloist} & Multi-head attention & Encoder-Decoder & GPT-2 & Taskmaster \\ \hline

\multirow{2}{*}{Machine Translation} & PIA \cite{translation1} & PIA attention & Encoder-Decoder & Custom & WMT 2014 \\ 
& Interacting-head attention \cite{translation2} & Multi-head attention & Encoder-Decoder & Custom & WMT 2014 \\ \hline

\multirow{5}{*}{Text Classification} & SCIBERT \cite{TC_7_beltagy2019scibert} & Multi-head attention & Encoder-only & BERT & PubMed \\ 
& ClinicalBERT \cite{TC_8_huang2019clinicalbert} & Multi-head attention & Encoder-only & BERT & MIMIC-III \\ 
& BioBERT \cite{TC_9_lee2020biobert} & Multi-head attention & Encoder-only & BERT & PubMed \\ 
& MalBERT \cite{TC_10_rahali2021malbert} & Multi-head attention & Encoder-only & BERT & MalSent \\ 
& BiTransformer \cite{TC_12_tezgider2022text} & Multi-head attention & Encoder-only & BERT & IMDb \\ \hline

\multirow{2}{*}{Text Summarizing} & Longformer \cite{TS_2beltagy2020longformer} & Sliding window attention & Encoder-Decoder & BERT & arXiv, PubMed \\ 
& Primer \cite{TS_4_xiao2021primera} & Sliding window attention & Encoder-Decoder & BigBird & arXiv \\ \hline

\multirow{5}{*}{Named Entity Recognition} & FLAT \cite{NER_2_li2020flat} & Self-attention & Encoder-only & BERT & OntoNotes 4 \\ 
& FinBERT-MRC \cite{NER_3_yang2020finbert} & Multi-head attention & Encoder-only & BERT & FiQA, Financial NER \\ 
& Wojood \cite{NER_5_jarrar2022wojood} & Multi-head attention & Encoder-only & BERT & Wojood Corpus \\ 
& GeoBERT \cite{NER_6_liu2022few} & Multi-head attention & Encoder-only & BERT & GeoCorpus \\ 
& ALBERT-BiLSTM-CRF \cite{NER_7_ren2022named} & BiLSTM-CRF & Encoder-only & ALBERT & CoNLL-2003 \\ \hline

\multirow{4}{*}{Sentiment Analysis} & KEAHT \cite{SA_1_Tiwari2022} & Multi-head attention & Encoder-only & RoBERTa & SST-2 \\ 
& TextGT \cite{SA_2_zhang2022textgt} & Multi-head attention & Encoder-only & Custom & SemEval 2014 \\ 
& RoBERTa-LSTM \cite{SA_3_chen2022roberta}  & Multi-head attention & Encoder-only & RoBERTa & IMDb \\ 
& BMT-Net \cite{SA_4_li2022bmt} & Multi-head attention & Encoder-only & Custom & Sentiment140 \\ \hline

\end{tabular}%
}
\end{table*}

\subsection{NLP downstream tasks}

NLP has many real-world applications. We will discuss language modeling, question answering, machine translation, text classification, and text summarization. Table \ref{text_app} categorizes these NLP tasks using transformer-based models, detailing the attention mechanism, transformer variant, and underlying model for each method.

\subsubsection{Language Modeling} 

    Language Modeling (LM) is a key task in NLP focused on predicting the next word in a text sequence based on the context of preceding words\cite{LM_4_krause2019dynamic}. It is essential for many NLP applications like machine translation, speech recognition, and text generation. Notable advancements in LM include Autoprompt\cite{LM_2_shin2020autoprompt}, Transformer-XL\cite{LM_3_dai2019transformer}, and Dynamic Evaluation\cite{LM_4_krause2019dynamic}. Autoprompt enhances knowledge extraction by automating prompt generation. Transformer-XL improves the handling of long-term dependencies with a recurrence mechanism and new positional encoding. Dynamic Evaluation adapts the model parameters dynamically to better suit domain-specific or stylistically varied content.
    
\subsubsection{Question Answering}

    In the dynamic field of question-answering (QA) models \cite{QA_3_zhu2018sdnet}, several key architectures have made significant contributions. SDNet (Semantic Decoding Network) \cite{QA_3_zhu2018sdnet} incorporates semantic parsing for better question comprehension, as demonstrated on the CoQA dataset \cite{QA_10_reddy2019coqa}. XLNet \cite{QA_5_yang2019xlnet}, introduced by Yang et al., uses a permutation-based training strategy to capture bidirectional context, outperforming earlier models on benchmarks like SQuAD \cite{QA_6_rajpurkar2016squad}. DIALOGPT \cite{QA_8_zhang2019dialogpt}, built on the GPT-2 architecture, is fine-tuned on extensive dialogue data to generate coherent conversational responses. The Reformer model \cite{QA_2_kitaev2020reformer} optimizes attention mechanisms for processing long sequences, aiding QA tasks with extensive contexts. TANDA (Transfer and Adapt) \cite{QA_4_garg2020tanda} improves pre-trained models like BERT for specific QA tasks through a two-step fine-tuning process. TOD-BERT \cite{QA_7_wu2020tod}, designed for task-oriented dialogue systems, enhances performance by fine-tuning on diverse dialogue datasets. SOLOIST \cite{QA_9_peng2020soloist} combines language generation and task completion in a unified framework, improving dialogue system robustness and accuracy.

    \subsubsection{Machine Translation}

    Machine translation (MT) involves the automatic translation of text from one language to another. MT models typically use an encoder-decoder structure that features a bidirectional encoder for effective context understanding and a decoder that produces text of variable lengths, based on the foundational design of the Transformer-based (TB) architecture. Elaffendi et al. introduced PIA \cite{translation1}, which converts natural language sentences into unique binary attention context vectors, capturing semantic context and word dependencies. Dongxing et al. \cite{translation2} refined TB for MT by introducing an interacting-head attention mechanism, overcoming the low-rank bottleneck by optimizing the number of attention heads and promoting extensive interactions among them through computations in low-dimensional subspaces across all tokens.

    \subsubsection{Text Classification}

    
    Text classification is an essential task in the field of natural language processing as it forms the baseline upon which other methodologies are constructed. TB models have emerged as the leading approach for text classification, boasting considerable success in recent years \cite{TC_12_tezgider2022text}. One notable model, SCIBERT \cite{TC_7_beltagy2019scibert}, leverages the BERT framework, pre-trained on a broad array of scientific literature, to overcome the challenges posed by the scarcity of high-caliber labeled data in the scientific domain. SCIBERT's pre-training enables enhanced performance on specialized scientific NLP tasks. Similarly, ClinicalBert \cite{TC_8_huang2019clinicalbert} applies BERT's bidirectional capabilities to the analysis of clinical notes, achieving superior results in predicting hospital readmission and discovering medical concept relationships. Moreover, BioBERT \cite{TC_9_lee2020biobert} is a domain-specific representation model pre-trained on biomedical texts, which surpasses BERT and other leading models across various biomedical text mining tasks. Beyond these domains, TB models find intriguing applications in cybersecurity, as seen with MalBERT \cite{TC_10_rahali2021malbert}, which leverages BERT's pre-trained model for malware classification using textual features extracted from application source codes. Additionally, Murat et al. \cite{TC_12_tezgider2022text} introduced the BiTransformer, a novel model utilizing dual Transformer encoder blocks with bidirectional position encoding to enhance text classification tasks by refining the attention mechanisms. 
    
    \subsubsection{Text Summarization}

    The field of text summarization has significantly progressed by adapting Transformer-based (TB) models to manage various text lengths and contexts \cite{TS_6_ghalandari2022efficient}. Recent efforts have focused on modifying existing text-to-text models for extended narratives. A key development in this area is the Longformer \cite{TS_2beltagy2020longformer}, a TB model optimized for long-document processing with an efficient attention mechanism capable of handling larger contexts. Xiao et al. \cite{TS_4_xiao2021primera} introduced PRIMERA, a pyramid-based pretraining technique for multi-document summarization that uses masked sentence pretraining to improve summary coherence and informativeness. For query-focused summarization, Xu et al. Additionally, Ghalandari et al. \cite{TS_6_ghalandari2022efficient} explored the use of reinforcement learning to fine-tune TB models for sentence compression, enhancing summarization efficiency by emphasizing brevity and content salience.
        
    \subsubsection{Sentiment Analysis}
    
    Sentiment analysis in NLP involves determining the emotional tone or attitude expressed in a piece of text. Dimple et al. \cite{SA_3_chen2022roberta} proposed KEAHT, a knowledge-enriched attention-based hybrid TB model for social sentiment analysis (SA), which enhances explicit knowledge using lexicalized domain ontology and latent Dirichlet allocation (LDA) topic modeling. BERT was utilized to train the corpus. This method effectively addresses complex text issues and incorporates an attention mechanism. Zhang et al.'s "TextGT" \cite{SA_2_zhang2022textgt} introduces a double-view graph transformer for aspect-based sentiment analysis, incorporating both syntactic and semantic structures for comprehensive sentiment understanding. Chen et al. combine RoBERTa and LSTM in "RoBERTa-LSTM \cite{SA_3_chen2022roberta}," utilizing the strengths of both models for enhanced sentiment analysis performance. Lastly, Li et al.'s "BMT-Net" \cite{SA_4_li2022bmt} demonstrates the power of multitask learning by employing a broad multitask transformer network for robust sentiment analysis.

    \subsubsection{Named Entity Recognition}

    Named Entity Recognition (NER) involves identifying and classifying entities within text into predefined categories, such as businesses, locations, dates, numbers, and people \cite{NER_6_liu2022few}. Li et al. \cite{NER_2_li2020flat} introduced FLAT, a TB model for Chinese NER, which converts lattice structures into flat spans where each span represents a character or latent word along with its position in the original lattice. Zhang et al. \cite{NER_3_yang2020finbert} developed FinBERT-MRC, a BERT-based financial NER model within the machine reading comprehension framework. Jarrar et al. \cite{NER_5_jarrar2022wojood} presented Wojood, an Arabic NER corpus recognized using BERT, employing the pre-trained ARaBERT to train a nested NER model through multi-task learning. Liu et al. \cite{NER_6_liu2022few} proposed a two-stage fine-tuning method for BERT tailored for NER in the geological domain, resulting in GeoBERT, which was initially fine-tuned on a pre-trained BERT model and then on a small dataset for geological reports. Kezhou et al. \cite{NER_7_ren2022named} introduced an ALBERT-based model, combining it with BiLSTM and Conditional Random Field (CRF) to create the ALBERT-BiLSTM-CRF model.

\subsection{Text to Vision}

\begin{figure}[t]
\centering
\includegraphics[width=3.5in]{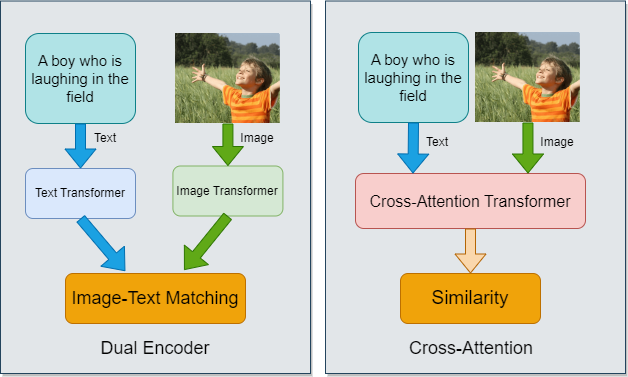}
\DeclareGraphicsExtensions.
\caption{Dual encoder and cross-attention models in Text to vision approaches.}
\label{Text2Vision}
\end{figure}

Text-to-vision TB models aim to take text input and produce a corresponding image or video. There are two main types of text-to-vision transformers dual-encoder and cross-attention. The dual-encoder methods, which involve separately mapping text and vision into a shared embedding space, are appealing for their scalability in retrieval and efficiency in handling billions of images through approximate nearest-neighbor searches. Fast models, referred to as dual encoders (as shown on the left side of Fig. \ref{Text2Vision}), evaluate the input image and text separately to calculate a similarity score using a single dot product. This score may be efficiently indexed, allowing for large-scale search. Conversely, slow models, which are also referred to as cross-attention models (as shown on the right side of Fig. \ref{Text2Vision}), simultaneously analyze the input image and text using cross-modal attention in order to calculate a similarity score.


There are three different applications of text-to-vision TB models: story visualization, text-to-image, and text-to-video. The summary of all the methods for these applications is provided in Table \ref{text-to-vision_table}.

\begin{table*}[ht]
\centering
\small
\caption{Categorization of TB methods in text-to-vision modality. Methods related to story visualization, text-to-image methods, and text-to-video methods.}
\label{text-to-vision_table}
\begin{tabular}{>{\centering\arraybackslash}m{2.5cm} >{\centering\arraybackslash}m{0.8cm} >{\centering\arraybackslash}m{1.2cm} >{\centering\arraybackslash}m{1.8cm} >{\centering\arraybackslash}m{3.8cm} >{\centering\arraybackslash}m{3cm}}
\toprule
\textbf{Method Name} & \textbf{Year} & \textbf{Base Model} & \textbf{Architecture Type} & \textbf{Key Features} & \textbf{Datasets} \\ \midrule
StoryGAN & 2019 & GAN & - & Global consistency with story-level discriminator & CLEVR-SV, PororoSV, FlintstonesSV \\
\addlinespace[0.5em]
CP-CSV & 2020 & GAN & - & Character coherence by separating figure and background & CLEVR-SV, PororoSV \\
\addlinespace[0.5em]
VP-CSV & 2022 & Transformer & - & Two-stage generation for characters and backgrounds & CLEVR-SV, PororoSV \\
\addlinespace[0.5em]
StoryDALL-E & 2022 & DALL-E & Cross-Attention & Conditioned on the first image, fine-tuned for story continuation & CLEVR-SV, PororoSV \\
\midrule
DALL-E & 2021 & Transformer & Cross-Attention & Zero-shot coherent image generation from text & Conceptual Captions, YFCC100M, custom 250M pairs \\
\addlinespace[0.5em]
DALL-E2 & 2022 & CLIP & Dual-Encoder & Improved quality with hierarchical text-conditional generation & Custom 650M pairs \\
\addlinespace[0.5em]
CogView & 2021 & Transformer & Cross-Attention & High-resolution generation with discrete VAE & Wukong dataset \\
\addlinespace[0.5em]
Make-A-Scene & 2022 & Conditional GAN & - & User-provided sketches for controlled generation & Custom sketch-based datasets \\
\addlinespace[0.5em]
Imagen & 2022 & Diffusion models & Cross-Attention & Photorealistic generation with state-of-the-art quality & Custom diverse pairs \\
\midrule
GODIVA & 2021 & Transformer & Cross-Attention & Efficient mapping of text to video tokens & Custom diverse pairs \\
\addlinespace[0.5em]
NUWA & 2022 & Diffusion models & Cross-Attention & High fidelity image and video generation from text & Custom large-scale datasets \\
\addlinespace[0.5em]
CogVideo & 2022 & Transformer & Cross-Attention & High-resolution video generation from text & Custom diverse pairs \\
\addlinespace[0.5em]
Make-A-Video & 2022 & Transformer & Cross-Attention & Video generation without paired text-video data & Custom leveraging existing models \\
\addlinespace[0.5em]
Phenaki & 2022 & Transformer & Cross-Attention & Long-duration video generation from paragraphs & Custom large-scale dataset \\
\bottomrule
\end{tabular}
\end{table*}

\textbf{Story Visualization:} Several recent methods have been proposed to enhance consistency and semantic matching in story-based image generation. StoryGAN \cite{TTV_1_li2019storygan} employed a story-level discriminator to improve global consistency, while CP-CSV \cite{TTV_2_song2020character} separated figure and background elements to boost character coherence. VP-CSV \cite{TTV_5_chen2022character} introduced a two-stage approach using a TB model to separately generate characters and complete backgrounds. However, most of these methods neglect the importance of encoding the story's narrative. Maharana et al. \cite{TTV_7_maharana2022storydall} recently introduced a new task setup called story continuation, where they conditioned on the first image and fine-tuned DALL-E \cite{TTV_6_ramesh2021zero} for the StoryDALL-E \cite{TTV_7_maharana2022storydall} model.

\textbf{Text-to-image generation}, a subset of story visualization, has traditionally focused on enhancing semantic relevance and resolution. Recent advances in text-driven image creation have been achieved through extensive training data and large-scale models like DALL-E \cite{TTV_6_ramesh2021zero}, its successor DALL-E2 \cite{TTV_11_ramesh2022hierarchical}, CogView \cite{TTV_8_ding2021cogview}, and Make-A-Scene \cite{TTV_9_gafni2022make}, which incorporate sketch input. Despite their ability to produce high-quality images, text-to-image models sometimes struggle to encode context, particularly with metaphorical phrases across multiple sentences. Furthermore, utilizing advanced models can be computationally challenging due to their large size. For instance, diffusion-based models such as Imagen \cite{TTV_10_saharia2022photorealistic} and DALL-E2 \cite{TTV_11_ramesh2022hierarchical} contain 2 billion and 3.5 billion parameters, respectively, limiting their use in resource-constrained inference scenarios. 

\textbf{Text to video}
Creating videos from text involves producing multiple frames based on textual input, like story visualization. Recent advancements have led to sophisticated video generation models like GODIVA \cite{TTV_15_wu2021godiva} and NUWA \cite{TTV_16_wu2022nuwa}. Moreover, contemporary research has succeeded in creating high-resolution videos with sequential frames of superior quality \cite{TTV_17_hong2022cogvideo, TTV_18_singer2022make}. Typically, cutting-edge text-to-video models (CogVideo \cite{TTV_17_hong2022cogvideo} and Make-A-Video \cite{ TTV_18_singer2022make}) create videos from a single sentence, often featuring uniform backgrounds. The Phenaki method \cite{TTV_20_villegas2022phenaki} now allows for video generation based on extensive paragraphs. This method, while capable of generating longer-duration videos, demands a very large model trained on extensive data and a detailed paragraph with closely timed scene descriptions. 

\subsection{Text to Speech}
Speech-to-text technologies convert written text into spoken words, usually producing mel-spectrum and phonemes. These technologies have numerous applications, including chatbots and voice assistants. Various neural architectures have been used for this purpose, but we will focus solely on transformer-based architecture. We will discuss Tacotron 1 \& 2 in section \ref{sec:sp} in detail. These two methods have been designed for speech-to-text tasks. In addition to Tacotron 1 \& 2, FastSpeech \cite{ren2019fastspeech} is another popular method for this problem. FastSpeech uses a parallel setting for mel-spectrogram generation based on transformer blocks, which reduces the inference time significantly. It also focuses on increasing the robustness of the generation. In the past, an autoregressive setting caused propagated errors in generation, possibly due to incorrect attention alignments between text and speech. To address this, FastSpeech utilizes a phoneme duration predictor to ensure accurate alignment between text and speech, thereby increasing robustness. 

FastSpeech was tested on the LJSpeech dataset \cite{ljspeech17}. It matches the quality of autoregressive models while significantly accelerating mel-spectrogram generation by 270 times and end-to-end speech synthesis by 38 times. FastSpeech 2 \cite{ren2020fastspeech} is a natural extension of FastSpeech. It uses a more straightforward training pipeline to reduce the training time. FastSpeech2 incorporates more information, including pitch and energy, to improve quality and accuracy. They also introduced FastSpeech 2s \cite{ren2020fastspeech}, the first system to convert waveform from text. Both FastSpeech 2 and 2s perform better than FastSpeech 2 and autoregressive models.

FASTPITCH \cite{lancucki2021fastpitch} is also based on the FastSpeech model. FASTPITCH aims to improve synthesized speech quality by integrating conditioning based on fundamental frequency estimation for each input symbol, eliminating the need for knowledge distillation of the mel-spectrogram. In addition to the methods mentioned above, there are other TB approaches, such as Durian \cite{yu2019durian}, MultiSpeech \cite{chen2020multispeech}, and s-Transformer \cite{wang2020s}, that have been utilized for this task. A summary of the discussed methods is presented in Table \ref{table:texttospeech}.

\begin{table}[ht]
\centering
\small
\caption{Categorization of TB methods in text-to-speech tasks.}
\label{table:texttospeech}
\begin{tabular}{>{\centering\arraybackslash}m{2.5cm} >{\centering\arraybackslash}m{1cm} >{\centering\arraybackslash}m{2cm} >{\centering\arraybackslash}m{3.5cm} >{\centering\arraybackslash}m{2.5cm}} 
\toprule
\textbf{Method Name} & \textbf{Year} & \textbf{Base Model} & \textbf{Advantages} & \textbf{Dataset} \\ 
\midrule
Tacotron 1 \cite{wang2017tacotron} & 2017 & Seq2Seq + Attention & High-quality speech generation from characters, faster generation using frames. & Internal US English  \\
Tacotron 2 \cite{shen2018natural} & 2018 & Seq2Seq + Attention & Improved speech synthesis quality with WaveNet vocoder. & Internal US English \cite{wang2017tacotron}  \\
FastSpeech \cite{ren2019fastspeech} & 2019 & Transformer & Parallel generation of mel-spectrogram using a transformer, reduced inference time by 270 times. & LJSpeech  \\
FastSpeech 2 \cite{ren2020fastspeech} & 2020 & Transformer & Enhanced quality with pitch and energy information, reduced training time. & LJSpeech \\
FastSpeech 2s \cite{ren2020fastspeech} & 2020 & Transformer & Direct waveform generation from text, improved over autoregressive models. & LJSpeech  \\
FASTPITCH \cite{lancucki2021fastpitch} & 2021 & Transformer & Fully parallel, conditioning on frequency contours for improved speech quality and better semantic match. & LJSpeech \\
Durian \cite{yu2019durian} & 2019 & WaveRNN + RNN & Enhanced alignment accuracy using duration-based attention, multimodal with face expression generation. & Internal datasets \\
MultiSpeech \cite{chen2020multispeech} & 2020 & Transformer & Improved robustness in multi-speaker text-to-speech synthesis. & VCTK Corpus \\
\bottomrule
\end{tabular}
\end{table}

\section{Computer Vision}
\subsection{Vision Transformer}
\begin{figure*}[!t]
\centering
\begin{subfigure}{0.35\textwidth}
    \centering
    \includegraphics[width=1in]{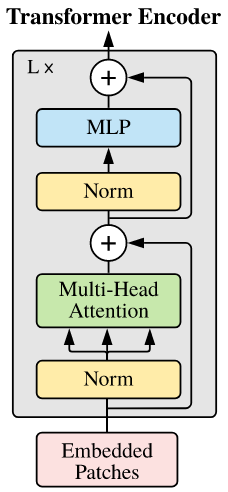}
    \caption{Vanilla vision transformer \cite{dosovitskiy2010image}}
    \label{vanilla_vit}
\end{subfigure}
\hfill
\begin{subfigure}{0.6\textwidth}
    \centering
    \includegraphics[width=1\textwidth]{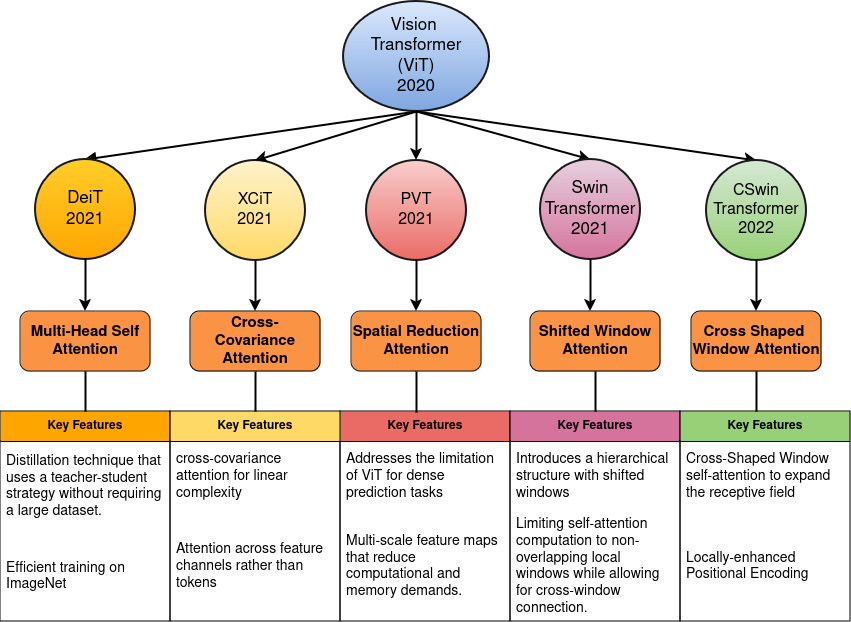}
    \caption{Different types of vision transformers and Attention mechanisms}
    \label{vit_survey}
\end{subfigure}
\caption{Vision Transformers}
\label{ref:vit}
\end{figure*}

After the success of the vanilla transformer in NLP tasks, the Vision Transformer (ViT) was introduced, catering specifically to image-based modalities \cite{dosovitskiy2010image}. This adaptation of transformers for image processing marked a significant shift in the approach to computer vision problems. An overview of the vision Transformer has been shown in (Fig \ref{vanilla_vit}).


The Vision Transformer re\-imagines an image not as a grid of pixels but as a sequence of flattened 2D patches. These patches are analogous to tokens (words) in NLP tasks. Unlike the vanilla transformer, where embeddings are directly related to the tokens of a textual sequence, in ViT, the embeddings represent the information contained in each image patch.

\begin{figure}[!t]
\centering
\includegraphics[width=3in]{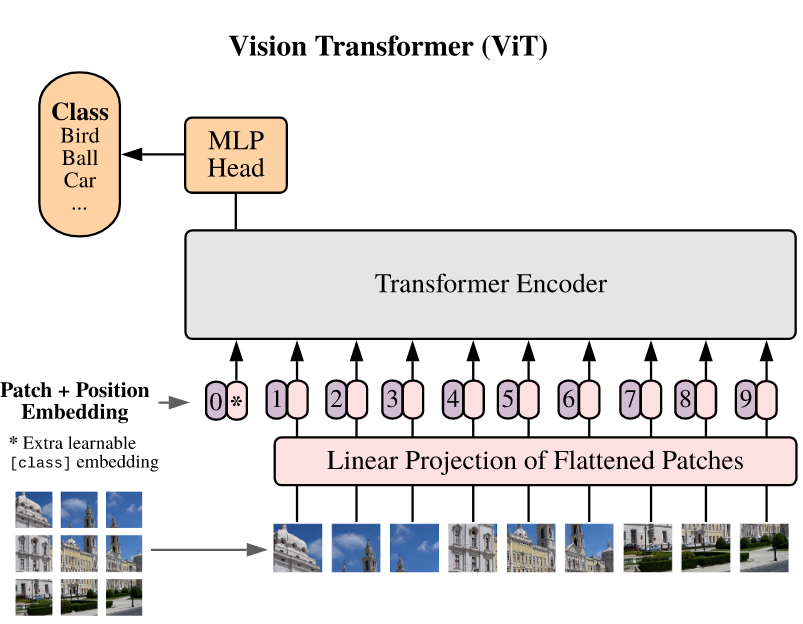}
\DeclareGraphicsExtensions.
\caption{Linear projection of flattened patch \cite{dosovitskiy2010image}}
\label{flat_patch}
\end{figure}

An input image $\mathbf{x} \in \mathbb{R}^{H \times W \times C}$ is first divided into a set of fixed-size patches $\mathbf{x}_p \in \mathbb{R}^{N \times (P^2 \cdot C)}$. Here, $(H, W)$ denotes the original image resolution, $C$ is the number of channels, $(P, P)$ is the size of each patch, and $N = HW / P^2$ is the number of patches \cite{dosovitskiy2010image}. These patches are then linearly embedded into a $D$-dimensional space to create patch embeddings. This embedding process is different from the embedding mechanism in the vanilla transformer, reflecting the adaptation required for image data. The sequence of patch embeddings then serves as the input for the subsequent layers of the TB model. The TB models utilize a multi-head self-attention mechanism. In the context of ViT, this mechanism is applied to the sequence of patch embeddings. Each head in the multi-head attention module can focus on different parts of the image, capturing diverse features from the patch embeddings. This approach allows the model to consider both local and global information from the image. ViT also explores a hybrid architecture where the input sequence is derived from the feature maps of a CNN \cite{lecun1989backpropagation}. In this configuration, the patch embedding projection, which is typically applied to raw image patches, is instead applied to patches extracted from CNN feature maps. In a particular instance, these patches can be as small as 1x1, indicating that the input sequence is created by flattening the spatial dimensions and then projecting them into the Transformer dimension, as shown in Figure \ref{flat_patch}. This hybrid approach allows for the incorporation of CNN's strengths in extracting local features and hierarchical representations while still leveraging the global self-attention mechanism of the transformer.

In computer vision, the foundational principles established by Vision Transformers (ViT) \cite{dosovitskiy2010image} have paved the way for innovative TB models. Each of these models introduces unique capabilities and addresses specific challenges (Figure \ref{vit_survey}) within the domain:

\paragraph{Training data-efficient image transformers \& distillation through attention (DeiT)} DeiT \cite{touvron2021training} specifically tackles the challenges of training efficiency and resource demands, issues that have been significant hurdles for the application of earlier TB models in image processing tasks. The main achievement of DeiT is demonstrating the feasibility of training high-performing, convolution-free transformers solely on the ImageNet dataset. This training is remarkably efficient, requiring only a single computer and less than three days to complete. This contrasts sharply with previous models that relied on pre-training with hundreds of millions of images and substantial computational resources, limiting their applicability.

\paragraph{Pyramid Vision Transformer (PVT)} Following the trail of the Vision Transformer (ViT) \cite{dosovitskiy2010image}, the Pyramid Vision Transformer (PVT) \cite{wang2021pyramid} emerges as a versatile backbone for various dense prediction tasks in computer vision to address the limitations of ViT in this domain. The architecture of PVT introduces a pyramid structure that generates multi-scale feature maps and incorporates a progressive shrinking pyramid. PVT has shown substantial performance in downstream tasks. When combined with RetinaNet, it achieves a 40.4 AP on the COCO dataset with a parameter count comparable to ResNet50+RetinaNet.

\paragraph{Swin Transformer} The Swin Transformer \cite{liu2021swin} has emerged as a successor to the Vision Transformer. The Swin Transformer introduces a hierarchical design with shifted windows which enhances the efficiency by limiting self-attention to local windows while enabling cross-window connection. This architecture allows for scalable representation at various resolutions by maintaining linear computational complexity regardless of image size. Swin Transformer outperforms the state-of-the-art models across multiple downstream task such as image classification, object detection, and semantic segmentation.

\paragraph{CSwin Transformer} The CSWin Transformer \cite{dong2022cswin} represents a significant advancement in the domain of TB backbones for vision tasks. Its novel Cross-Shaped Window self-attention mechanism computes self-attention across horizontal and vertical stripes in parallel to form a cross-shaped window. This approach outperforms existing encoding schemes by better handling local positional information, empowers the CSWin Transformer to support various input resolutions. CSWin Transformer achieves new state-of-the-art performances on benchmarks like ImageNet-1K, COCO, and ADE20K without relying on extra data or labels. It also exhibits superior speed-accuracy trade-offs compare to prior architectures like Swin Transformer.

\paragraph{Cross-Covariance Image Transformers (XCiT)} The Cross-Covariance Image Transformer (XCiT) \cite{ali2021xcit} introduces an innovative approach to address the computational inefficiencies of TB models. XCiT implements a novel cross-covariance attention (XCA) mechanism that operates across feature channels rather than tokens. This leads to linear complexity with respect to the number of tokens and facilitates efficient processing of high-resolution images. XCiT delivers excellent results on diverse benchmarks, including image classification, object detection, and semantic segmentation. The strategic combination of cross-covariance attention with other models marks a significant advancement in applying TB models to high-resolution image processing.

\begin{table*}[ht]
\centering
\small
\caption{Categorization of TB Models for Image Processing}
\label{table:image_processing_transformers}
\begin{tabular}{>{\centering\arraybackslash}m{1.5cm} >{\centering\arraybackslash}m{2.5cm} >{\centering\arraybackslash}m{2cm} >{\centering\arraybackslash}m{2.5cm} >{\centering\arraybackslash}m{2.5cm} >{\centering\arraybackslash}m{2cm}}
\toprule
\textbf{Application} & \textbf{Transformer Model} & \textbf{Base Model} & \textbf{Key Innovation} & \textbf{Main Contribution} & \textbf{Datasets Used} \\ 
\midrule
Edge Detection & EDTER \cite{pu2022edter} & ViT & Two-stage process with BiMLA and FFM & Combines long-range dependencies and local cues & BSDS500, NYUDv2, Multicue \\ 
\midrule
\multirow{2}{=}{Semantic Segmentation} & Segmenter \cite{strudel2021segmenter} & Swin Transformer & Utilizes global context and mask transformer decoder & Directly translates patch embeddings into class labels & ADE20K, Pascal Context, Cityscapes \\ 
 & SeMask \cite{xie2021segformer} & Swin Transformer & Incorporates semantic context into pretrained backbones & Enhances performance with minimal additional computation & Cityscapes, ADE20K \\ 
\midrule
Object Detection & DETR \cite{carion2020end} & ViT & Direct set prediction, bipartite matching loss & Simplifies detection pipeline, removes need for hand-designed components & COCO \\ 
\midrule
\multirow{2}{=}{Image Super-Resolution} & TTSR \cite{yang2020learning} & ViT & Texture transfer using attention mechanisms & Transfers HR textures from reference images & RefSR, CUFED5 \\ 
 & DAT \cite{chen2023dual} & ViT & Alternating spatial and channel self-attention & Comprehensive context capture and feature aggregation & DIV2K, Flickr2K, Urban100, Manga109 \\ 
\bottomrule
\end{tabular}
\end{table*}

\subsection{Image Processing}
In the domain of Image Processing, TB models are emerging as a transformative force to reshape how visual data is interpreted and utilized. The advancements in transformer architectures, such as Vision Transformer (ViT) \cite{dosovitskiy2010image}, Data-efficient Image Transformer (DeiT) \cite{touvron2021training}, Pyramid Vision Transformer (PVT) \cite{wang2021pyramid}, Cross-Shaped Window Transformer (CSWin) \cite{dong2022cswin}, and Cross-Covariance Image Transformer (XCiT) \cite{ali2021xcit} have provided robust frameworks that significantly outperform traditional convolutional and recurrent networks \cite{khan2022transformers}\cite{shrestha2019review}. These transformer-based models, originally excels in natural language processing, and pave the way for sophisticated image processing applications, from image classification to object detection and beyond \cite{touvron2021training} \cite{hu2021istr}.

A well-known innovation in this domain is the development of the \textit{Image Processing Transformer (IPT)} \cite{chen2021pre}, which leverages the power of pre-training on large-scale datasets. The IPT model utilizes the representation ability of TB models, enhanced with multi-heads, multi-tails, and further augmented by contrastive learning to adapt to different image processing tasks efficiently. The IPT's ability to generalize across tasks, even with limited task-specific data, addresses the challenges of dataset variability and the need for multiple processing modules. With this approach, a single pre-trained model can be fine-tuned to outperform state-of-the-art methods across various low-level benchmarks.

This section aims to discuss the role of TB architectures in image processing domain and study their implementation in various sub-domains. The following sections will further explore the specific contributions and implementations of TB models in image processing tasks.

\subsubsection{Edge Detection} The \textit{Edge Detection TransformER (EDTER)} \cite{pu2022edter} use a novel approach for edge detection and addresses the limitation of traditional convolutional neural networks (CNNs). EDTER mitigate this limitation by combining the transformer's proficiency in capturing long-range dependencies with a two-stage process that preserves detailed local cues. The first stage of EDTER employs a global transformer encoder to take in long-range global context from coarse-grained image patches. The second stage refines this with a local transformer encoder that targets short-range local cues from fine-grained patches. EDTER produces crisp and less noisy edge maps. EDTER outperforms state-of-the-art methods on benchmarks like BSDS500 \cite{amfm_pami2011}, NYUDv2 \cite{Silberman:ECCV12}, and Multicue \cite{mely2016systematic}.

\subsubsection{Semantic Segmentation} The Swin Transformer \cite{liu2021swin}, with its hierarchical structure utilizing shifted windows, demonstrates exceptional adeptness in modeling at various scales, and achieves a new state-of-the-art performance with significant margins. Segmenter \cite{strudel2021segmenter} is built upon the Vision Transformer (ViT) that leverages global context from the first layer, and uses a mask transformer decoder or a point-wise linear decoder to translate patch embeddings directly into class labels. SegFormer \cite{xie2021segformer} introduces a hierarchically structured Transformer encoder and a lightweight MLP decoder to create a system that balances efficiency with high accuracy. SeMask \cite{jain2023semask} further refines this approach by infusing semantic context into pre-trained TB backbones during fine-tuning. It also enhances the performance with minimal additional computational cost. 

\subsubsection{Object Detection} The \textit{End-to-End Object Detection with Transformers (DETR)}'s architecture incorporates a transformer encoder-decoder and a novel set-based global loss enforced by bipartite matching. It is designed to output a unique set of predictions in parallel to address the challenge of duplicate predictions inherent in traditional methods \cite{NLP_2_devlin2018bert, oord2018parallel}. It also captures the intricate relationships between objects and their context within the image to simplify the detection pipeline. DETR demonstrates comparable, if not superior, performance to well-established detection systems like Faster R-CNN \cite{ren2015faster} on benchmarks such as COCO \cite{lin2014microsoft}.

\subsubsection{Image resolution enhancement} The \textit{Learning Texture Transformer Network for Image Super-Resolution (TTSR)} \cite{yang2020learning} introduces a novel approach to image super-resolution (SR) by transferring high-resolution (HR) textures from reference images to low-resolution (LR) ones. This novel Texture Transformer Network (TTSR) utilizes a set of interconnected modules designed for image generation tasks by facilitating the transfer and synthesis of textures through attention mechanisms. Building upon the foundations laid by TB applications in SR, the \textit{Dual Aggregation Transformer for Image Super-Resolution (DAT)} \cite{chen2023dual} further innovates by merging spatial and channel dimensions within a TB framework for enhanced representation capability. DAT's unique strategy of alternating spatial and channel self-attention across consecutive transformer blocks allows for comprehensive context capture and feature aggregation in both inter- and intra-block.

\subsection{Image to Text}
The integration of vision and language has fostered a novel paradigm for understanding and generating rich textual descriptions from images, commonly known as "Image to Text" translation. The Contrastive Captioner (CoCa) a foundational model for image caption generation is anexample of such a model \cite{yu2022coca}. By training jointly with contrastive and captioning losses, it adeptly merges the capabilities of both generative and contrastive methods which leads to state-of-the-art performances across a multiple tasks like visual recognition, cross-modal retrieval, and image captioning. Notably, on ImageNet, CoCa's architecture excludes cross-attention in the initial decoder layers to focus on unimodal text representations and achieved a remarkable 86.3\% zero-shot top-1 accuracy. When fine-tuned, it further cemented its dominance with a new peak of 91.0\% top-1 accuracy \cite{yu2022coca}.

Another innovative approach, NLIP (Noise-robust Language-Image Pre-training), aims to counteract the inherent noise in web-crawled data, such as incorrect or irrelevant content \cite{huang2023nlip}. It introduces a principled framework to stabilize pre-training through two schemes: noise-harmonization and noise-completion. With noise-harmonization, it adjusts the cross-modal alignments by estimating noise probability. While noise-completion enhances text with missing object information, guided by visual concepts. NLIP thus promises an efficient mitigation of noise impacts during pre-training \cite{huang2023nlip}.

The PaLI model \cite{chen2022pali} further exemplifies the synergy of large-scale encoders and decoders from language \cite{intro_1vaswani2017attention} and vision \cite{dosovitskiy2010image} pathways. By processing both modalities, it performs a broad array of tasks in many languages with its simplicity, modularity, and scalability. PaLI notably benefits from scaling the vision and language components jointly, utilizing a vast 10B image-text dataset covering over 100 languages, which pushes it to achieve impressive results in tasks like captioning, visual question-answering, and scene-text understanding.

FUSECAP \cite{rotstein2024fusecap}, meanwhile, use "frozen" vision experts to enrich image captions with salient details that is often overlooked in traditional datasets. By fusing outputs from object detectors \cite{ren2015faster}, attribute recognizers \cite{zhang2021vinvl}, and OCR with original captions using a large language model, FUSECAP delivers detailed and precise descriptions, pushing the boundaries of the state-of-the-art in caption generation \cite{rotstein2024fusecap}. This data-centric approach reflects the potential of leveraging sophisticated language models for enriched content creation.

Collectively, these advancements reflect the trajectory of Image to Text translation toward models that not only comprehend visual content with exceptional capability but also articulate it with an increasingly nuanced understanding of language.

\begin{table*}[ht]
\centering
\small
\caption{Categorization of TB Models for Multi-Modal Data Conversion}
\label{table:multi_modal_transformers}
\begin{tabularx}{\textwidth}{>{\centering\arraybackslash}m{1.9cm} >{\centering\arraybackslash}m{2.2cm} >{\centering\arraybackslash}m{1.6cm} >{\centering\arraybackslash}m{2.5cm} >{\centering\arraybackslash}m{2.8cm} >{\centering\arraybackslash}m{2.2cm}}
\toprule
\textbf{Modality Conversion} & \textbf{Transformer Model} & \textbf{Base Model} & \textbf{Key Innovation} & \textbf{Main Contribution} & \textbf{Datasets Used} \\ 
\midrule

\multirow{4}{*}{Image to Text} 
& CoCa \cite{yu2022coca} & Custom & Joint training with contrastive and captioning losses & State-of-the-art performance in visual recognition and image captioning & ImageNet, MSCOCO, Flickr30K \\ 
& NLIP \cite{huang2023nlip} & Custom & Noise-harmonization and noise-completion schemes & Efficient mitigation of noise impacts during pre-training & MSCOCO, Flicker-Audio, SPEECH-COCO \\ 
& PaLI \cite{chen2022pali} & ViT + Language Model & Joint scaling of vision and language components & Impressive results in captioning and scene-text understanding & MSCOCO \\ 
& FUSECAP \cite{rotstein2024fusecap} & ViT + LLM & Use of "frozen" vision experts to enrich captions & Detailed and precise image descriptions & MSCOCO \\ 
\midrule

\multirow{4}{*}{Video to Text} 
& Bidirectional Attentive Fusion \cite{Wang_2018_CVPR} & Custom & Utilizes both past and future video contexts & Accurate event proposal predictions & ActivityNet Captions \\ 
& SMAN \cite{zheng2021stacked} & Custom & Integrates historical visual and textual information & Enhanced captioning results with reinforcement learning & MSVD, MSR-VTT \\ 
& CTAN \cite{lei2019channel} & Custom & Channel-wise attention generation & Enhances fine-grained informative features for action recognition & UCF101, HMDB51 \\ 
& Bi-modal Transformer \cite{iashin2020better} & Custom & Processes both audio and visual modalities & Superior performance in dense video captioning & ActivityNet Captions \\ 
\midrule

\multirow{4}{*}{Vision to Speech} 
& Image2speech \cite{hasegawa2017image2speech} & VGG16 + Clustergen & Sequence-to-sequence with attention & Intelligible spoken sentences from images & Flickr8k, MSCOCO \\ 
& Text-Free Image-to-Speech \cite{hsu2020text} & ResDAVEnet-VQ & Eliminates need for text as intermediate representation & Captures diverse visual semantics effectively & Flickr8k, MSCOCO \\ 
& Practical Im2Sp \cite{kim2024towards} & ViT + HuBERT & Vision-language pre-training with multi-modal tokens & Reduces data storage and computational requirements & COCO, Flickr8k \\ 
& IMAGEBIND \cite{girdhar2023imagebind} & Custom & Joint embedding across multiple modalities & Zero-shot capabilities across new modalities & Multiple datasets (image, text, audio, depth, thermal, IMU) \\ 
\bottomrule

\end{tabularx}
\end{table*}

\subsection{Video to Text} Video to text generation is a pivotal area in the intersection of computer vision and natural language processing that focuses on interpreting and translating visual content into descriptive language. This task not only encapsulates the challenge of understanding visual cues from static images and dynamic sequences but also demands the generation of coherent and contextually relevant textual descriptions. Within this domain, two key tasks have emerged as significant research avenues: Temporal Action Localization and Description (TALD) and Dense Video Captioning.

Wang et al. present an innovative approach in \textit{Bidirectional Attentive Fusion with Context Gating for Dense Video Captioning} \cite{Wang_2018_CVPR}. They tackle the challenges of utilizing both past and future video contexts for accurate event proposal predictions and constructing informative inputs for the decoder to generate natural event descriptions. This bidirectional proposal method, exploits past and future contexts with a novel context gating mechanism that dynamically balances contributions from the current event and its surrounding contexts. By fusing hidden states attentively, it outperforms state-of-the-art models on the ActivityNet Captions dataset, and shows significant improvement in dense video captioning tasks.

Zheng et al. introduce the Stacked Multimodal Attention Network (SMAN) for context-aware video captioning \cite{zheng2021stacked}. They focus on integrating historical visual and textual information into the caption generation process, which has often been overlooked in previous models. By utilizing a stacked architecture to gradually process different features and applying reinforcement learning for refinement, SMAN leverages historical context effectively to enhance captioning results.

Lei et al. propose the Channel-wise Temporal Attention Network (CTAN) for video action recognition \cite{lei2019channel}. Although it primarily targets action recognition, CTAN's methodology is notable for its focus on channel-wise attention generation and its exploitation of interactions across video frames. The paper assumes that enhancing fine-grained informative features via channel-wise attention could be beneficial for TALD as well, hinting at broader applications for their proposed network.

Lastly, \textit{A Better Use of Audio-Visual Cues: Dense Video Captioning with Bi-modal Transformer}  \cite{iashin2020better} addresses the untapped potential of audio cues in dense video captioning. By integrating a novel Bi-modal Transformer that processes both audio and visual modalities, the authors demonstrate superior performance on the ActivityNet Captions dataset. Their work suggests that leveraging multimodal inputs can lead to richer and more descriptive video captions.

\subsection{Vision to Speech}
The evolution of artificial intelligence has brought about significant advancements in the translation of visual data into spoken descriptions, a task referred to as "Vision to Speech." This section explores the methodologies and innovations in the domain of generating speech from images to highlight the advancements made by various TB models.

One of the foundational study in this domain is the "Image2speech" model, which aims to generate spoken descriptions from images \cite{hasegawa2017image2speech}. This model utilizes a sequence-to-sequence architecture with attention mechanisms to convert image features extracted from the VGG16 network into sequences of speech units. These units are then synthesized into audio using Clustergen. The innovation here lies in the use of different segmentation methods, including words, first-language phones, pseudo-phones, and second-language phones, to facilitate speech generation in both languages with and without known orthography. The model's performance is validated using BLEU scores and token error rates, demonstrating its ability to produce intelligible and coherent spoken sentences \cite{hasegawa2017image2speech}.

Building on this concept, the "Text-Free Image-to-Speech Synthesis Using Learned Segmental Units" model takes a novel approach by eliminating the need for text as an intermediate representation \cite{hsu2020text}. This method connects image captioning and speech synthesis through a set of discrete, sub-word speech units discovered via a self-supervised visual grounding task. The model employs ResDAVEnet-VQ to learn these units and generates spoken audio captions directly from images. This approach is validated on the Flickr8k and MSCOCO datasets, showing that the generated captions capture diverse visual semantics effectively \cite{hsu2020text}.

The "Towards Practical and Efficient Image-to-Speech Captioning with Vision-Language Pre-training and Multi-Modal Tokens" model introduces a powerful and efficient framework for image-to-speech captioning \cite{kim2024towards}. It leverages a pre-trained vision-language model, GiT, to transfer rich image comprehension and language modeling knowledge into the image-to-speech task. The model generates discretized speech units using HuBERT and converts images into compressed image units through vector quantization. This method significantly reduces data storage and computational requirements, setting new state-of-the-art performance on the COCO and Flickr8k datasets \cite{kim2024towards}.

Finally, the "IMAGEBIND: One Embedding Space To Bind Them All" model aims to learn a joint embedding across multiple modalities, including images, text, audio, depth, thermal, and IMU data \cite{girdhar2023imagebind}. By using images to bind these modalities together, IMAGEBIND enables zero-shot capabilities across new modalities without requiring extensive paired data for training. This model demonstrates emergent capabilities such as cross-modal retrieval and zero-shot recognition, outperforming specialist supervised models in various tasks. IMAGEBIND leverages the natural pairing of modalities with images to create a unified embedding space, showcasing strong few-shot recognition results and setting a new benchmark for evaluating vision models across both visual and non-visual tasks \cite{girdhar2023imagebind}.

\section{Speech Processing}\label{sec:sp}
Sequence processing is one of the areas where TB architectures \cite{vaswani2021scaling} have shown great impact. In speech processing tasks such as automatic speech recognition (ASR), speech translation (ST), and text-to-speech (TTS), algorithms deal with sequences, making TB architectures attractive for integration. While long short-term memory (LSTM) has been the go-to sequence processing module in the past, recent studies have highlighted the superiority of transformers over recurrent-based approaches \cite{li2020comparison,wang2021transformer,zeyer2019comparison}. The TB processing mechanism allows for smoother information flow through sequential steps, leading to improved gradient flow and streamlined training. Moreover, the parallel calculation capability of transformers accelerates the training process significantly \cite{zeyer2019comparison}.

\begin{figure}[h]
    \centering
    \includegraphics[width=0.6\linewidth]{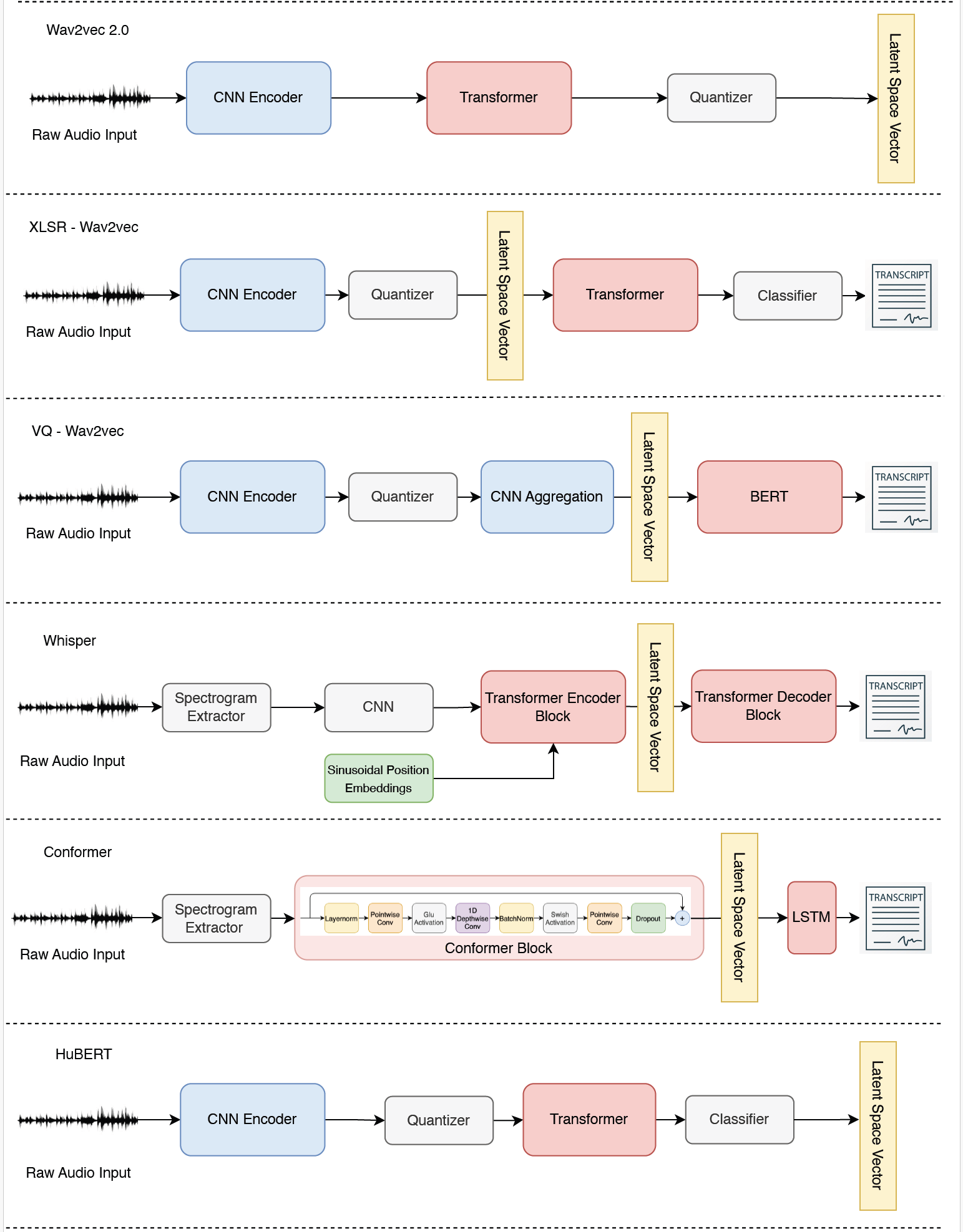}
    \caption{This diagram shows different models used for learning and recognizing speech representation. The models include Conformer Block \cite{gulati2020conformer}, UniSpeech \cite{wang2021unispeech}, Wav2vec 2.0 \cite{baevski2020wav2vec}, HuBERT \cite{hsu2021hubert}, Whisper \cite{radford2023robust}, Speech-Transformer \cite{dong2018speech}, VQ-Wav2vec \cite{baevski2019vq}, and XLSR-Wav2vec \cite{conneau2020unsupervised}. Each model takes raw audio input and processes it through encoders, quantizers, and Transformer blocks to generate latent space vectors and speech recognition outputs.}
    \label{tab:summ-tf-res}
\end{figure}

In \cite{zeyer2019comparison,li2020comparison,wang2021transformer}, researchers have shown how TB and recurrent-based approaches performed across various speech processing tasks and datasets. The empirical results demonstrate the superiority of TB approaches. In \cite{karita2019comparative}, Karita et al. have provided instructions for the community to make utilization of TB approaches easier (an attempt for the community to use TB models). Additionally, they work on an open, community-driven project for end-to-end speech applications using both TB and RNN models. Albert Zeyer et al. \cite{zeyer2019comparison} validate the efficiency of gradient flow and parallel computing, asserting that TB models require less training time than a similarly performing LSTM model. However, the TB models are prone to overfitting. To address this issue, the authors demonstrate the effectiveness of data augmentation. In \cite{zeyer2019comparison,chen2018best}, they have investigated the performance of the combined architecture. The forthcoming section will discuss this domain's most popular TB architecture alongside various prominent tasks. Tables \ref{tab:summ-tf} and Fig. \ref{tab:summ-tf-res} present a summary of the overall structure and performance of the methods. Subsequently, we will dedicate two sections to speech conversion to X—text, image, or video.

\subsection{Speech TB models}

\begin{table*}[ht]
    \small
    \centering
    \caption{A comprehensive overview of prominent speech processing techniques. In the "Tasks" column, TTS stands for Text-to-Speech, ST denotes Speech Translation, and STT represents speech-to-text conversion.}
    \begin{tabular}{lcccccc}
        \toprule
        \textbf{Method} & \textbf{Year} & \textbf{Tasks} & \makecell{\textbf{Transformer} \\ \textbf{Type}} & \makecell{\textbf{Training} \\ \textbf{Approach}} & \makecell{\textbf{Unique} \\ \textbf{Features}} \\
        \midrule
        VALL-E \cite{wang2023neural}& 2023 & TTS & Encoder-Decoder & Semi-Supervised & \makecell{In-Context Learning Enabled \\ Personalized Synthesis\\ Cross-Lingual } \\
        \cmidrule(lr){1-6}
        Whisper \cite{radford2023robust} & 202 & STT - ST & Encoder-Decoder & Weak Supervision & \makecell{Trained on Large-Scale Datasets, \\ Multilingual and Multitask, \\ Zero-Shot Transfer} \\
        \cmidrule(lr){1-6}
        WavLM \cite{chen2022wavlm} & 2022 & STT & Encoder & \makecell{Self-Supervised \\ + Fine-Tuning} & \makecell{Using Masked Speech Prediction/Denoising \\ 
                    Stabilized Transformer Training \\ 
                    Versatile for Various Tasks (beyond SST)}\\
        \cmidrule(lr){1-6}
        HuBERT \cite{hsu2021hubert}& 2021 & STT & Encoder & \makecell{Self-Supervised \\ + Fine-Tuning} & \makecell{Intermediate Clustering \\ Using Pseudo-labeling} \\
        \cmidrule(lr){1-6}
        Conformer \cite{gulati2020conformer} & 2020 & STT - ST & Encoder-Decoder & \makecell{Self-Supervised \\ + Fine-Tuning} & \makecell{Combind CNN \& Transformer \\ for Global \& Local Feature Extraction} \\
        \cmidrule(lr){1-6}
        UniSpeech \cite{wang2021unispeech} & 2021 & STT & Encoder & \makecell{Self-Supervised \\ + Fine-Tuning} & \makecell{Cross-Lingual \\ Multi-Task \\
        CTC Loss} \\
        \cmidrule(lr){1-6}
        wav2vec 2.0 \cite{baevski2019vq} & 2020 & STT & Encoder & \makecell{Self-Supervised \\ + Fine-Tuning} & \makecell{Backbone of Popular Methods} \\
        \cmidrule(lr){1-6}
        Tacotron \cite{wang2017tacotron} & 2017 & TTS & --- & Supervised & \makecell{Seq2Seq Architecture with, \\ Attention Mechanism} \\
        \bottomrule
    \end{tabular}
   \label{tab:summ-tf}
\end{table*}

\paragraph{wav2vec 2.0} This framework uses self-supervised learning to convert raw audio data into a latent vector space. It consists of feature encoding and contextual representation learning. The raw waveform is fed to the encoder, which includes temporal convolutional neural networks and quantization \cite{baevski2019vq}. The output is input into a TB context network to extract contextual relations. Pre-training involves using masked waveforms to predict masked segments. Fine-tuning for specific speech recognition tasks involves labeled data with CTC loss \cite{graves2006connectionist} utilized \cite{baevski2020wav2vec}.


\paragraph{Conformer} Anmol Gulati et al. \cite{gulati2020conformer} aims to capture global and fine-grained features by integrating TB and convolution networks. Convolutional neural networks excel at extracting detailed features, but requiring multiple layers can lead to computational challenges during training and inference. TB models were used to understand the global context better, but they may need assistance identifying detailed local feature patterns. Consequently, Anmol Gulati et al. adopted a collaborative approach with convolution to leverage the strengths of both techniques. The method they proposed demonstrated superiority over the state-of-the-art models in the LibriSpeech \cite{panayotov2015librispeech} and AISHELL-1 \cite{bu2017aishell} benchmarks, as indicated by the Word Error Rate (WER) metric.


\paragraph{HuBERT} HuBERT (Hidden Unit BERT) \cite{hsu2021hubert} uses intermediate clustering for latent embedding vector learning in a self-supervised setting, particularly for ASR tasks. HuBERT uses a K-means clustering algorithm to assign each 25-millisecond segment of input audio to a cluster. HuBERT uses Mel-Frequency Cepstral Coefficients (MFCCs) \cite{aghajan2009human} as the feature vectors for the first clustering step. The audio feature vectors are input into a network comprising convolution and transformer layers in the second phase. Half of the feature vectors are masked, and this network is employed to generate a meaningful representation for each segment. The cosine similarity function assigns each segment to its respective cluster. The training takes place using cross-entropy loss. HuBERT outperformed the state-of-the-art wav2vec 2.0 on both the Librispeech \cite{panayotov2015librispeech} and Libri-light \cite{kahn2020libri} benchmarks.

\paragraph{Whisper}
Whisper is a multi-language, general-purpose framework for speech processing in resource-limited settings. It employs a standard transformer encoder-decoder architecture. The primary focus of this study is to explore the capabilities of large-scale supervised pre-training for speech recognition. Whisper utilizes weak supervision and adopts a minimalist approach to data pre-processing, avoiding significant standardization. Training on a large and diverse audio dataset enables Whisper's multitasking capabilities, allowing it to perform well on various speech-related tasks. Additionally, Whisper can be applied to different applications such as education, voice assistants, and translation \cite{radford2023robust}.

\paragraph{WavLM}
WavLM \cite{chen2022wavlm} is a pre-trained model for speech processing, trained on 94,000 hours of audio using self-supervised learning. Sanyuan Chen et al. proposed WavLM as a comprehensive backbone network for all speech processing. The model uses TB architecture to understand long-range dependencies and employs Gated Relative Positioning to efficiently order tokens within the audio signal. WavLM experiences pre-training in a setting involving masked speech prediction and denoising. WavLM expands HuBERT framework to handle masked speech processing and noise reduction modeling. WavLM is evaluated on nineteen different tasks in different settings, such as ASR, speech verification, speech separation, and speech diarization. It achieves state-of-the-art results in various settings.

\paragraph{Tecotron}
Tacotron aims to integrate attention mechanisms into end-to-end speech synthesis. It directly converts input characters into corresponding spectrograms. The backbone network is a sequence-to-sequence model called CBHG, which combines 1-D convolutional neural networks and GRUs. Utilizing a simple vocoder and waveform synthesis module achieves good and natural outcomes regarding the Mean Opinion Score (MOS) for US English. The MOS score is a rating humans give to judge audio quality \cite{wang2017tacotron}.

\paragraph{VALL-E}
Wang et al. \cite{wang2023neural} proposed a text-to-speech (TTS) synthesis method by treating it as conditional language modeling tasks rather than focusing on continuous signal prediction. This approach marks the first attempt to use language models for TTS tasks. Wang et al. (2023) used a large, diverse, and multi-speaker speech dataset of 60k hours of English speech for pre-training and natural waveform generation. VALL-E converts phonemes directly to discrete code, eliminating the need for intermediate feature engineering and manipulation, resulting in an end-to-end system. VALL-E generates personalized acoustic tokens using the 3-second enrolled recording and phoneme prompt to tailor the resulting waveform. Leveraging diverse and extensive datasets empowers VALL-E to generate waveforms resembling the 3-second input prompt seamlessly. Experiments on LintiSpeech and VCTK \cite{Yamagishi2019-ti} showed significant improvement over state-of-the-art zero-shot TTS systems in speech naturalness and speaker similarity. Furthermore, VALL-E demonstrates the ability to preserve the speaker's emotion and acoustic environment of the acoustic prompt in synthesis. VALL-E X is an extension of VALL-E designed for cross-lingual speech generation \cite{zhang2023speak}.

\paragraph{UniSpeech}
UniSpeech \cite{wang2021unispeech} is a pre-trained model proposed to address the challenge of learning speech representation in low-resource languages. The need for more training samples, even in high-resource languages, is problematic. Domain shifts and background noises can significantly reduce the performance of the model. UniSpeech uses labeled high-resource and unlabeled low-resource data to train a model that captures phoneme identities and remains robust to small details like background noise. The model architecture is based on Wav2vec 2.0, using convolutional neural networks for feature extraction, a TB model to learn contextual representation, and quantization to discretize the latent representation. During training, they pre-trained the model on labeled high-resource and low-resource data. Then, they conducted another training phase using sequence-level CTC loss and a contrastive task over masked contextual representations. Evaluation of cross-lingual speech processing using CommonVoice \cite{ardila2019common} and Librispeech datasets shows that Unispeech outperforms other approaches in low-resource settings and reduces the Word Error Rate in domain-shifted scenarios compared to baselines.

\begin{table}[ht]
    \scriptsize
    \centering
    \caption{Performance metrics for the speech-processing methods. Metrics include Word Error Rate (WER), Mean Opinion Score (MOS), and Speaker Similarity (SPK).}
    \begin{tabularx}{\linewidth}{l X}
        \toprule
        \textbf{Method} & \textbf{Performance Benchmark} \\
        \midrule
        VALL-E & \texttt{VALL-E Continual}: LibriSpeech (test-clean): 3.8 WER, 0.508 SPK; \texttt{VALL-E}: LibriSpeech (test-clean): 5.9 WER, 0.589 SPK \\
        \cmidrule(lr){1-2}
        Whisper & \texttt{Whisper large-v2}: LibriSpeech (test-clean): 2.5 WER\footnote{For English transcription}, LibriSpeech (test-other): 4.9 WER, WSJ: 2.6 WER, CommonVoice5.1: 8.2 WER, TED-LIUM3: 3.7 WER \\
        \cmidrule(lr){1-2}
        WavLM & \texttt{WavLM Large}: LibriSpeech (1-hour labeled and test-clean): 2.9 WER, (1-hour labeled and test-other): 5.1 WER; (10-hour labeled and test-clean): 2.4 WER, (10-hour labeled and test-other): 4.6 WER; (100-hour labeled and test-clean): 2.1 WER, (100-hour labeled and test-other): 4.0 WER; (960-hour labeled and test-clean): 1.8 WER, (960-hour labeled and test-other): 3.2 WER \\
        \cmidrule(lr){1-2}
        HuBERT & \texttt{HuBERT X-Large}: LibriSpeech (1-hour labeled and test-clean): 2.8 WER, (1-hour labeled and test-other): 4.8 WER; (10-hour labeled and test-clean): 2.0 WER, (10-hour labeled and test-other): 4.0 WER; (100-hour labeled and test-clean): 1.9 WER, (100-hour labeled and test-other): 3.5 WER; (960-hour labeled and test-clean): 1.8 WER, (960-hour labeled and test-other): 2.9 WER \\
        \cmidrule(lr){1-2}
        UniSpeech & \texttt{UniSpeech-L}: CommonVoice (cv) average PER\footnote{Average performance when trained in English and tested in Spanish, French, Italian, Kyrgyz, Dutch, Russian, Swedish, and Tatar.}: 13.6 \\
        \cmidrule(lr){1-2}
        Conformer & \texttt{Conformer (L)}: LibriSpeech (test-clean): 2.1 WER Without LM, (test-other): 4.3 WER Without LM; LibriSpeech (test-clean): 1.9 WER With LM, (test-other): 3.9 WER With LM \\
        \cmidrule(lr){1-2}
        wav2vec 2.0 & \texttt{Wav2vec LARGE}: LibriSpeech (test-clean): 1.8 WER, (test-other): 3.3 WER \\
        \cmidrule(lr){1-2}
        Tacotron & \texttt{Tacotron}: 5-scale mean opinion score: 3.82 ± 0.085 \\
        \bottomrule
    \end{tabularx}
    \label{tab:performance-sp}
\end{table}

\subsection{Speech Processing Downstream Tasks}
This section will explore the various speech tasks addressed using TB architecture.
\subsubsection{Speech Translation}
Speech translation (ST) involves translating spoken language into another language or text while maintaining the original meaning, context, and tone. Di Gangi et al. \cite{di2019adapting} made one of the initial efforts to customize TB models for speech translation. They employed downsampling using convolution networks and introduced bidirectionality for TB models. The performance improvement and computational efficiency validate these modifications. In a different study \cite{li2020multilingual}, researchers used wav2vec 2.0 and mBART \cite{chipman2022mbart} - pre-trained single-modality modules in a multi-modal speech-to-text setup. They incorporated an adaptor for translation tasks to address discrepancies in length between audio and text sequences. Furthermore, they improved transfer learning efficiency by finetuning only LayerNorm and Attention (LNA) parameters. The evaluation results on CoVoST 2 (Wang et al., 2020) showed that their proposed solution surpasses state-of-the-art (SOTA) models. FAIRSEQ S2T \cite{wang2020fairseq} extends FAIRSEQ \cite{ott2019fairseq} for speech-to-text (ST) tasks. FAIRSEQ is an open-source toolkit developed by Facebook AI Research and Google Brain for tasks like text summarization. FAIRSEQ S2T incorporates methods such as attention-based RNN (Chan et al., 2016) \cite{chan2016listen}, TB, and Conformer models. For speech translation, evaluation was conducted using two multilingual datasets: MuST-C \cite{di-gangi-etal-2019-must} and CoVoST 2. The experimental results demonstrate competitiveness. 


\subsubsection{Speech Enhancement}
Speech enhancement aims to extract clear speech from a noisy input signal. T-GSA (Transformed Gaussian-weighted Self-attention) \cite{kim2020t} is a modified TB method used for speech enhancement. It utilizes a variation of multi-head attention and employs a Gaussian weighting matrix to adjust the score matrix, which measures the distance between the target frame and the attended symbols. The evaluation of this approach using two datasets, QUT-NOISE-TIMIT  \cite{dean2010qut} and VoiceBank-DEMAND \cite{valentini2016investigating}, showed significant improvement. TSTNN \cite{wang2021tstnn} also is a modified variant of the TB architecture, featuring a two-stage design. The initial TB model extracts local information, while the second one integrates information from the local transformer, thus capturing both local and global contexts. Notably, only the encoder part of the TB model is utilized, with additional modifications applied. TSTNN omits positional encoding and replaces the first fully connected layer of the feed-forward network with a GRU \cite{cho2014learning} layer. Experimental results underscore the effectiveness of these modifications, improving both performance and efficiency.


\subsubsection{Speech Separation}
Speech separation involves separating the undesirable components of a mixed audio signal. These components could include background noise and uninterested speakers. This task is also often referred to as the cocktail party problem \cite{haykin2005cocktail}. To address this task, dual-path TB approaches have become popular. From this family of models, SepFormer \cite{subakan2021attention} employs a sandwich architecture comprising an Encoder, Masking Net., and Decoder. The encoder utilizes convolutional networks to transform the time-domain mixed signal into a new representation. Subsequently, the masking nets employ two transformer block models to capture both short-term and long-term dependencies. Finally, the decoder, consisting of a transposed convolution layer, generates the separated signal. SepFormer was evaluated on the WSJ0-2/3mix \cite{hershey2016deep} datasets, achieving superior performance compared to state-of-the-art models at that time. Additionally, DPT-Net \cite{chen2020dual} follows a similar framework to SepFormer. There is a challenge with dual-path TB models such as SepFormer and DPT-Net. They encounter difficulty in effectively capturing long-range interactions alongside local patterns directly. MossFormer \cite{zhao2023mossformer} addresses this limitation by implementing a joint attention framework, allowing for the direct capture of all sequence element interactions. MossFormer resolves this inefficiency by employing parallel computation of self-attention on local segments and lightweight self-attention on the entire sequence. These modifications empower MossFormer to outperform state-of-the-art models such as SepFormer and DPT-Net on benchmarks including WSJ0-2/3mix and WHAM!/WHAMR \cite{wichern2019wham, maciejewski2020whamr}.


\subsection{Speech To Text}

Speech-to-Text (STT) or Automatic Speech Recognition (ASR) enables machines to transcribe spoken language into text. This process involves preprocessing audio signals and extracting relevant features. These features typically include spectral characteristics such as Mel-frequency cepstral coefficients (MFCCs), which capture speech signal frequency content, as well as fundamental frequency (pitch), formants, and other important acoustic properties. In the past, approaches required more manual feature design, but nowadays, all proposed approaches are end-to-end. In terms of architecture, there has been a shift in the field from relying on recurrent-based methods to embracing TB approaches. Among these methods, Speech-Transformer \cite{dong2018speech} stands as one of the early attempts to employ a TB architecture for ASR tasks. Its objective is to address the slow training associated with recurrent-based approaches by utilizing a sequence-to-sequence TB architecture along with a 2D attention mechanism. Notably, Speech-Transformer achieves competitive results within a relatively short training time.

Shifting the focus to another line of research within a self-supervised setting, alternative methods such as Wav2vec have been explored previously. Two other proposed solutions in this setting include VQ-Wav2vec \cite{baevski2019vq} and XLSR-Wav2vec \cite{conneau2020unsupervised}, which extend the architecture of Wav2vec. VQ-Wav2vec introduces a novel quantization module for constructing discrete representations, improved the then-state-of-the-art results in WSJ \cite{Garofolo1993-rq} and TIMIT \cite{Garofolo1993-bf} benchmarks. On the other hand, XLSR-Wav2vec pioneers multilingual cross-language representation learning, converting raw waveforms into discrete speech representations. Utilizing techniques such as Gumbel softmax \cite{jang2016categorical} for discrete codebook selection and modifying BERT with relative positional embeddings in the context network, XLSR-Wav2vec extends the capabilities of its predecessor. Results demonstrate the efficacy of multilingual cross-language learning, particularly benefiting low-resource languages through enhanced representation vectors. These findings are validated by a series of experiments using datasets such as CommonVoice \cite{ardila2019common} and BABEL \cite{gales2014speech}. This research unlocks the potential of multilingual approaches in speech processing.

The recent success of models capable of handling multiple languages has stimulated researchers' interest. A notable example is the Universal Speech Model (USM) developed by Google \cite{zhang2023google}. This model represents the first endeavor to construct a comprehensive model that can effectively process all languages. However, a major obstacle lies in the scarcity of available data for less widely spoken languages. To address this challenge, USM leverages diverse and substantial datasets in three distinct phases to maximize the utilization of available data. In the initial phase, 12 million hours of audio spanning 300 languages were utilized. Subsequently, the model incorporated 28 billion sentences across 1140 languages. Finally, approximately 200,000 hours of labeled multilingual data were used to train the model in text-audio association. The architecture employed for USM was the 2-billion-parameter Conformer. USM has achieved state-of-the-art performance in multilingual Automatic Speech Recognition (ASR) and Automatic Speech Translation (AST) across various benchmarks. Another area of research, as outlined in the Transformer Transducer paper \cite{zhang2020transformer}, involves the supervised use of TB models to encode audio and labels within the RNN-T architecture and the corresponding loss function. The model's performance was evaluated using the LibriSpeech dataset. Moreover, alternative methodologies such as Whisper and Conformer have also been applied in ASR tasks. Table \ref{speech-to-text_table} provides a summary of the discussed methods.

\begin{figure}[H]
    \centering
    \includegraphics[width=0.6\linewidth]{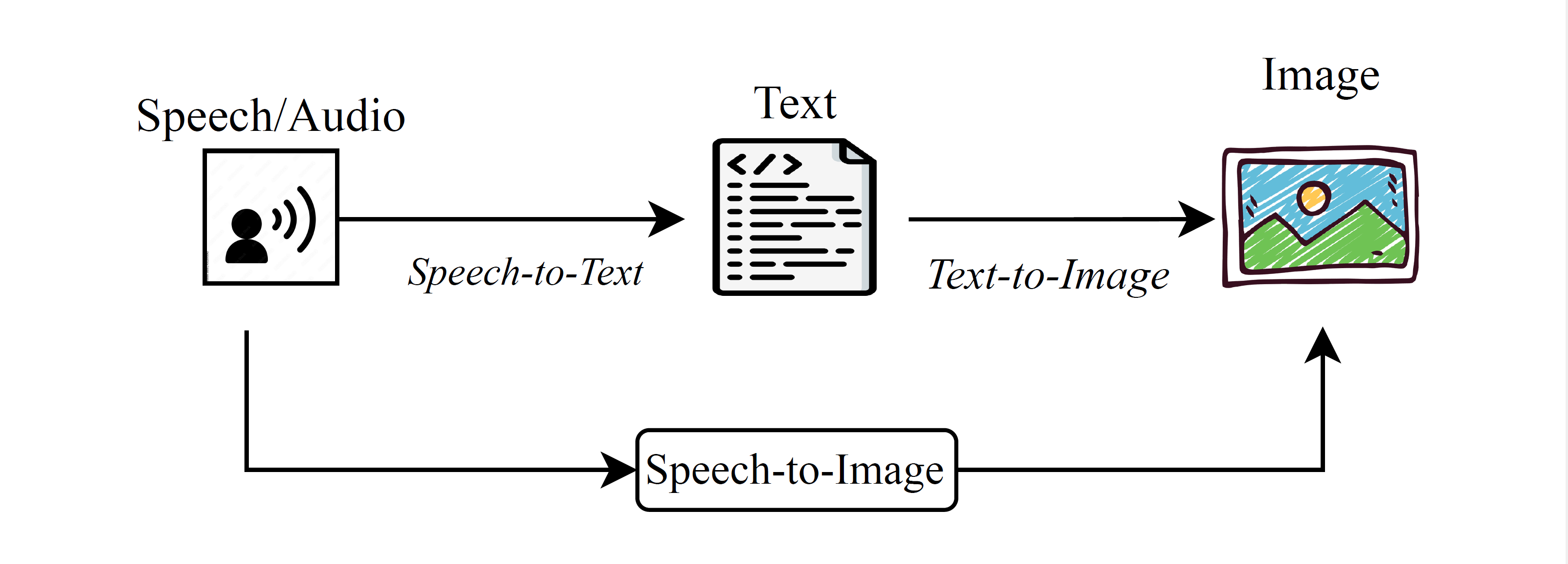}
    \caption{Two approaches to generating images from speech: (top) converting speech to text before generating the image, (bottom) generating the image directly from speech without using text transcripts.}
    \label{fig:speech-to-text-diagram}
\end{figure}

\begin{table*}[ht]
\centering
\small
\caption{Categorization of TB methods in speech-to-text modality conversion.}
\label{speech-to-text_table}
\resizebox{\textwidth}{!}{
\begin{tabular}{>{\centering\arraybackslash}m{3cm} >{\centering\arraybackslash}m{1cm} >{\centering\arraybackslash}m{2cm} >{\centering\arraybackslash}m{5.5cm} >{\centering\arraybackslash}m{3cm}} 
\toprule
\textbf{Method Name} & \textbf{Year} & \textbf{Base Model} & \textbf{Novelty} & \textbf{Dataset} \\ \midrule
Speech-Transformer \cite{dong2018speech} & 2018 & Transformer & Using a non-recurrent sequence-to-sequence + 2D-attention mechanisms for faster training and joint attention to both time and frequency axes inputs, resulting in more expressive representations. & Wall Street Journal (WSJ)  \\ \addlinespace[0.5em]

VQ-Wav2vec \cite{baevski2019vq} & 2019 & Wav2vec & Utilizing Gumbel-Softmax or online k-means clustering to quantize dense representations. It enables the use of NLP algorithms that require discrete inputs. & WSJ, TIMIT \\ \addlinespace[0.5em]

XLSR-Wav2vec \cite{conneau2020unsupervised} & 2020 & Wav2vec 2.0 & XLSR learns cross-lingual speech representations by pretraining on raw speech waveforms in multiple languages, utilizing a shared quantization across languages. When fine-tuned on labeled data, it significantly outperforms monolingual models. & CommonVoice, BABEL \\ \addlinespace[0.5em]

Universal Speech Model (USM) \cite{zhang2023google} & 2023 & Conformer & USM operates in 100+ languages using multilingual pre-training with random-projection quantization and speech-text modality matching. The encoder is pre-trained on a large unlabeled multilingual dataset of 12 million hours spanning over 300 languages. & Custom datasets \\ \addlinespace[0.5em]

Transformer Transducer \cite{zhang2020transformer} & 2020 & Transformer & Using RNNs for information encoding instead of Transformer encoders. & LibriSpeech \\ \addlinespace[0.5em]

Squeezeformer \cite{kim2022squeezeformer} & 2022 & Modified Conformer & Using a Temporal U-Net structure to reduce the cost of multi-head attention. It simplifies convolution module activations in the convolutional block, removes redundant Layer Normalization operations, and introduces an efficient depthwise down-sampling layer. & LibriSpeech \\ \addlinespace[0.5em]

SpeechStew \cite{chan2021speechstew} & 2021 & Conformer + RNN-T & Combines multiple pre-trained models to improve ASR performance using diverse datasets. & AMI-IHM, WSJ, Switchboard, CallHome \\ \bottomrule
\end{tabular}
}
\end{table*}



\subsection{Speech to Vision}

\subsubsection{Speech To Images}
Speech-to-image conversion is a challenging cross-modal task. Here, speech-to-image means no text modality in the middle, as shown in Figure \ref{fig:speech-to-text-diagram}. Despite its complexity, there are compelling motivations to pursue this task. Firstly, there are approximately 7,000 languages without a written form \cite{campbell2008ethnologue}, rendering them unable to utilize speech processing technologies that predominantly rely on scripts. Direct conversion allows speakers of these languages to access the technology. Secondly, text modality causes information to be lost. Speech has distinct features, e.g., tone of voice and emotion. Lastly, many researchers in cognitive science have acknowledged that infants learn to communicate by associating visual information, such as facial expressions of parents and shapes of objects, with auditory stimuli without the need for words \cite{bergelson20126}. This insight can be a source of inspiration for the AI research community.

The predominant approach to this task is through Generative Adversarial Networks (GANs) \cite{li2022speech,zhao2022generating,sung2023sound}. As far as we know, no TB architecture has been proposed despite its potential. Only a few of the generative methods have utilized the attention mechanism. The work by Scharenborg et al. in 2020 \cite{scharenborg2020speech} was one of the pioneering efforts in direct speech-to-image conversion, utilizing the attention mechanism. Their results showed promise. Another notable effort is the S2IGAN (Speech-to-Image Generative Adversarial Network) \cite{wang2021generating}, which consists of a Speech Embedding Network (SEN) and a Relation-Supervised Densely-Stacked Generative Model (RDG). SEN contains a combined architecture of convolution, gated recurrent units, and a self-attention mechanism to transform speech into a latent space. RDG generates high-quality images according to the speech latent space. Evaluation using four datasets showed good performance of S2GAN. Fusion-s2igan \cite{zhang2024fusion} is another GAN-based method that uses the attention mechanism. Pixel-attention submodule (PAM) used the attention mechanism. PAM models spatial correlations between visual pixels. Doing so allows the generator to focus more on crucial and informative areas. Fusion-S2iGan has been assessed using four datasets (CUB bird \cite{wah2011caltech}, Oxford-102 \cite{nilsback2008automated}, Flickr8k \cite{hodosh2013framing}, and Places-subset \cite{zhou2014learning}) and has demonstrated superior performance to current speech-to-image methods.


\subsubsection{Speech To Video}
The field of direct speech-to-video conversion without text involvement is currently under-researched. Speech and video data are highly detailed and intricate, making efficient encoding of speech to correspond to video challenging yet impactful. Existing works in this field often focus on specific applications, such as face or face-animation detection. These methods are commonly based on diffusion and GAN (generative adversarial network) models \cite{wei2024aniportrait} \cite{hogue2024diffted} \cite{xu2024vasa} \cite{zhang2023sadtalker} \cite{liu2024padvg} \cite{stypulkowski2024diffused}, with some incorporating TB components for sequential feature extraction and generation. One notable work is AniPortrait \cite{wei2024aniportrait}, which aims to generate high-quality animated portraits driven by audio and a reference image using diffusion models and a TB architecture for fine-grained audio feature extraction. Another work, DiffTED \cite{hogue2024diffted}, focuses on generating one-shot audio-driven TED-style talking videos using a diffusion-based architecture and a TB network for temporal modeling. Similarly, DreamTalk \cite{ma2023dreamtalk} endeavors to create a video of a talking head synchronized with a given speech and reflect the speaking style of a reference video. This method utilizes a diffusion-based model and a TB architecture as the denoising network. VASA \cite{xu2024vasa} shares the same goal as DreamTalk but with differences in the input, where it accepts a single static image and a speech audio clip. In the architecture, a TB model is used to handle sequence generation.


\section{Conclusion and future directions}

In this survey, we provided a comprehensive review of transformer-based models applied to data modality conversion, focusing on text, vision, and speech modalities. Our analysis highlights the versatility and scalability of transformers, showcasing their ability to handle complex data representations and conversions effectively. We discussed the architecture, conversion methodologies, and applications of various transformer variants, underscoring their significant advancements in AI-driven content generation and understanding. Despite these advancements, several challenges remain. The computational complexity and resource requirements for training large transformer models are substantial. Additionally, the integration of multimodal data poses unique challenges in terms of aligning and processing diverse data types.


The future of transformer-based models in data modality conversion looks promising, with several exciting directions for research and development:

\textbf{Efficiency Improvements:} Developing more efficient transformer models that require less computational power and training time is crucial. Techniques such as model pruning, quantization, and knowledge distillation can help create lightweight models suitable for deployment on edge devices.

\textbf{Multimodal Integration:} Enhancing the ability of transformers to integrate and process multimodal data seamlessly remains a key area of research. Future models should focus on improving the alignment and fusion of different data types to enable more accurate and contextually relevant conversions.

\textbf{Scalability and Generalization:} Ensuring that transformer models can scale effectively to handle large datasets and generalize well across diverse applications is essential. Research should explore methods to enhance the robustness and adaptability of these models.

\textbf{Real-time Applications:} As transformers continue to advance, their application in real-time scenarios, such as live speech translation and real-time video captioning, will become increasingly feasible. Optimizing models for low-latency performance will be critical for these applications.

\textbf{Ethical and Fair AI:} Addressing ethical concerns and ensuring fairness in AI models is paramount. Researchers must focus on developing transformers that are unbiased and transparent, with mechanisms to detect and mitigate potential biases in data and model outputs.

\textbf{Cross-lingual and Multilingual Capabilities:} Expanding the cross-lingual and multilingual capabilities of transformers will be vital for global applications. Future models should aim to perform high-quality translations and conversions across a wide range of languages and dialects.

By focusing on these future directions, the research community can continue to advance transformer-based models, unlocking new possibilities and applications in data modality conversion and beyond.

\bibliographystyle{ACM-Reference-Format}
\bibliography{sample-manuscript}


\begin{thebibliography}{200}


\ifx \showCODEN    \undefined \def \showCODEN     #1{\unskip}     \fi
\ifx \showDOI      \undefined \def \showDOI       #1{#1}\fi
\ifx \showISBNx    \undefined \def \showISBNx     #1{\unskip}     \fi
\ifx \showISBNxiii \undefined \def \showISBNxiii  #1{\unskip}     \fi
\ifx \showISSN     \undefined \def \showISSN      #1{\unskip}     \fi
\ifx \showLCCN     \undefined \def \showLCCN      #1{\unskip}     \fi
\ifx \shownote     \undefined \def \shownote      #1{#1}          \fi
\ifx \showarticletitle \undefined \def \showarticletitle #1{#1}   \fi
\ifx \showURL      \undefined \def \showURL       {\relax}        \fi
\providecommand\bibfield[2]{#2}
\providecommand\bibinfo[2]{#2}
\providecommand\natexlab[1]{#1}
\providecommand\showeprint[2][]{arXiv:#2}

\bibitem[Aghajan et~al\mbox{.}(2009)]%
        {aghajan2009human}
\bibfield{author}{\bibinfo{person}{Hamid Aghajan}, \bibinfo{person}{Juan~Carlos Augusto}, {and} \bibinfo{person}{Ram{\'o}n L{\'o}pez-C{\'o}zar Delgado}.} \bibinfo{year}{2009}\natexlab{}.
\newblock \bibinfo{booktitle}{\emph{Human-centric interfaces for ambient intelligence}}.
\newblock \bibinfo{publisher}{Academic Press}.
\newblock


\bibitem[Ali et~al\mbox{.}(2021)]%
        {ali2021xcit}
\bibfield{author}{\bibinfo{person}{Alaaeldin Ali}, \bibinfo{person}{Hugo Touvron}, \bibinfo{person}{Mathilde Caron}, \bibinfo{person}{Piotr Bojanowski}, \bibinfo{person}{Matthijs Douze}, \bibinfo{person}{Armand Joulin}, \bibinfo{person}{Ivan Laptev}, \bibinfo{person}{Natalia Neverova}, \bibinfo{person}{Gabriel Synnaeve}, \bibinfo{person}{Jakob Verbeek}, {et~al\mbox{.}}} \bibinfo{year}{2021}\natexlab{}.
\newblock \showarticletitle{Xcit: Cross-covariance image transformers}.
\newblock \bibinfo{journal}{\emph{Advances in neural information processing systems}}  \bibinfo{volume}{34} (\bibinfo{year}{2021}), \bibinfo{pages}{20014--20027}.
\newblock


\bibitem[Ammus et~al\mbox{.}(2021)]%
        {back_1_ammus2021survey}
\bibfield{author}{\bibinfo{person}{A. Ammus} {et~al\mbox{.}}} \bibinfo{year}{2021}\natexlab{}.
\newblock \showarticletitle{A survey of transformer based pretrained models in natural language processing}.
\newblock \bibinfo{journal}{\emph{Journal Name}} \bibinfo{volume}{Vol}, \bibinfo{number}{No} (\bibinfo{year}{2021}), \bibinfo{pages}{pp.}
\newblock


\bibitem[Arbelaez et~al\mbox{.}(2011)]%
        {amfm_pami2011}
\bibfield{author}{\bibinfo{person}{Pablo Arbelaez}, \bibinfo{person}{Michael Maire}, \bibinfo{person}{Charless Fowlkes}, {and} \bibinfo{person}{Jitendra Malik}.} \bibinfo{year}{2011}\natexlab{}.
\newblock \showarticletitle{Contour Detection and Hierarchical Image Segmentation}.
\newblock \bibinfo{journal}{\emph{IEEE Trans. Pattern Anal. Mach. Intell.}} \bibinfo{volume}{33}, \bibinfo{number}{5} (\bibinfo{date}{May} \bibinfo{year}{2011}), \bibinfo{pages}{898--916}.
\newblock
\showISSN{0162-8828}
\urldef\tempurl%
\url{https://doi.org/10.1109/TPAMI.2010.161}
\showDOI{\tempurl}


\bibitem[Ardila et~al\mbox{.}(2019)]%
        {ardila2019common}
\bibfield{author}{\bibinfo{person}{Rosana Ardila}, \bibinfo{person}{Megan Branson}, \bibinfo{person}{Kelly Davis}, \bibinfo{person}{Michael Henretty}, \bibinfo{person}{Michael Kohler}, \bibinfo{person}{Josh Meyer}, \bibinfo{person}{Reuben Morais}, \bibinfo{person}{Lindsay Saunders}, \bibinfo{person}{Francis~M Tyers}, {and} \bibinfo{person}{Gregor Weber}.} \bibinfo{year}{2019}\natexlab{}.
\newblock \showarticletitle{Common voice: A massively-multilingual speech corpus}.
\newblock \bibinfo{journal}{\emph{arXiv preprint arXiv:1912.06670}} (\bibinfo{year}{2019}).
\newblock


\bibitem[Baevski et~al\mbox{.}(2019)]%
        {baevski2019vq}
\bibfield{author}{\bibinfo{person}{Alexei Baevski}, \bibinfo{person}{Steffen Schneider}, {and} \bibinfo{person}{Michael Auli}.} \bibinfo{year}{2019}\natexlab{}.
\newblock \showarticletitle{vq-wav2vec: Self-supervised learning of discrete speech representations}.
\newblock \bibinfo{journal}{\emph{arXiv preprint arXiv:1910.05453}} (\bibinfo{year}{2019}).
\newblock


\bibitem[Baevski et~al\mbox{.}(2020)]%
        {baevski2020wav2vec}
\bibfield{author}{\bibinfo{person}{Alexei Baevski}, \bibinfo{person}{Yuhao Zhou}, \bibinfo{person}{Abdelrahman Mohamed}, {and} \bibinfo{person}{Michael Auli}.} \bibinfo{year}{2020}\natexlab{}.
\newblock \showarticletitle{wav2vec 2.0: A framework for self-supervised learning of speech representations}.
\newblock \bibinfo{journal}{\emph{Advances in neural information processing systems}}  \bibinfo{volume}{33} (\bibinfo{year}{2020}), \bibinfo{pages}{12449--12460}.
\newblock


\bibitem[Beltagy et~al\mbox{.}(2019)]%
        {TC_7_beltagy2019scibert}
\bibfield{author}{\bibinfo{person}{Iz Beltagy}, \bibinfo{person}{Kyle Lo}, {and} \bibinfo{person}{Arman Cohan}.} \bibinfo{year}{2019}\natexlab{}.
\newblock \showarticletitle{SciBERT: A pretrained language model for scientific text}.
\newblock \bibinfo{journal}{\emph{arXiv preprint arXiv:1903.10676}} (\bibinfo{year}{2019}).
\newblock


\bibitem[Beltagy et~al\mbox{.}(2020)]%
        {TS_2beltagy2020longformer}
\bibfield{author}{\bibinfo{person}{Iz Beltagy}, \bibinfo{person}{Matthew~E Peters}, {and} \bibinfo{person}{Arman Cohan}.} \bibinfo{year}{2020}\natexlab{}.
\newblock \showarticletitle{Longformer: The long-document transformer}.
\newblock \bibinfo{journal}{\emph{arXiv preprint arXiv:2004.05150}} (\bibinfo{year}{2020}).
\newblock


\bibitem[Bergelson and Swingley(2012)]%
        {bergelson20126}
\bibfield{author}{\bibinfo{person}{Elika Bergelson} {and} \bibinfo{person}{Daniel Swingley}.} \bibinfo{year}{2012}\natexlab{}.
\newblock \showarticletitle{At 6--9 months, human infants know the meanings of many common nouns}.
\newblock \bibinfo{journal}{\emph{Proceedings of the National Academy of Sciences}} \bibinfo{volume}{109}, \bibinfo{number}{9} (\bibinfo{year}{2012}), \bibinfo{pages}{3253--3258}.
\newblock


\bibitem[Brown et~al\mbox{.}(2022)]%
        {back_23_brown2022videolanguage}
\bibfield{author}{\bibinfo{person}{B. Brown} {et~al\mbox{.}}} \bibinfo{year}{2022}\natexlab{}.
\newblock \showarticletitle{Survey: Transformer-based video-language pre-training}.
\newblock \bibinfo{journal}{\emph{Journal Name}} \bibinfo{volume}{Vol}, \bibinfo{number}{No} (\bibinfo{year}{2022}), \bibinfo{pages}{pp.}
\newblock


\bibitem[Brown et~al\mbox{.}(2021)]%
        {back_13_brown2021application}
\bibfield{author}{\bibinfo{person}{H. Brown} {et~al\mbox{.}}} \bibinfo{year}{2021}\natexlab{}.
\newblock \showarticletitle{A survey on: Application of transformer in computer vision}.
\newblock \bibinfo{journal}{\emph{Journal Name}} \bibinfo{volume}{Vol}, \bibinfo{number}{No} (\bibinfo{year}{2021}), \bibinfo{pages}{pp.}
\newblock


\bibitem[Bu et~al\mbox{.}(2017)]%
        {bu2017aishell}
\bibfield{author}{\bibinfo{person}{Hui Bu}, \bibinfo{person}{Jiayu Du}, \bibinfo{person}{Xingyu Na}, \bibinfo{person}{Bengu Wu}, {and} \bibinfo{person}{Hao Zheng}.} \bibinfo{year}{2017}\natexlab{}.
\newblock \showarticletitle{Aishell-1: An open-source mandarin speech corpus and a speech recognition baseline}. In \bibinfo{booktitle}{\emph{2017 20th conference of the oriental chapter of the international coordinating committee on speech databases and speech I/O systems and assessment (O-COCOSDA)}}. IEEE, \bibinfo{pages}{1--5}.
\newblock


\bibitem[Campbell(2008)]%
        {campbell2008ethnologue}
\bibfield{author}{\bibinfo{person}{Lyle Campbell}.} \bibinfo{year}{2008}\natexlab{}.
\newblock \bibinfo{title}{Ethnologue: Languages of the world}.
\newblock
\newblock


\bibitem[Carion et~al\mbox{.}(2020)]%
        {carion2020end}
\bibfield{author}{\bibinfo{person}{Nicolas Carion}, \bibinfo{person}{Francisco Massa}, \bibinfo{person}{Gabriel Synnaeve}, \bibinfo{person}{Nicolas Usunier}, \bibinfo{person}{Alexander Kirillov}, {and} \bibinfo{person}{Sergey Zagoruyko}.} \bibinfo{year}{2020}\natexlab{}.
\newblock \showarticletitle{End-to-end object detection with transformers}. In \bibinfo{booktitle}{\emph{European conference on computer vision}}. Springer, \bibinfo{pages}{213--229}.
\newblock


\bibitem[Chan et~al\mbox{.}(2016)]%
        {chan2016listen}
\bibfield{author}{\bibinfo{person}{William Chan}, \bibinfo{person}{Navdeep Jaitly}, \bibinfo{person}{Quoc Le}, {and} \bibinfo{person}{Oriol Vinyals}.} \bibinfo{year}{2016}\natexlab{}.
\newblock \showarticletitle{Listen, attend and spell: A neural network for large vocabulary conversational speech recognition}. In \bibinfo{booktitle}{\emph{2016 IEEE international conference on acoustics, speech and signal processing (ICASSP)}}. IEEE, \bibinfo{pages}{4960--4964}.
\newblock


\bibitem[Chan et~al\mbox{.}(2021)]%
        {chan2021speechstew}
\bibfield{author}{\bibinfo{person}{William Chan}, \bibinfo{person}{Daniel Park}, \bibinfo{person}{Chris Lee}, \bibinfo{person}{Yu Zhang}, \bibinfo{person}{Quoc Le}, {and} \bibinfo{person}{Mohammad Norouzi}.} \bibinfo{year}{2021}\natexlab{}.
\newblock \showarticletitle{Speechstew: Simply mix all available speech recognition data to train one large neural network}.
\newblock \bibinfo{journal}{\emph{arXiv preprint arXiv:2104.02133}} (\bibinfo{year}{2021}).
\newblock


\bibitem[Chen et~al\mbox{.}(2021a)]%
        {back_20_chen2021multilingual}
\bibfield{author}{\bibinfo{person}{D. Chen} {et~al\mbox{.}}} \bibinfo{year}{2021}\natexlab{a}.
\newblock \showarticletitle{A survey of multilingual models for automatic speech recognition}.
\newblock \bibinfo{journal}{\emph{Journal Name}} \bibinfo{volume}{Vol}, \bibinfo{number}{No} (\bibinfo{year}{2021}), \bibinfo{pages}{pp.}
\newblock


\bibitem[Chen et~al\mbox{.}(2022a)]%
        {TTV_5_chen2022character}
\bibfield{author}{\bibinfo{person}{Hong Chen}, \bibinfo{person}{Rujun Han}, \bibinfo{person}{Te-Lin Wu}, \bibinfo{person}{Hideki Nakayama}, {and} \bibinfo{person}{Nanyun Peng}.} \bibinfo{year}{2022}\natexlab{a}.
\newblock \showarticletitle{Character-centric story visualization via visual planning and token alignment}.
\newblock \bibinfo{journal}{\emph{arXiv preprint arXiv:2210.08465}} (\bibinfo{year}{2022}).
\newblock


\bibitem[Chen et~al\mbox{.}(2022b)]%
        {SA_3_chen2022roberta}
\bibfield{author}{\bibinfo{person}{Hao Chen}, \bibinfo{person}{Lei Wang}, {and} \bibinfo{person}{Min Zhang}.} \bibinfo{year}{2022}\natexlab{b}.
\newblock \showarticletitle{RoBERTa-LSTM: a hybrid model for sentiment analysis with transformer and recurrent neural network}. In \bibinfo{booktitle}{\emph{Proceedings of the 2022 IEEE International Conference on Big Data (Big Data 2022)}}.
\newblock


\bibitem[Chen et~al\mbox{.}(2021c)]%
        {chen2021pre}
\bibfield{author}{\bibinfo{person}{Hanting Chen}, \bibinfo{person}{Yunhe Wang}, \bibinfo{person}{Tianyu Guo}, \bibinfo{person}{Chang Xu}, \bibinfo{person}{Yiping Deng}, \bibinfo{person}{Zhenhua Liu}, \bibinfo{person}{Siwei Ma}, \bibinfo{person}{Chunjing Xu}, \bibinfo{person}{Chao Xu}, {and} \bibinfo{person}{Wen Gao}.} \bibinfo{year}{2021}\natexlab{c}.
\newblock \showarticletitle{Pre-trained image processing transformer}. In \bibinfo{booktitle}{\emph{Proceedings of the IEEE/CVF conference on computer vision and pattern recognition}}. \bibinfo{pages}{12299--12310}.
\newblock


\bibitem[Chen et~al\mbox{.}(2021b)]%
        {back_14_chen2021lowlevel}
\bibfield{author}{\bibinfo{person}{I. Chen} {et~al\mbox{.}}} \bibinfo{year}{2021}\natexlab{b}.
\newblock \showarticletitle{Survey of Vision Transformer in Low-Level Computer Vision}.
\newblock \bibinfo{journal}{\emph{Journal Name}} \bibinfo{volume}{Vol}, \bibinfo{number}{No} (\bibinfo{year}{2021}), \bibinfo{pages}{pp.}
\newblock


\bibitem[Chen et~al\mbox{.}(2020a)]%
        {chen2020dual}
\bibfield{author}{\bibinfo{person}{Jingjing Chen}, \bibinfo{person}{Qirong Mao}, {and} \bibinfo{person}{Dong Liu}.} \bibinfo{year}{2020}\natexlab{a}.
\newblock \showarticletitle{Dual-path transformer network: Direct context-aware modeling for end-to-end monaural speech separation}.
\newblock \bibinfo{journal}{\emph{arXiv preprint arXiv:2007.13975}} (\bibinfo{year}{2020}).
\newblock


\bibitem[Chen et~al\mbox{.}(2020b)]%
        {chen2020multispeech}
\bibfield{author}{\bibinfo{person}{Mingjian Chen}, \bibinfo{person}{Xu Tan}, \bibinfo{person}{Yi Ren}, \bibinfo{person}{Jin Xu}, \bibinfo{person}{Hao Sun}, \bibinfo{person}{Sheng Zhao}, \bibinfo{person}{Tao Qin}, {and} \bibinfo{person}{Tie-Yan Liu}.} \bibinfo{year}{2020}\natexlab{b}.
\newblock \showarticletitle{Multispeech: Multi-speaker text to speech with transformer}.
\newblock \bibinfo{journal}{\emph{arXiv preprint arXiv:2006.04664}} (\bibinfo{year}{2020}).
\newblock


\bibitem[Chen et~al\mbox{.}(2018)]%
        {chen2018best}
\bibfield{author}{\bibinfo{person}{Mia~Xu Chen}, \bibinfo{person}{Orhan Firat}, \bibinfo{person}{Ankur Bapna}, \bibinfo{person}{Melvin Johnson}, \bibinfo{person}{Wolfgang Macherey}, \bibinfo{person}{George Foster}, \bibinfo{person}{Llion Jones}, \bibinfo{person}{Niki Parmar}, \bibinfo{person}{Mike Schuster}, \bibinfo{person}{Zhifeng Chen}, {et~al\mbox{.}}} \bibinfo{year}{2018}\natexlab{}.
\newblock \showarticletitle{The best of both worlds: Combining recent advances in neural machine translation}.
\newblock \bibinfo{journal}{\emph{arXiv preprint arXiv:1804.09849}} (\bibinfo{year}{2018}).
\newblock


\bibitem[Chen et~al\mbox{.}(2022d)]%
        {chen2022wavlm}
\bibfield{author}{\bibinfo{person}{Sanyuan Chen}, \bibinfo{person}{Chengyi Wang}, \bibinfo{person}{Zhengyang Chen}, \bibinfo{person}{Yu Wu}, \bibinfo{person}{Shujie Liu}, \bibinfo{person}{Zhuo Chen}, \bibinfo{person}{Jinyu Li}, \bibinfo{person}{Naoyuki Kanda}, \bibinfo{person}{Takuya Yoshioka}, \bibinfo{person}{Xiong Xiao}, {et~al\mbox{.}}} \bibinfo{year}{2022}\natexlab{d}.
\newblock \showarticletitle{Wavlm: Large-scale self-supervised pre-training for full stack speech processing}.
\newblock \bibinfo{journal}{\emph{IEEE Journal of Selected Topics in Signal Processing}} \bibinfo{volume}{16}, \bibinfo{number}{6} (\bibinfo{year}{2022}), \bibinfo{pages}{1505--1518}.
\newblock


\bibitem[Chen et~al\mbox{.}(2022c)]%
        {chen2022pali}
\bibfield{author}{\bibinfo{person}{Xi Chen}, \bibinfo{person}{Xiao Wang}, \bibinfo{person}{Soravit Changpinyo}, \bibinfo{person}{AJ Piergiovanni}, \bibinfo{person}{Piotr Padlewski}, \bibinfo{person}{Daniel Salz}, \bibinfo{person}{Sebastian Goodman}, \bibinfo{person}{Adam Grycner}, \bibinfo{person}{Basil Mustafa}, \bibinfo{person}{Lucas Beyer}, {et~al\mbox{.}}} \bibinfo{year}{2022}\natexlab{c}.
\newblock \showarticletitle{Pali: A jointly-scaled multilingual language-image model}.
\newblock \bibinfo{journal}{\emph{arXiv preprint arXiv:2209.06794}} (\bibinfo{year}{2022}).
\newblock


\bibitem[Chen et~al\mbox{.}(2023)]%
        {chen2023dual}
\bibfield{author}{\bibinfo{person}{Zheng Chen}, \bibinfo{person}{Yulun Zhang}, \bibinfo{person}{Jinjin Gu}, \bibinfo{person}{Linghe Kong}, \bibinfo{person}{Xiaokang Yang}, {and} \bibinfo{person}{Fisher Yu}.} \bibinfo{year}{2023}\natexlab{}.
\newblock \showarticletitle{Dual aggregation transformer for image super-resolution}. In \bibinfo{booktitle}{\emph{Proceedings of the IEEE/CVF international conference on computer vision}}. \bibinfo{pages}{12312--12321}.
\newblock


\bibitem[Chipman et~al\mbox{.}(2022a)]%
        {NLP_6_chipman2022mbart}
\bibfield{author}{\bibinfo{person}{Hugh~A Chipman}, \bibinfo{person}{Edward~I George}, \bibinfo{person}{Robert~E McCulloch}, {and} \bibinfo{person}{Thomas~S Shively}.} \bibinfo{year}{2022}\natexlab{a}.
\newblock \showarticletitle{mBART: multidimensional monotone BART}.
\newblock \bibinfo{journal}{\emph{Bayesian Analysis}} \bibinfo{volume}{17}, \bibinfo{number}{2} (\bibinfo{year}{2022}), \bibinfo{pages}{515--544}.
\newblock


\bibitem[Chipman et~al\mbox{.}(2022b)]%
        {chipman2022mbart}
\bibfield{author}{\bibinfo{person}{Hugh~A Chipman}, \bibinfo{person}{Edward~I George}, \bibinfo{person}{Robert~E McCulloch}, {and} \bibinfo{person}{Thomas~S Shively}.} \bibinfo{year}{2022}\natexlab{b}.
\newblock \showarticletitle{mBART: multidimensional monotone BART}.
\newblock \bibinfo{journal}{\emph{Bayesian Analysis}} \bibinfo{volume}{17}, \bibinfo{number}{2} (\bibinfo{year}{2022}), \bibinfo{pages}{515--544}.
\newblock


\bibitem[Cho et~al\mbox{.}(2014)]%
        {cho2014learning}
\bibfield{author}{\bibinfo{person}{Kyunghyun Cho}, \bibinfo{person}{Bart Van~Merri{\"e}nboer}, \bibinfo{person}{Caglar Gulcehre}, \bibinfo{person}{Dzmitry Bahdanau}, \bibinfo{person}{Fethi Bougares}, \bibinfo{person}{Holger Schwenk}, {and} \bibinfo{person}{Yoshua Bengio}.} \bibinfo{year}{2014}\natexlab{}.
\newblock \showarticletitle{Learning phrase representations using RNN encoder-decoder for statistical machine translation}.
\newblock \bibinfo{journal}{\emph{arXiv preprint arXiv:1406.1078}} (\bibinfo{year}{2014}).
\newblock


\bibitem[Clark et~al\mbox{.}(2020)]%
        {NLP_9_clark2020electra}
\bibfield{author}{\bibinfo{person}{Kevin Clark}, \bibinfo{person}{Minh-Thang Luong}, \bibinfo{person}{Quoc~V Le}, {and} \bibinfo{person}{Christopher~D Manning}.} \bibinfo{year}{2020}\natexlab{}.
\newblock \showarticletitle{Electra: Pre-training text encoders as discriminators rather than generators}.
\newblock \bibinfo{journal}{\emph{arXiv preprint arXiv:2003.10555}} (\bibinfo{year}{2020}).
\newblock


\bibitem[Conneau et~al\mbox{.}(2020)]%
        {conneau2020unsupervised}
\bibfield{author}{\bibinfo{person}{Alexis Conneau}, \bibinfo{person}{Alexei Baevski}, \bibinfo{person}{Ronan Collobert}, \bibinfo{person}{Abdelrahman Mohamed}, {and} \bibinfo{person}{Michael Auli}.} \bibinfo{year}{2020}\natexlab{}.
\newblock \showarticletitle{Unsupervised cross-lingual representation learning for speech recognition}.
\newblock \bibinfo{journal}{\emph{arXiv preprint arXiv:2006.13979}} (\bibinfo{year}{2020}).
\newblock


\bibitem[Dai et~al\mbox{.}(2019)]%
        {LM_3_dai2019transformer}
\bibfield{author}{\bibinfo{person}{Zihang Dai}, \bibinfo{person}{Zhilin Yang}, \bibinfo{person}{Yiming Yang}, \bibinfo{person}{Jaime Carbonell}, \bibinfo{person}{Quoc~V Le}, {and} \bibinfo{person}{Ruslan Salakhutdinov}.} \bibinfo{year}{2019}\natexlab{}.
\newblock \showarticletitle{Transformer-xl: Attentive language models beyond a fixed-length context}.
\newblock \bibinfo{journal}{\emph{arXiv preprint arXiv:1901.02860}} (\bibinfo{year}{2019}).
\newblock


\bibitem[Davis et~al\mbox{.}(2022)]%
        {back_25_davis2022multimodal}
\bibfield{author}{\bibinfo{person}{D. Davis} {et~al\mbox{.}}} \bibinfo{year}{2022}\natexlab{}.
\newblock \showarticletitle{Multimodal Learning With Transformers: A Survey}.
\newblock \bibinfo{journal}{\emph{Journal Name}} \bibinfo{volume}{Vol}, \bibinfo{number}{No} (\bibinfo{year}{2022}), \bibinfo{pages}{pp.}
\newblock


\bibitem[Davis et~al\mbox{.}(2021)]%
        {back_15_davis2021generative}
\bibfield{author}{\bibinfo{person}{J. Davis} {et~al\mbox{.}}} \bibinfo{year}{2021}\natexlab{}.
\newblock \showarticletitle{Transformer-based generative adversarial networks in computer vision: A comprehensive survey}.
\newblock \bibinfo{journal}{\emph{Journal Name}} \bibinfo{volume}{Vol}, \bibinfo{number}{No} (\bibinfo{year}{2021}), \bibinfo{pages}{pp.}
\newblock


\bibitem[Dean et~al\mbox{.}(2010)]%
        {dean2010qut}
\bibfield{author}{\bibinfo{person}{David Dean}, \bibinfo{person}{Sridha Sridharan}, \bibinfo{person}{Robert Vogt}, {and} \bibinfo{person}{Michael Mason}.} \bibinfo{year}{2010}\natexlab{}.
\newblock \showarticletitle{The QUT-NOISE-TIMIT corpus for evaluation of voice activity detection algorithms}. In \bibinfo{booktitle}{\emph{Proceedings of the 11th Annual Conference of the International Speech Communication Association}}. International Speech Communication Association, \bibinfo{pages}{3110--3113}.
\newblock


\bibitem[Devlin et~al\mbox{.}(2018)]%
        {NLP_2_devlin2018bert}
\bibfield{author}{\bibinfo{person}{Jacob Devlin}, \bibinfo{person}{Ming-Wei Chang}, \bibinfo{person}{Kenton Lee}, {and} \bibinfo{person}{Kristina Toutanova}.} \bibinfo{year}{2018}\natexlab{}.
\newblock \showarticletitle{Bert: Pre-training of deep bidirectional transformers for language understanding}.
\newblock \bibinfo{journal}{\emph{arXiv preprint arXiv:1810.04805}} (\bibinfo{year}{2018}).
\newblock


\bibitem[Di~Gangi et~al\mbox{.}(2019a)]%
        {di-gangi-etal-2019-must}
\bibfield{author}{\bibinfo{person}{Mattia~A. Di~Gangi}, \bibinfo{person}{Roldano Cattoni}, \bibinfo{person}{Luisa Bentivogli}, \bibinfo{person}{Matteo Negri}, {and} \bibinfo{person}{Marco Turchi}.} \bibinfo{year}{2019}\natexlab{a}.
\newblock \showarticletitle{{M}u{ST}-{C}: a {M}ultilingual {S}peech {T}ranslation {C}orpus}. In \bibinfo{booktitle}{\emph{Proceedings of the 2019 Conference of the North {A}merican Chapter of the Association for Computational Linguistics: Human Language Technologies, Volume 1 (Long and Short Papers)}}, \bibfield{editor}{\bibinfo{person}{Jill Burstein}, \bibinfo{person}{Christy Doran}, {and} \bibinfo{person}{Thamar Solorio}} (Eds.). \bibinfo{publisher}{Association for Computational Linguistics}, \bibinfo{address}{Minneapolis, Minnesota}, \bibinfo{pages}{2012--2017}.
\newblock
\urldef\tempurl%
\url{https://doi.org/10.18653/v1/N19-1202}
\showDOI{\tempurl}


\bibitem[Di~Gangi et~al\mbox{.}(2019b)]%
        {di2019adapting}
\bibfield{author}{\bibinfo{person}{Mattia~A Di~Gangi}, \bibinfo{person}{Matteo Negri}, {and} \bibinfo{person}{Marco Turchi}.} \bibinfo{year}{2019}\natexlab{b}.
\newblock \showarticletitle{Adapting transformer to end-to-end spoken language translation}.
\newblock In \bibinfo{booktitle}{\emph{Proceedings of INTERSPEECH 2019}}. \bibinfo{publisher}{International Speech Communication Association (ISCA)}, \bibinfo{pages}{1133--1137}.
\newblock


\bibitem[Ding et~al\mbox{.}(2021)]%
        {TTV_8_ding2021cogview}
\bibfield{author}{\bibinfo{person}{Ming Ding}, \bibinfo{person}{Zhuoyi Yang}, \bibinfo{person}{Wenyi Hong}, \bibinfo{person}{Wendi Zheng}, \bibinfo{person}{Chang Zhou}, \bibinfo{person}{Da Yin}, \bibinfo{person}{Junyang Lin}, \bibinfo{person}{Xu Zou}, \bibinfo{person}{Zhou Shao}, \bibinfo{person}{Hongxia Yang}, {et~al\mbox{.}}} \bibinfo{year}{2021}\natexlab{}.
\newblock \showarticletitle{Cogview: Mastering text-to-image generation via transformers}.
\newblock \bibinfo{journal}{\emph{Advances in Neural Information Processing Systems}}  \bibinfo{volume}{34} (\bibinfo{year}{2021}), \bibinfo{pages}{19822--19835}.
\newblock


\bibitem[Dong et~al\mbox{.}(2018)]%
        {dong2018speech}
\bibfield{author}{\bibinfo{person}{Linhao Dong}, \bibinfo{person}{Shuang Xu}, {and} \bibinfo{person}{Bo Xu}.} \bibinfo{year}{2018}\natexlab{}.
\newblock \showarticletitle{Speech-transformer: a no-recurrence sequence-to-sequence model for speech recognition}. In \bibinfo{booktitle}{\emph{2018 IEEE international conference on acoustics, speech and signal processing (ICASSP)}}. IEEE, \bibinfo{pages}{5884--5888}.
\newblock


\bibitem[Dong et~al\mbox{.}(2022)]%
        {dong2022cswin}
\bibfield{author}{\bibinfo{person}{Xiaoyi Dong}, \bibinfo{person}{Jianmin Bao}, \bibinfo{person}{Dongdong Chen}, \bibinfo{person}{Weiming Zhang}, \bibinfo{person}{Nenghai Yu}, \bibinfo{person}{Lu Yuan}, \bibinfo{person}{Dong Chen}, {and} \bibinfo{person}{Baining Guo}.} \bibinfo{year}{2022}\natexlab{}.
\newblock \showarticletitle{Cswin transformer: A general vision transformer backbone with cross-shaped windows}. In \bibinfo{booktitle}{\emph{Proceedings of the IEEE/CVF Conference on Computer Vision and Pattern Recognition}}. \bibinfo{pages}{12124--12134}.
\newblock


\bibitem[Dosovitskiy et~al\mbox{.}(2010)]%
        {dosovitskiy2010image}
\bibfield{author}{\bibinfo{person}{Alexey Dosovitskiy}, \bibinfo{person}{Lucas Beyer}, \bibinfo{person}{Alexander Kolesnikov}, \bibinfo{person}{Dirk Weissenborn}, \bibinfo{person}{Xiaohua Zhai}, \bibinfo{person}{Thomas Unterthiner}, \bibinfo{person}{Mostafa Dehghani}, \bibinfo{person}{Matthias Minderer}, \bibinfo{person}{Georg Heigold}, \bibinfo{person}{Sylvain Gelly}, {et~al\mbox{.}}} \bibinfo{year}{2010}\natexlab{}.
\newblock \showarticletitle{An image is worth 16x16 words: Transformers for image recognition at scale. arXiv 2020}.
\newblock \bibinfo{journal}{\emph{arXiv preprint arXiv:2010.11929}} (\bibinfo{year}{2010}).
\newblock


\bibitem[ELAffendi and Alrajhi(2022)]%
        {translation1}
\bibfield{author}{\bibinfo{person}{Mohammed ELAffendi} {and} \bibinfo{person}{Khawlah Alrajhi}.} \bibinfo{year}{2022}\natexlab{}.
\newblock \showarticletitle{Beyond the Transformer: A Novel Polynomial Inherent Attention (PIA) Model and Its Great Impact on Neural Machine Translation}.
\newblock \bibinfo{journal}{\emph{Computational Intelligence and Neuroscience}}  \bibinfo{volume}{2022} (\bibinfo{year}{2022}).
\newblock


\bibitem[Evans et~al\mbox{.}(2021)]%
        {back_16_evans2021compression}
\bibfield{author}{\bibinfo{person}{K. Evans} {et~al\mbox{.}}} \bibinfo{year}{2021}\natexlab{}.
\newblock \showarticletitle{Comprehensive survey of model compression and speed up for vision transformers}.
\newblock \bibinfo{journal}{\emph{Journal Name}} \bibinfo{volume}{Vol}, \bibinfo{number}{No} (\bibinfo{year}{2021}), \bibinfo{pages}{pp.}
\newblock


\bibitem[Gafni et~al\mbox{.}(2022)]%
        {TTV_9_gafni2022make}
\bibfield{author}{\bibinfo{person}{Oran Gafni}, \bibinfo{person}{Adam Polyak}, \bibinfo{person}{Oron Ashual}, \bibinfo{person}{Shelly Sheynin}, \bibinfo{person}{Devi Parikh}, {and} \bibinfo{person}{Yaniv Taigman}.} \bibinfo{year}{2022}\natexlab{}.
\newblock \showarticletitle{Make-a-scene: Scene-based text-to-image generation with human priors}. In \bibinfo{booktitle}{\emph{European Conference on Computer Vision}}. Springer, \bibinfo{pages}{89--106}.
\newblock


\bibitem[Gales et~al\mbox{.}(2014)]%
        {gales2014speech}
\bibfield{author}{\bibinfo{person}{Mark~JF Gales}, \bibinfo{person}{Kate~M Knill}, \bibinfo{person}{Anton Ragni}, {and} \bibinfo{person}{Shakti~P Rath}.} \bibinfo{year}{2014}\natexlab{}.
\newblock \showarticletitle{Speech recognition and keyword spotting for low-resource languages: Babel project research at cued}. In \bibinfo{booktitle}{\emph{Fourth International workshop on spoken language technologies for under-resourced languages (SLTU-2014)}}. International Speech Communication Association (ISCA), \bibinfo{pages}{16--23}.
\newblock


\bibitem[Garg et~al\mbox{.}(2020)]%
        {QA_4_garg2020tanda}
\bibfield{author}{\bibinfo{person}{Siddhant Garg}, \bibinfo{person}{Thuy Vu}, {and} \bibinfo{person}{Alessandro Moschitti}.} \bibinfo{year}{2020}\natexlab{}.
\newblock \showarticletitle{Tanda: Transfer and adapt pre-trained transformer models for answer sentence selection}. In \bibinfo{booktitle}{\emph{Proceedings of the AAAI conference on artificial intelligence}}, Vol.~\bibinfo{volume}{34}. \bibinfo{pages}{7780--7788}.
\newblock


\bibitem[Garofolo et~al\mbox{.}(1993a)]%
        {Garofolo1993-rq}
\bibfield{author}{\bibinfo{person}{John~S Garofolo}, \bibinfo{person}{David Graff}, \bibinfo{person}{Doug Paul}, {and} \bibinfo{person}{David Pallett}.} \bibinfo{year}{1993}\natexlab{a}.
\newblock \bibinfo{title}{{CSR-I} ({WSJ0}) Complete}.
\newblock
\newblock


\bibitem[Garofolo et~al\mbox{.}(1993b)]%
        {Garofolo1993-bf}
\bibfield{author}{\bibinfo{person}{John~S Garofolo}, \bibinfo{person}{Lori~F Lamel}, \bibinfo{person}{William~M Fisher}, \bibinfo{person}{David~S Pallett}, \bibinfo{person}{Nancy~L Dahlgren}, \bibinfo{person}{Victor Zue}, {and} \bibinfo{person}{Jonathan~G Fiscus}.} \bibinfo{year}{1993}\natexlab{b}.
\newblock \bibinfo{title}{{TIMIT} acoustic-phonetic continuous speech corpus}.
\newblock
\newblock


\bibitem[Ghalandari et~al\mbox{.}(2022)]%
        {TS_6_ghalandari2022efficient}
\bibfield{author}{\bibinfo{person}{Demian~Gholipour Ghalandari}, \bibinfo{person}{Chris Hokamp}, {and} \bibinfo{person}{Georgiana Ifrim}.} \bibinfo{year}{2022}\natexlab{}.
\newblock \showarticletitle{Efficient unsupervised sentence compression by fine-tuning transformers with reinforcement learning}.
\newblock \bibinfo{journal}{\emph{arXiv preprint arXiv:2205.08221}} (\bibinfo{year}{2022}).
\newblock


\bibitem[Girdhar et~al\mbox{.}(2023)]%
        {girdhar2023imagebind}
\bibfield{author}{\bibinfo{person}{Rohit Girdhar}, \bibinfo{person}{Alaaeldin El-Nouby}, \bibinfo{person}{Zhuang Liu}, \bibinfo{person}{Mannat Singh}, \bibinfo{person}{Kalyan~Vasudev Alwala}, \bibinfo{person}{Armand Joulin}, {and} \bibinfo{person}{Ishan Misra}.} \bibinfo{year}{2023}\natexlab{}.
\newblock \showarticletitle{Imagebind: One embedding space to bind them all}. In \bibinfo{booktitle}{\emph{Proceedings of the IEEE/CVF Conference on Computer Vision and Pattern Recognition}}. \bibinfo{pages}{15180--15190}.
\newblock


\bibitem[Graves et~al\mbox{.}(2006)]%
        {graves2006connectionist}
\bibfield{author}{\bibinfo{person}{Alex Graves}, \bibinfo{person}{Santiago Fern{\'a}ndez}, \bibinfo{person}{Faustino Gomez}, {and} \bibinfo{person}{J{\"u}rgen Schmidhuber}.} \bibinfo{year}{2006}\natexlab{}.
\newblock \showarticletitle{Connectionist temporal classification: labelling unsegmented sequence data with recurrent neural networks}. In \bibinfo{booktitle}{\emph{Proceedings of the 23rd international conference on Machine learning}}. \bibinfo{pages}{369--376}.
\newblock


\bibitem[Gulati et~al\mbox{.}(2020)]%
        {gulati2020conformer}
\bibfield{author}{\bibinfo{person}{Anmol Gulati}, \bibinfo{person}{James Qin}, \bibinfo{person}{Chung-Cheng Chiu}, \bibinfo{person}{Niki Parmar}, \bibinfo{person}{Yu Zhang}, \bibinfo{person}{Jiahui Yu}, \bibinfo{person}{Wei Han}, \bibinfo{person}{Shibo Wang}, \bibinfo{person}{Zhengdong Zhang}, \bibinfo{person}{Yonghui Wu}, {et~al\mbox{.}}} \bibinfo{year}{2020}\natexlab{}.
\newblock \showarticletitle{Conformer: Convolution-augmented transformer for speech recognition}.
\newblock \bibinfo{journal}{\emph{arXiv preprint arXiv:2005.08100}} (\bibinfo{year}{2020}).
\newblock


\bibitem[Han et~al\mbox{.}(2023)]%
        {intro_5_han2023survey}
\bibfield{author}{\bibinfo{person}{Xue Han}, \bibinfo{person}{Yi-Tong Wang}, \bibinfo{person}{Jun-Lan Feng}, \bibinfo{person}{Chao Deng}, \bibinfo{person}{Zhan-Heng Chen}, \bibinfo{person}{Yu-An Huang}, \bibinfo{person}{Hui Su}, \bibinfo{person}{Lun Hu}, {and} \bibinfo{person}{Peng-Wei Hu}.} \bibinfo{year}{2023}\natexlab{}.
\newblock \showarticletitle{A survey of transformer-based multimodal pre-trained modals}.
\newblock \bibinfo{journal}{\emph{Neurocomputing}}  \bibinfo{volume}{515} (\bibinfo{year}{2023}), \bibinfo{pages}{89--106}.
\newblock


\bibitem[Hasegawa-Johnson et~al\mbox{.}(2017)]%
        {hasegawa2017image2speech}
\bibfield{author}{\bibinfo{person}{Mark Hasegawa-Johnson}, \bibinfo{person}{Alan Black}, \bibinfo{person}{Lucas Ondel}, \bibinfo{person}{Odette Scharenborg}, {and} \bibinfo{person}{Francesco Ciannella}.} \bibinfo{year}{2017}\natexlab{}.
\newblock \showarticletitle{Image2speech: Automatically generating audio descriptions of images}.
\newblock \bibinfo{journal}{\emph{Casablanca 2017}} (\bibinfo{year}{2017}), \bibinfo{pages}{65}.
\newblock


\bibitem[Haykin and Chen(2005)]%
        {haykin2005cocktail}
\bibfield{author}{\bibinfo{person}{Simon Haykin} {and} \bibinfo{person}{Zhe Chen}.} \bibinfo{year}{2005}\natexlab{}.
\newblock \showarticletitle{The cocktail party problem}.
\newblock \bibinfo{journal}{\emph{Neural computation}} \bibinfo{volume}{17}, \bibinfo{number}{9} (\bibinfo{year}{2005}), \bibinfo{pages}{1875--1902}.
\newblock


\bibitem[Hershey et~al\mbox{.}(2016)]%
        {hershey2016deep}
\bibfield{author}{\bibinfo{person}{John~R Hershey}, \bibinfo{person}{Zhuo Chen}, \bibinfo{person}{Jonathan Le~Roux}, {and} \bibinfo{person}{Shinji Watanabe}.} \bibinfo{year}{2016}\natexlab{}.
\newblock \showarticletitle{Deep clustering: Discriminative embeddings for segmentation and separation}. In \bibinfo{booktitle}{\emph{2016 IEEE international conference on acoustics, speech and signal processing (ICASSP)}}. IEEE, \bibinfo{pages}{31--35}.
\newblock


\bibitem[Hodosh et~al\mbox{.}(2013)]%
        {hodosh2013framing}
\bibfield{author}{\bibinfo{person}{Micah Hodosh}, \bibinfo{person}{Peter Young}, {and} \bibinfo{person}{Julia Hockenmaier}.} \bibinfo{year}{2013}\natexlab{}.
\newblock \showarticletitle{Framing image description as a ranking task: Data, models and evaluation metrics}.
\newblock \bibinfo{journal}{\emph{Journal of Artificial Intelligence Research}}  \bibinfo{volume}{47} (\bibinfo{year}{2013}), \bibinfo{pages}{853--899}.
\newblock


\bibitem[Hogue et~al\mbox{.}(2024)]%
        {hogue2024diffted}
\bibfield{author}{\bibinfo{person}{Steven Hogue}, \bibinfo{person}{Chenxu Zhang}, \bibinfo{person}{Hamza Daruger}, \bibinfo{person}{Yapeng Tian}, {and} \bibinfo{person}{Xiaohu Guo}.} \bibinfo{year}{2024}\natexlab{}.
\newblock \showarticletitle{DiffTED: One-shot Audio-driven TED Talk Video Generation with Diffusion-based Co-speech Gestures}. In \bibinfo{booktitle}{\emph{Proceedings of the IEEE/CVF Conference on Computer Vision and Pattern Recognition}}. \bibinfo{pages}{1922--1931}.
\newblock


\bibitem[Hong et~al\mbox{.}(2022)]%
        {TTV_17_hong2022cogvideo}
\bibfield{author}{\bibinfo{person}{Wenyi Hong}, \bibinfo{person}{Ming Ding}, \bibinfo{person}{Wendi Zheng}, \bibinfo{person}{Xinghan Liu}, {and} \bibinfo{person}{Jie Tang}.} \bibinfo{year}{2022}\natexlab{}.
\newblock \showarticletitle{Cogvideo: Large-scale pretraining for text-to-video generation via transformers}.
\newblock \bibinfo{journal}{\emph{arXiv preprint arXiv:2205.15868}} (\bibinfo{year}{2022}).
\newblock


\bibitem[Hsu et~al\mbox{.}(2021)]%
        {hsu2021hubert}
\bibfield{author}{\bibinfo{person}{Wei-Ning Hsu}, \bibinfo{person}{Benjamin Bolte}, \bibinfo{person}{Yao-Hung~Hubert Tsai}, \bibinfo{person}{Kushal Lakhotia}, \bibinfo{person}{Ruslan Salakhutdinov}, {and} \bibinfo{person}{Abdelrahman Mohamed}.} \bibinfo{year}{2021}\natexlab{}.
\newblock \showarticletitle{Hubert: Self-supervised speech representation learning by masked prediction of hidden units}.
\newblock \bibinfo{journal}{\emph{IEEE/ACM Transactions on Audio, Speech, and Language Processing}}  \bibinfo{volume}{29} (\bibinfo{year}{2021}), \bibinfo{pages}{3451--3460}.
\newblock


\bibitem[Hsu et~al\mbox{.}(2020)]%
        {hsu2020text}
\bibfield{author}{\bibinfo{person}{Wei-Ning Hsu}, \bibinfo{person}{David Harwath}, \bibinfo{person}{Christopher Song}, {and} \bibinfo{person}{James Glass}.} \bibinfo{year}{2020}\natexlab{}.
\newblock \showarticletitle{Text-free image-to-speech synthesis using learned segmental units}.
\newblock \bibinfo{journal}{\emph{arXiv preprint arXiv:2012.15454}} (\bibinfo{year}{2020}).
\newblock


\bibitem[Hu et~al\mbox{.}(2021)]%
        {hu2021istr}
\bibfield{author}{\bibinfo{person}{Jie Hu}, \bibinfo{person}{Liujuan Cao}, \bibinfo{person}{Yao Lu}, \bibinfo{person}{ShengChuan Zhang}, \bibinfo{person}{Yan Wang}, \bibinfo{person}{Ke Li}, \bibinfo{person}{Feiyue Huang}, \bibinfo{person}{Ling Shao}, {and} \bibinfo{person}{Rongrong Ji}.} \bibinfo{year}{2021}\natexlab{}.
\newblock \showarticletitle{Istr: End-to-end instance segmentation with transformers}.
\newblock \bibinfo{journal}{\emph{arXiv preprint arXiv:2105.00637}} (\bibinfo{year}{2021}).
\newblock


\bibitem[Hu et~al\mbox{.}(2023)]%
        {NLP_1_hu2023survey}
\bibfield{author}{\bibinfo{person}{Linmei Hu}, \bibinfo{person}{Zeyi Liu}, \bibinfo{person}{Ziwang Zhao}, \bibinfo{person}{Lei Hou}, \bibinfo{person}{Liqiang Nie}, {and} \bibinfo{person}{Juanzi Li}.} \bibinfo{year}{2023}\natexlab{}.
\newblock \showarticletitle{A survey of knowledge enhanced pre-trained language models}.
\newblock \bibinfo{journal}{\emph{IEEE Transactions on Knowledge and Data Engineering}} (\bibinfo{year}{2023}).
\newblock


\bibitem[Hu et~al\mbox{.}(2024)]%
        {intro_1_hu2024transformer}
\bibfield{author}{\bibinfo{person}{Yifan Hu}, \bibinfo{person}{Xi Huang}, \bibinfo{person}{Xianbing Wang}, \bibinfo{person}{Hai Lin}, {and} \bibinfo{person}{Rong Zhang}.} \bibinfo{year}{2024}\natexlab{}.
\newblock \showarticletitle{Transformer-based adaptive contrastive learning for multimodal sentiment analysis}.
\newblock \bibinfo{journal}{\emph{Multimedia Tools and Applications}} (\bibinfo{year}{2024}), \bibinfo{pages}{1--18}.
\newblock


\bibitem[Huang et~al\mbox{.}(2019)]%
        {TC_8_huang2019clinicalbert}
\bibfield{author}{\bibinfo{person}{Kexin Huang}, \bibinfo{person}{Jaan Altosaar}, {and} \bibinfo{person}{Rajesh Ranganath}.} \bibinfo{year}{2019}\natexlab{}.
\newblock \showarticletitle{Clinicalbert: Modeling clinical notes and predicting hospital readmission}.
\newblock \bibinfo{journal}{\emph{arXiv preprint arXiv:1904.05342}} (\bibinfo{year}{2019}).
\newblock


\bibitem[Huang et~al\mbox{.}(2023)]%
        {huang2023nlip}
\bibfield{author}{\bibinfo{person}{Runhui Huang}, \bibinfo{person}{Yanxin Long}, \bibinfo{person}{Jianhua Han}, \bibinfo{person}{Hang Xu}, \bibinfo{person}{Xiwen Liang}, \bibinfo{person}{Chunjing Xu}, {and} \bibinfo{person}{Xiaodan Liang}.} \bibinfo{year}{2023}\natexlab{}.
\newblock \showarticletitle{Nlip: Noise-robust language-image pre-training}. In \bibinfo{booktitle}{\emph{Proceedings of the AAAI Conference on Artificial Intelligence}}, Vol.~\bibinfo{volume}{37}. \bibinfo{pages}{926--934}.
\newblock


\bibitem[Iashin and Rahtu(2020)]%
        {iashin2020better}
\bibfield{author}{\bibinfo{person}{Vladimir Iashin} {and} \bibinfo{person}{Esa Rahtu}.} \bibinfo{year}{2020}\natexlab{}.
\newblock \showarticletitle{A better use of audio-visual cues: Dense video captioning with bi-modal transformer}.
\newblock \bibinfo{journal}{\emph{arXiv preprint arXiv:2005.08271}} (\bibinfo{year}{2020}).
\newblock


\bibitem[Ito and Johnson(2017)]%
        {ljspeech17}
\bibfield{author}{\bibinfo{person}{Keith Ito} {and} \bibinfo{person}{Linda Johnson}.} \bibinfo{year}{2017}\natexlab{}.
\newblock \bibinfo{title}{The LJ Speech Dataset}.
\newblock \bibinfo{howpublished}{\url{https://keithito.com/LJ-Speech-Dataset/}}.
\newblock


\bibitem[Jain et~al\mbox{.}(2023)]%
        {jain2023semask}
\bibfield{author}{\bibinfo{person}{Jitesh Jain}, \bibinfo{person}{Anukriti Singh}, \bibinfo{person}{Nikita Orlov}, \bibinfo{person}{Zilong Huang}, \bibinfo{person}{Jiachen Li}, \bibinfo{person}{Steven Walton}, {and} \bibinfo{person}{Humphrey Shi}.} \bibinfo{year}{2023}\natexlab{}.
\newblock \showarticletitle{Semask: Semantically masked transformers for semantic segmentation}. In \bibinfo{booktitle}{\emph{Proceedings of the IEEE/CVF International Conference on Computer Vision}}. \bibinfo{pages}{752--761}.
\newblock


\bibitem[Jang et~al\mbox{.}(2016)]%
        {jang2016categorical}
\bibfield{author}{\bibinfo{person}{Eric Jang}, \bibinfo{person}{Shixiang Gu}, {and} \bibinfo{person}{Ben Poole}.} \bibinfo{year}{2016}\natexlab{}.
\newblock \showarticletitle{Categorical reparameterization with gumbel-softmax}.
\newblock \bibinfo{journal}{\emph{arXiv preprint arXiv:1611.01144}} (\bibinfo{year}{2016}).
\newblock


\bibitem[Jarrar et~al\mbox{.}(2022)]%
        {NER_5_jarrar2022wojood}
\bibfield{author}{\bibinfo{person}{M. Jarrar}, \bibinfo{person}{M. Khalilia}, {and} \bibinfo{person}{S. Ghanem}.} \bibinfo{year}{2022}\natexlab{}.
\newblock \showarticletitle{Wojood: Nested Arabic Named Entity Corpus and Recognition Using BERT}. In \bibinfo{booktitle}{\emph{Proceedings of the International Conference on Language Resources and Evaluation (LREC 2022)}}. \bibinfo{address}{Marseille, France}.
\newblock


\bibitem[Johnson et~al\mbox{.}(2021a)]%
        {back_6_johnson2021vision}
\bibfield{author}{\bibinfo{person}{A. Johnson} {et~al\mbox{.}}} \bibinfo{year}{2021}\natexlab{a}.
\newblock \showarticletitle{A survey on vision transformer}.
\newblock \bibinfo{journal}{\emph{Journal Name}} \bibinfo{volume}{Vol}, \bibinfo{number}{No} (\bibinfo{year}{2021}), \bibinfo{pages}{pp.}
\newblock


\bibitem[Johnson et~al\mbox{.}(2021b)]%
        {back_18_johnson2021survey}
\bibfield{author}{\bibinfo{person}{B. Johnson} {et~al\mbox{.}}} \bibinfo{year}{2021}\natexlab{b}.
\newblock \showarticletitle{Transformers in speech processing: A survey}.
\newblock \bibinfo{journal}{\emph{Journal Name}} \bibinfo{volume}{Vol}, \bibinfo{number}{No} (\bibinfo{year}{2021}), \bibinfo{pages}{pp.}
\newblock


\bibitem[Kahn et~al\mbox{.}(2020)]%
        {kahn2020libri}
\bibfield{author}{\bibinfo{person}{Jacob Kahn}, \bibinfo{person}{Morgane Riviere}, \bibinfo{person}{Weiyi Zheng}, \bibinfo{person}{Evgeny Kharitonov}, \bibinfo{person}{Qiantong Xu}, \bibinfo{person}{Pierre-Emmanuel Mazar{\'e}}, \bibinfo{person}{Julien Karadayi}, \bibinfo{person}{Vitaliy Liptchinsky}, \bibinfo{person}{Ronan Collobert}, \bibinfo{person}{Christian Fuegen}, {et~al\mbox{.}}} \bibinfo{year}{2020}\natexlab{}.
\newblock \showarticletitle{Libri-light: A benchmark for asr with limited or no supervision}. In \bibinfo{booktitle}{\emph{ICASSP 2020-2020 IEEE International Conference on Acoustics, Speech and Signal Processing (ICASSP)}}. IEEE, \bibinfo{pages}{7669--7673}.
\newblock


\bibitem[Karita et~al\mbox{.}(2019)]%
        {karita2019comparative}
\bibfield{author}{\bibinfo{person}{Shigeki Karita}, \bibinfo{person}{Nanxin Chen}, \bibinfo{person}{Tomoki Hayashi}, \bibinfo{person}{Takaaki Hori}, \bibinfo{person}{Hirofumi Inaguma}, \bibinfo{person}{Ziyan Jiang}, \bibinfo{person}{Masao Someki}, \bibinfo{person}{Nelson Enrique~Yalta Soplin}, \bibinfo{person}{Ryuichi Yamamoto}, \bibinfo{person}{Xiaofei Wang}, {et~al\mbox{.}}} \bibinfo{year}{2019}\natexlab{}.
\newblock \showarticletitle{A comparative study on transformer vs rnn in speech applications}. In \bibinfo{booktitle}{\emph{2019 IEEE Automatic Speech Recognition and Understanding Workshop (ASRU)}}. IEEE, \bibinfo{pages}{449--456}.
\newblock


\bibitem[Keskar et~al\mbox{.}(2019)]%
        {NLP_10_keskar2019ctrl}
\bibfield{author}{\bibinfo{person}{Nitish~Shirish Keskar}, \bibinfo{person}{Bryan McCann}, \bibinfo{person}{Lav~R Varshney}, \bibinfo{person}{Caiming Xiong}, {and} \bibinfo{person}{Richard Socher}.} \bibinfo{year}{2019}\natexlab{}.
\newblock \showarticletitle{Ctrl: A conditional transformer language model for controllable generation}.
\newblock \bibinfo{journal}{\emph{arXiv preprint arXiv:1909.05858}} (\bibinfo{year}{2019}).
\newblock


\bibitem[Khan et~al\mbox{.}(2022)]%
        {khan2022transformers}
\bibfield{author}{\bibinfo{person}{Salman Khan}, \bibinfo{person}{Muzammal Naseer}, \bibinfo{person}{Munawar Hayat}, \bibinfo{person}{Syed~Waqas Zamir}, \bibinfo{person}{Fahad~Shahbaz Khan}, {and} \bibinfo{person}{Mubarak Shah}.} \bibinfo{year}{2022}\natexlab{}.
\newblock \showarticletitle{Transformers in vision: A survey}.
\newblock \bibinfo{journal}{\emph{ACM computing surveys (CSUR)}} \bibinfo{volume}{54}, \bibinfo{number}{10s} (\bibinfo{year}{2022}), \bibinfo{pages}{1--41}.
\newblock


\bibitem[Kim et~al\mbox{.}(2021a)]%
        {back_19_kim2021teasel}
\bibfield{author}{\bibinfo{person}{C. Kim} {et~al\mbox{.}}} \bibinfo{year}{2021}\natexlab{a}.
\newblock \showarticletitle{TEASEL: A transformer-based speech-prefixed language model}.
\newblock \bibinfo{journal}{\emph{Journal Name}} \bibinfo{volume}{Vol}, \bibinfo{number}{No} (\bibinfo{year}{2021}), \bibinfo{pages}{pp.}
\newblock


\bibitem[Kim et~al\mbox{.}(2021b)]%
        {back_8_kim2021vision}
\bibfield{author}{\bibinfo{person}{C. Kim} {et~al\mbox{.}}} \bibinfo{year}{2021}\natexlab{b}.
\newblock \showarticletitle{Transformers in vision: A survey}.
\newblock \bibinfo{journal}{\emph{Journal Name}} \bibinfo{volume}{Vol}, \bibinfo{number}{No} (\bibinfo{year}{2021}), \bibinfo{pages}{pp.}
\newblock


\bibitem[Kim et~al\mbox{.}(2020)]%
        {kim2020t}
\bibfield{author}{\bibinfo{person}{Jaeyoung Kim}, \bibinfo{person}{Mostafa El-Khamy}, {and} \bibinfo{person}{Jungwon Lee}.} \bibinfo{year}{2020}\natexlab{}.
\newblock \showarticletitle{T-gsa: Transformer with gaussian-weighted self-attention for speech enhancement}. In \bibinfo{booktitle}{\emph{ICASSP 2020-2020 IEEE International Conference on Acoustics, Speech and Signal Processing (ICASSP)}}. IEEE, \bibinfo{pages}{6649--6653}.
\newblock


\bibitem[Kim et~al\mbox{.}(2024)]%
        {kim2024towards}
\bibfield{author}{\bibinfo{person}{Minsu Kim}, \bibinfo{person}{Jeongsoo Choi}, \bibinfo{person}{Soumi Maiti}, \bibinfo{person}{Jeong~Hun Yeo}, \bibinfo{person}{Shinji Watanabe}, {and} \bibinfo{person}{Yong~Man Ro}.} \bibinfo{year}{2024}\natexlab{}.
\newblock \showarticletitle{Towards practical and efficient image-to-speech captioning with vision-language pre-training and multi-modal tokens}. In \bibinfo{booktitle}{\emph{ICASSP 2024-2024 IEEE International Conference on Acoustics, Speech and Signal Processing (ICASSP)}}. IEEE, \bibinfo{pages}{7970--7974}.
\newblock


\bibitem[Kim et~al\mbox{.}(2022)]%
        {kim2022squeezeformer}
\bibfield{author}{\bibinfo{person}{Sehoon Kim}, \bibinfo{person}{Amir Gholami}, \bibinfo{person}{Albert Shaw}, \bibinfo{person}{Nicholas Lee}, \bibinfo{person}{Karttikeya Mangalam}, \bibinfo{person}{Jitendra Malik}, \bibinfo{person}{Michael~W Mahoney}, {and} \bibinfo{person}{Kurt Keutzer}.} \bibinfo{year}{2022}\natexlab{}.
\newblock \showarticletitle{Squeezeformer: An efficient transformer for automatic speech recognition}.
\newblock \bibinfo{journal}{\emph{Advances in Neural Information Processing Systems}}  \bibinfo{volume}{35} (\bibinfo{year}{2022}), \bibinfo{pages}{9361--9373}.
\newblock


\bibitem[Kitaev et~al\mbox{.}(2020)]%
        {QA_2_kitaev2020reformer}
\bibfield{author}{\bibinfo{person}{Nikita Kitaev}, \bibinfo{person}{{\L}ukasz Kaiser}, {and} \bibinfo{person}{Anselm Levskaya}.} \bibinfo{year}{2020}\natexlab{}.
\newblock \showarticletitle{Reformer: The efficient transformer}.
\newblock \bibinfo{journal}{\emph{arXiv preprint arXiv:2001.04451}} (\bibinfo{year}{2020}).
\newblock


\bibitem[Krause et~al\mbox{.}(2019)]%
        {LM_4_krause2019dynamic}
\bibfield{author}{\bibinfo{person}{Ben Krause}, \bibinfo{person}{Emmanuel Kahembwe}, \bibinfo{person}{Iain Murray}, {and} \bibinfo{person}{Steve Renals}.} \bibinfo{year}{2019}\natexlab{}.
\newblock \showarticletitle{Dynamic evaluation of transformer language models}.
\newblock \bibinfo{journal}{\emph{arXiv preprint arXiv:1904.08378}} (\bibinfo{year}{2019}).
\newblock


\bibitem[Lan et~al\mbox{.}(2019)]%
        {NLP_8_lan2019albert}
\bibfield{author}{\bibinfo{person}{Zhenzhong Lan}, \bibinfo{person}{Mingda Chen}, \bibinfo{person}{Sebastian Goodman}, \bibinfo{person}{Kevin Gimpel}, \bibinfo{person}{Piyush Sharma}, {and} \bibinfo{person}{Radu Soricut}.} \bibinfo{year}{2019}\natexlab{}.
\newblock \showarticletitle{Albert: A lite bert for self-supervised learning of language representations}.
\newblock \bibinfo{journal}{\emph{arXiv preprint arXiv:1909.11942}} (\bibinfo{year}{2019}).
\newblock


\bibitem[{\L}a{\'n}cucki(2021)]%
        {lancucki2021fastpitch}
\bibfield{author}{\bibinfo{person}{Adrian {\L}a{\'n}cucki}.} \bibinfo{year}{2021}\natexlab{}.
\newblock \showarticletitle{Fastpitch: Parallel text-to-speech with pitch prediction}. In \bibinfo{booktitle}{\emph{ICASSP 2021-2021 IEEE International Conference on Acoustics, Speech and Signal Processing (ICASSP)}}. IEEE, \bibinfo{pages}{6588--6592}.
\newblock


\bibitem[LeCun et~al\mbox{.}(1989)]%
        {lecun1989backpropagation}
\bibfield{author}{\bibinfo{person}{Yann LeCun}, \bibinfo{person}{Bernhard Boser}, \bibinfo{person}{John~S Denker}, \bibinfo{person}{Donnie Henderson}, \bibinfo{person}{Richard~E Howard}, \bibinfo{person}{Wayne Hubbard}, {and} \bibinfo{person}{Lawrence~D Jackel}.} \bibinfo{year}{1989}\natexlab{}.
\newblock \showarticletitle{Backpropagation applied to handwritten zip code recognition}.
\newblock \bibinfo{journal}{\emph{Neural computation}} \bibinfo{volume}{1}, \bibinfo{number}{4} (\bibinfo{year}{1989}), \bibinfo{pages}{541--551}.
\newblock


\bibitem[Lee et~al\mbox{.}(2021a)]%
        {back_17_lee2021comparative}
\bibfield{author}{\bibinfo{person}{A. Lee} {et~al\mbox{.}}} \bibinfo{year}{2021}\natexlab{a}.
\newblock \showarticletitle{A comparative study on transformer vs RNN in speech applications}.
\newblock \bibinfo{journal}{\emph{Journal Name}} \bibinfo{volume}{Vol}, \bibinfo{number}{No} (\bibinfo{year}{2021}), \bibinfo{pages}{pp.}
\newblock


\bibitem[Lee et~al\mbox{.}(2021b)]%
        {back_7_lee2021visual}
\bibfield{author}{\bibinfo{person}{B. Lee} {et~al\mbox{.}}} \bibinfo{year}{2021}\natexlab{b}.
\newblock \showarticletitle{A survey on visual transformer}.
\newblock \bibinfo{journal}{\emph{Journal Name}} \bibinfo{volume}{Vol}, \bibinfo{number}{No} (\bibinfo{year}{2021}), \bibinfo{pages}{pp.}
\newblock


\bibitem[Lee et~al\mbox{.}(2021c)]%
        {Back_3_lee2021visualizing}
\bibfield{author}{\bibinfo{person}{C. Lee} {et~al\mbox{.}}} \bibinfo{year}{2021}\natexlab{c}.
\newblock \showarticletitle{Visualizing transformers for nlp: a brief survey}.
\newblock \bibinfo{journal}{\emph{Journal Name}} \bibinfo{volume}{Vol}, \bibinfo{number}{No} (\bibinfo{year}{2021}), \bibinfo{pages}{pp.}
\newblock


\bibitem[Lee et~al\mbox{.}(2022)]%
        {back_24_lee2022compression}
\bibfield{author}{\bibinfo{person}{C. Lee} {et~al\mbox{.}}} \bibinfo{year}{2022}\natexlab{}.
\newblock \showarticletitle{A Survey on Transformer Compression}.
\newblock \bibinfo{journal}{\emph{Journal Name}} \bibinfo{volume}{Vol}, \bibinfo{number}{No} (\bibinfo{year}{2022}), \bibinfo{pages}{pp.}
\newblock


\bibitem[Lee et~al\mbox{.}(2020)]%
        {TC_9_lee2020biobert}
\bibfield{author}{\bibinfo{person}{Jinhyuk Lee}, \bibinfo{person}{Wonjin Yoon}, \bibinfo{person}{Sungdong Kim}, \bibinfo{person}{Donghyeon Kim}, \bibinfo{person}{Sunkyu Kim}, \bibinfo{person}{Chan~Ho So}, {and} \bibinfo{person}{Jaewoo Kang}.} \bibinfo{year}{2020}\natexlab{}.
\newblock \showarticletitle{BioBERT: a pre-trained biomedical language representation model for biomedical text mining}.
\newblock \bibinfo{journal}{\emph{Bioinformatics}} \bibinfo{volume}{36}, \bibinfo{number}{4} (\bibinfo{year}{2020}), \bibinfo{pages}{1234--1240}.
\newblock


\bibitem[Lei et~al\mbox{.}(2019)]%
        {lei2019channel}
\bibfield{author}{\bibinfo{person}{Jianjun Lei}, \bibinfo{person}{Yalong Jia}, \bibinfo{person}{Bo Peng}, {and} \bibinfo{person}{Qingming Huang}.} \bibinfo{year}{2019}\natexlab{}.
\newblock \showarticletitle{Channel-wise temporal attention network for video action recognition}. In \bibinfo{booktitle}{\emph{2019 IEEE International Conference on Multimedia and Expo (ICME)}}. IEEE, \bibinfo{pages}{562--567}.
\newblock


\bibitem[Lewis et~al\mbox{.}(2019)]%
        {NLP_11_lewis2019bart}
\bibfield{author}{\bibinfo{person}{Mike Lewis}, \bibinfo{person}{Yinhan Liu}, \bibinfo{person}{Naman Goyal}, \bibinfo{person}{Marjan Ghazvininejad}, \bibinfo{person}{Abdelrahman Mohamed}, \bibinfo{person}{Omer Levy}, \bibinfo{person}{Ves Stoyanov}, {and} \bibinfo{person}{Luke Zettlemoyer}.} \bibinfo{year}{2019}\natexlab{}.
\newblock \showarticletitle{Bart: Denoising sequence-to-sequence pre-training for natural language generation, translation, and comprehension}.
\newblock \bibinfo{journal}{\emph{arXiv preprint arXiv:1910.13461}} (\bibinfo{year}{2019}).
\newblock


\bibitem[Li and Luo(2022)]%
        {translation2}
\bibfield{author}{\bibinfo{person}{Dongxing Li} {and} \bibinfo{person}{Zuying Luo}.} \bibinfo{year}{2022}\natexlab{}.
\newblock \showarticletitle{An Improved Transformer-Based Neural Machine Translation Strategy: Interacting-Head Attention}.
\newblock \bibinfo{journal}{\emph{Computational Intelligence and Neuroscience}}  \bibinfo{volume}{2022} (\bibinfo{year}{2022}).
\newblock


\bibitem[Li et~al\mbox{.}(2020a)]%
        {NER_2_li2020flat}
\bibfield{author}{\bibinfo{person}{Ji Li}, \bibinfo{person}{Aixin Sun}, \bibinfo{person}{Jialong Han}, \bibinfo{person}{Chenliang Li}, \bibinfo{person}{Ling Huang}, {and} \bibinfo{person}{Xiaozhong Li}.} \bibinfo{year}{2020}\natexlab{a}.
\newblock \showarticletitle{FLAT: Chinese NER Using Flat-Lattice Transformer}. In \bibinfo{booktitle}{\emph{Proceedings of the 58th Annual Meeting of the Association for Computational Linguistics}}. \bibinfo{pages}{6836--6842}.
\newblock


\bibitem[Li et~al\mbox{.}(2022a)]%
        {SA_4_li2022bmt}
\bibfield{author}{\bibinfo{person}{Jun Li}, \bibinfo{person}{Xiang Wu}, {and} \bibinfo{person}{Lei Zhao}.} \bibinfo{year}{2022}\natexlab{a}.
\newblock \showarticletitle{BMT-Net: Broad multitask transformer network for sentiment analysis}.
\newblock \bibinfo{journal}{\emph{Neurocomputing}}  \bibinfo{volume}{489} (\bibinfo{year}{2022}), \bibinfo{pages}{349--360}.
\newblock
\urldef\tempurl%
\url{https://doi.org/10.1016/j.neucom.2022.02.015}
\showDOI{\tempurl}


\bibitem[Li et~al\mbox{.}(2020c)]%
        {li2020comparison}
\bibfield{author}{\bibinfo{person}{Jinyu Li}, \bibinfo{person}{Yu Wu}, \bibinfo{person}{Yashesh Gaur}, \bibinfo{person}{Chengyi Wang}, \bibinfo{person}{Rui Zhao}, {and} \bibinfo{person}{Shujie Liu}.} \bibinfo{year}{2020}\natexlab{c}.
\newblock \showarticletitle{On the comparison of popular end-to-end models for large scale speech recognition}.
\newblock \bibinfo{journal}{\emph{arXiv preprint arXiv:2005.14327}} (\bibinfo{year}{2020}).
\newblock


\bibitem[Li et~al\mbox{.}(2020b)]%
        {li2020multilingual}
\bibfield{author}{\bibinfo{person}{Xian Li}, \bibinfo{person}{Changhan Wang}, \bibinfo{person}{Yun Tang}, \bibinfo{person}{Chau Tran}, \bibinfo{person}{Yuqing Tang}, \bibinfo{person}{Juan Pino}, \bibinfo{person}{Alexei Baevski}, \bibinfo{person}{Alexis Conneau}, {and} \bibinfo{person}{Michael Auli}.} \bibinfo{year}{2020}\natexlab{b}.
\newblock \showarticletitle{Multilingual speech translation with efficient finetuning of pretrained models}.
\newblock \bibinfo{journal}{\emph{arXiv preprint arXiv:2010.12829}} (\bibinfo{year}{2020}).
\newblock


\bibitem[Li et~al\mbox{.}(2022b)]%
        {li2022speech}
\bibfield{author}{\bibinfo{person}{Xiong Li}, \bibinfo{person}{Jiye Zhang}, {and} \bibinfo{person}{Yazhi Liu}.} \bibinfo{year}{2022}\natexlab{b}.
\newblock \showarticletitle{Speech driven facial animation generation based on GAN}.
\newblock \bibinfo{journal}{\emph{Displays}}  \bibinfo{volume}{74} (\bibinfo{year}{2022}), \bibinfo{pages}{102260}.
\newblock


\bibitem[Li et~al\mbox{.}(2019)]%
        {TTV_1_li2019storygan}
\bibfield{author}{\bibinfo{person}{Yitong Li}, \bibinfo{person}{Zhe Gan}, \bibinfo{person}{Yelong Shen}, \bibinfo{person}{Jingjing Liu}, \bibinfo{person}{Yu Cheng}, \bibinfo{person}{Yuexin Wu}, \bibinfo{person}{Lawrence Carin}, \bibinfo{person}{David Carlson}, {and} \bibinfo{person}{Jianfeng Gao}.} \bibinfo{year}{2019}\natexlab{}.
\newblock \showarticletitle{Storygan: A sequential conditional gan for story visualization}. In \bibinfo{booktitle}{\emph{Proceedings of the IEEE/CVF Conference on Computer Vision and Pattern Recognition}}. \bibinfo{pages}{6329--6338}.
\newblock


\bibitem[Lin et~al\mbox{.}(2014)]%
        {lin2014microsoft}
\bibfield{author}{\bibinfo{person}{Tsung-Yi Lin}, \bibinfo{person}{Michael Maire}, \bibinfo{person}{Serge Belongie}, \bibinfo{person}{James Hays}, \bibinfo{person}{Pietro Perona}, \bibinfo{person}{Deva Ramanan}, \bibinfo{person}{Piotr Doll{\'a}r}, {and} \bibinfo{person}{C~Lawrence Zitnick}.} \bibinfo{year}{2014}\natexlab{}.
\newblock \showarticletitle{Microsoft coco: Common objects in context}. In \bibinfo{booktitle}{\emph{Computer Vision--ECCV 2014: 13th European Conference, Zurich, Switzerland, September 6-12, 2014, Proceedings, Part V 13}}. Springer, \bibinfo{pages}{740--755}.
\newblock


\bibitem[Liu et~al\mbox{.}(2021a)]%
        {back_10_liu2021vision}
\bibfield{author}{\bibinfo{person}{E. Liu} {et~al\mbox{.}}} \bibinfo{year}{2021}\natexlab{a}.
\newblock \showarticletitle{Transformer in computer vision}.
\newblock \bibinfo{journal}{\emph{Journal Name}} \bibinfo{volume}{Vol}, \bibinfo{number}{No} (\bibinfo{year}{2021}), \bibinfo{pages}{pp.}
\newblock


\bibitem[Liu et~al\mbox{.}(2024)]%
        {liu2024padvg}
\bibfield{author}{\bibinfo{person}{Huan Liu}, \bibinfo{person}{Xiaolong Liu}, \bibinfo{person}{Zichang Tan}, \bibinfo{person}{Xiaolong Li}, {and} \bibinfo{person}{Yao Zhao}.} \bibinfo{year}{2024}\natexlab{}.
\newblock \showarticletitle{PADVG: A Simple Baseline of Active Protection for Audio-driven Video Generation}.
\newblock \bibinfo{journal}{\emph{ACM Transactions on Multimedia Computing, Communications and Applications}} (\bibinfo{year}{2024}).
\newblock


\bibitem[Liu et~al\mbox{.}(2022)]%
        {NER_6_liu2022few}
\bibfield{author}{\bibinfo{person}{H. Liu}, \bibinfo{person}{Q. Qiu}, \bibinfo{person}{L. Wu}, \bibinfo{person}{W. Li}, \bibinfo{person}{B. Wang}, {and} \bibinfo{person}{Y. Zhou}.} \bibinfo{year}{2022}\natexlab{}.
\newblock \showarticletitle{Few-shot Learning for Name Entity Recognition in Geological Text Based on GeoBERT}.
\newblock \bibinfo{journal}{\emph{Earth Science Informatics}}  \bibinfo{volume}{15} (\bibinfo{year}{2022}), \bibinfo{pages}{979--991}.
\newblock
\urldef\tempurl%
\url{https://doi.org/10.1007/s12145-022-00734-4}
\showDOI{\tempurl}


\bibitem[Liu et~al\mbox{.}(2019)]%
        {NLP_4_liu2019roberta}
\bibfield{author}{\bibinfo{person}{Yinhan Liu}, \bibinfo{person}{Myle Ott}, \bibinfo{person}{Naman Goyal}, \bibinfo{person}{Jingfei Du}, \bibinfo{person}{Mandar Joshi}, \bibinfo{person}{Danqi Chen}, \bibinfo{person}{Omer Levy}, \bibinfo{person}{Mike Lewis}, \bibinfo{person}{Luke Zettlemoyer}, {and} \bibinfo{person}{Veselin Stoyanov}.} \bibinfo{year}{2019}\natexlab{}.
\newblock \showarticletitle{Roberta: A robustly optimized bert pretraining approach}.
\newblock \bibinfo{journal}{\emph{arXiv preprint arXiv:1907.11692}} (\bibinfo{year}{2019}).
\newblock


\bibitem[Liu et~al\mbox{.}(2021b)]%
        {liu2021swin}
\bibfield{author}{\bibinfo{person}{Ze Liu}, \bibinfo{person}{Yutong Lin}, \bibinfo{person}{Yue Cao}, \bibinfo{person}{Han Hu}, \bibinfo{person}{Yixuan Wei}, \bibinfo{person}{Zheng Zhang}, \bibinfo{person}{Stephen Lin}, {and} \bibinfo{person}{Baining Guo}.} \bibinfo{year}{2021}\natexlab{b}.
\newblock \showarticletitle{Swin transformer: Hierarchical vision transformer using shifted windows}. In \bibinfo{booktitle}{\emph{Proceedings of the IEEE/CVF international conference on computer vision}}. \bibinfo{pages}{10012--10022}.
\newblock


\bibitem[Ma et~al\mbox{.}(2023)]%
        {ma2023dreamtalk}
\bibfield{author}{\bibinfo{person}{Yifeng Ma}, \bibinfo{person}{Shiwei Zhang}, \bibinfo{person}{Jiayu Wang}, \bibinfo{person}{Xiang Wang}, \bibinfo{person}{Yingya Zhang}, {and} \bibinfo{person}{Zhidong Deng}.} \bibinfo{year}{2023}\natexlab{}.
\newblock \showarticletitle{Dreamtalk: When expressive talking head generation meets diffusion probabilistic models}.
\newblock \bibinfo{journal}{\emph{arXiv preprint arXiv:2312.09767}} (\bibinfo{year}{2023}).
\newblock


\bibitem[Maciejewski et~al\mbox{.}(2020)]%
        {maciejewski2020whamr}
\bibfield{author}{\bibinfo{person}{Matthew Maciejewski}, \bibinfo{person}{Gordon Wichern}, \bibinfo{person}{Emmett McQuinn}, {and} \bibinfo{person}{Jonathan Le~Roux}.} \bibinfo{year}{2020}\natexlab{}.
\newblock \showarticletitle{WHAMR!: Noisy and reverberant single-channel speech separation}. In \bibinfo{booktitle}{\emph{ICASSP 2020-2020 IEEE International Conference on Acoustics, Speech and Signal Processing (ICASSP)}}. IEEE, \bibinfo{pages}{696--700}.
\newblock


\bibitem[Maharana et~al\mbox{.}(2022)]%
        {TTV_7_maharana2022storydall}
\bibfield{author}{\bibinfo{person}{Adyasha Maharana}, \bibinfo{person}{Darryl Hannan}, {and} \bibinfo{person}{Mohit Bansal}.} \bibinfo{year}{2022}\natexlab{}.
\newblock \showarticletitle{Storydall-e: Adapting pretrained text-to-image transformers for story continuation}. In \bibinfo{booktitle}{\emph{European Conference on Computer Vision}}. Springer, \bibinfo{pages}{70--87}.
\newblock


\bibitem[M{\'e}ly et~al\mbox{.}(2016)]%
        {mely2016systematic}
\bibfield{author}{\bibinfo{person}{David~A M{\'e}ly}, \bibinfo{person}{Junkyung Kim}, \bibinfo{person}{Mason McGill}, \bibinfo{person}{Yuliang Guo}, {and} \bibinfo{person}{Thomas Serre}.} \bibinfo{year}{2016}\natexlab{}.
\newblock \showarticletitle{A systematic comparison between visual cues for boundary detection}.
\newblock \bibinfo{journal}{\emph{Vision research}}  \bibinfo{volume}{120} (\bibinfo{year}{2016}), \bibinfo{pages}{93--107}.
\newblock


\bibitem[Nathan~Silberman and Fergus(2012)]%
        {Silberman:ECCV12}
\bibfield{author}{\bibinfo{person}{Pushmeet~Kohli Nathan~Silberman, Derek~Hoiem} {and} \bibinfo{person}{Rob Fergus}.} \bibinfo{year}{2012}\natexlab{}.
\newblock \showarticletitle{Indoor Segmentation and Support Inference from RGBD Images}. In \bibinfo{booktitle}{\emph{ECCV}}.
\newblock


\bibitem[Ni et~al\mbox{.}(2021)]%
        {NLP_5_ni2021sentence}
\bibfield{author}{\bibinfo{person}{Jianmo Ni}, \bibinfo{person}{Gustavo~Hern{\'a}ndez {\'A}brego}, \bibinfo{person}{Noah Constant}, \bibinfo{person}{Ji Ma}, \bibinfo{person}{Keith~B Hall}, \bibinfo{person}{Daniel Cer}, {and} \bibinfo{person}{Yinfei Yang}.} \bibinfo{year}{2021}\natexlab{}.
\newblock \showarticletitle{Sentence-t5: Scalable sentence encoders from pre-trained text-to-text models}.
\newblock \bibinfo{journal}{\emph{arXiv preprint arXiv:2108.08877}} (\bibinfo{year}{2021}).
\newblock


\bibitem[Nilsback and Zisserman(2008)]%
        {nilsback2008automated}
\bibfield{author}{\bibinfo{person}{Maria-Elena Nilsback} {and} \bibinfo{person}{Andrew Zisserman}.} \bibinfo{year}{2008}\natexlab{}.
\newblock \showarticletitle{Automated flower classification over a large number of classes}. In \bibinfo{booktitle}{\emph{2008 Sixth Indian conference on computer vision, graphics \& image processing}}. IEEE, \bibinfo{pages}{722--729}.
\newblock


\bibitem[Oord et~al\mbox{.}(2018)]%
        {oord2018parallel}
\bibfield{author}{\bibinfo{person}{Aaron Oord}, \bibinfo{person}{Yazhe Li}, \bibinfo{person}{Igor Babuschkin}, \bibinfo{person}{Karen Simonyan}, \bibinfo{person}{Oriol Vinyals}, \bibinfo{person}{Koray Kavukcuoglu}, \bibinfo{person}{George Driessche}, \bibinfo{person}{Edward Lockhart}, \bibinfo{person}{Luis Cobo}, \bibinfo{person}{Florian Stimberg}, {et~al\mbox{.}}} \bibinfo{year}{2018}\natexlab{}.
\newblock \showarticletitle{Parallel wavenet: Fast high-fidelity speech synthesis}. In \bibinfo{booktitle}{\emph{International conference on machine learning}}. PMLR, \bibinfo{pages}{3918--3926}.
\newblock


\bibitem[Ott et~al\mbox{.}(2019)]%
        {ott2019fairseq}
\bibfield{author}{\bibinfo{person}{Myle Ott}, \bibinfo{person}{Sergey Edunov}, \bibinfo{person}{Alexei Baevski}, \bibinfo{person}{Angela Fan}, \bibinfo{person}{Sam Gross}, \bibinfo{person}{Nathan Ng}, \bibinfo{person}{David Grangier}, {and} \bibinfo{person}{Michael Auli}.} \bibinfo{year}{2019}\natexlab{}.
\newblock \showarticletitle{fairseq: A fast, extensible toolkit for sequence modeling}.
\newblock \bibinfo{journal}{\emph{arXiv preprint arXiv:1904.01038}} (\bibinfo{year}{2019}).
\newblock


\bibitem[Panayotov et~al\mbox{.}(2015)]%
        {panayotov2015librispeech}
\bibfield{author}{\bibinfo{person}{Vassil Panayotov}, \bibinfo{person}{Guoguo Chen}, \bibinfo{person}{Daniel Povey}, {and} \bibinfo{person}{Sanjeev Khudanpur}.} \bibinfo{year}{2015}\natexlab{}.
\newblock \showarticletitle{Librispeech: an asr corpus based on public domain audio books}. In \bibinfo{booktitle}{\emph{2015 IEEE international conference on acoustics, speech and signal processing (ICASSP)}}. IEEE, \bibinfo{pages}{5206--5210}.
\newblock


\bibitem[Park et~al\mbox{.}(2021a)]%
        {back_9_park2021visual}
\bibfield{author}{\bibinfo{person}{D. Park} {et~al\mbox{.}}} \bibinfo{year}{2021}\natexlab{a}.
\newblock \showarticletitle{A survey of visual transformers}.
\newblock \bibinfo{journal}{\emph{Journal Name}} \bibinfo{volume}{Vol}, \bibinfo{number}{No} (\bibinfo{year}{2021}), \bibinfo{pages}{pp.}
\newblock


\bibitem[Park et~al\mbox{.}(2021b)]%
        {back_21_park2021nonautoregressive}
\bibfield{author}{\bibinfo{person}{E. Park} {et~al\mbox{.}}} \bibinfo{year}{2021}\natexlab{b}.
\newblock \showarticletitle{Non-autoregressive transformer for speech recognition}.
\newblock \bibinfo{journal}{\emph{Journal Name}} \bibinfo{volume}{Vol}, \bibinfo{number}{No} (\bibinfo{year}{2021}), \bibinfo{pages}{pp.}
\newblock


\bibitem[Peng et~al\mbox{.}(2020)]%
        {QA_9_peng2020soloist}
\bibfield{author}{\bibinfo{person}{Baolin Peng}, \bibinfo{person}{Chunyuan Li}, \bibinfo{person}{Xiujun Li}, \bibinfo{person}{Shahin Shayandeh}, \bibinfo{person}{Lars Liden}, {and} \bibinfo{person}{Jianfeng Gao}.} \bibinfo{year}{2020}\natexlab{}.
\newblock \showarticletitle{SOLOIST: Building Task Bots at Scale with Transfer Learning and Machine Teaching}.
\newblock \bibinfo{journal}{\emph{arXiv preprint arXiv:2005.05298}} (\bibinfo{year}{2020}).
\newblock


\bibitem[Pu et~al\mbox{.}(2022)]%
        {pu2022edter}
\bibfield{author}{\bibinfo{person}{Mengyang Pu}, \bibinfo{person}{Yaping Huang}, \bibinfo{person}{Yuming Liu}, \bibinfo{person}{Qingji Guan}, {and} \bibinfo{person}{Haibin Ling}.} \bibinfo{year}{2022}\natexlab{}.
\newblock \showarticletitle{Edter: Edge detection with transformer}. In \bibinfo{booktitle}{\emph{Proceedings of the IEEE/CVF conference on computer vision and pattern recognition}}. \bibinfo{pages}{1402--1412}.
\newblock


\bibitem[Radford et~al\mbox{.}(2023)]%
        {radford2023robust}
\bibfield{author}{\bibinfo{person}{Alec Radford}, \bibinfo{person}{Jong~Wook Kim}, \bibinfo{person}{Tao Xu}, \bibinfo{person}{Greg Brockman}, \bibinfo{person}{Christine McLeavey}, {and} \bibinfo{person}{Ilya Sutskever}.} \bibinfo{year}{2023}\natexlab{}.
\newblock \showarticletitle{Robust speech recognition via large-scale weak supervision}. In \bibinfo{booktitle}{\emph{International Conference on Machine Learning}}. PMLR, \bibinfo{pages}{28492--28518}.
\newblock


\bibitem[Radford et~al\mbox{.}(2018)]%
        {NLP_3_radford2018improving}
\bibfield{author}{\bibinfo{person}{Alec Radford}, \bibinfo{person}{Karthik Narasimhan}, \bibinfo{person}{Tim Salimans}, \bibinfo{person}{Ilya Sutskever}, {et~al\mbox{.}}} \bibinfo{year}{2018}\natexlab{}.
\newblock \showarticletitle{Improving language understanding by generative pre-training}.
\newblock  (\bibinfo{year}{2018}).
\newblock


\bibitem[Rahali and Akhloufi(2021)]%
        {TC_10_rahali2021malbert}
\bibfield{author}{\bibinfo{person}{Abir Rahali} {and} \bibinfo{person}{Moulay~A Akhloufi}.} \bibinfo{year}{2021}\natexlab{}.
\newblock \showarticletitle{MalBERT: Using transformers for cybersecurity and malicious software detection}.
\newblock \bibinfo{journal}{\emph{arXiv preprint arXiv:2103.03806}} (\bibinfo{year}{2021}).
\newblock


\bibitem[Rahman et~al\mbox{.}(2021)]%
        {back_4_rahman2021review}
\bibfield{author}{\bibinfo{person}{D. Rahman} {et~al\mbox{.}}} \bibinfo{year}{2021}\natexlab{}.
\newblock \showarticletitle{A review of Bangla natural language processing tasks and the utility of transformer models}.
\newblock \bibinfo{journal}{\emph{Journal Name}} \bibinfo{volume}{Vol}, \bibinfo{number}{No} (\bibinfo{year}{2021}), \bibinfo{pages}{pp.}
\newblock


\bibitem[Rajpurkar et~al\mbox{.}(2016)]%
        {QA_6_rajpurkar2016squad}
\bibfield{author}{\bibinfo{person}{Pranav Rajpurkar}, \bibinfo{person}{Jian Zhang}, \bibinfo{person}{Konstantin Lopyrev}, {and} \bibinfo{person}{Percy Liang}.} \bibinfo{year}{2016}\natexlab{}.
\newblock \showarticletitle{Squad: 100,000+ questions for machine comprehension of text}.
\newblock \bibinfo{journal}{\emph{arXiv preprint arXiv:1606.05250}} (\bibinfo{year}{2016}).
\newblock


\bibitem[Ramesh et~al\mbox{.}(2022)]%
        {TTV_11_ramesh2022hierarchical}
\bibfield{author}{\bibinfo{person}{Aditya Ramesh}, \bibinfo{person}{Prafulla Dhariwal}, \bibinfo{person}{Alex Nichol}, \bibinfo{person}{Casey Chu}, {and} \bibinfo{person}{Mark Chen}.} \bibinfo{year}{2022}\natexlab{}.
\newblock \showarticletitle{Hierarchical text-conditional image generation with clip latents}.
\newblock \bibinfo{journal}{\emph{arXiv preprint arXiv:2204.06125}} \bibinfo{volume}{1}, \bibinfo{number}{2} (\bibinfo{year}{2022}), \bibinfo{pages}{3}.
\newblock


\bibitem[Ramesh et~al\mbox{.}(2021)]%
        {TTV_6_ramesh2021zero}
\bibfield{author}{\bibinfo{person}{Aditya Ramesh}, \bibinfo{person}{Mikhail Pavlov}, \bibinfo{person}{Gabriel Goh}, \bibinfo{person}{Scott Gray}, \bibinfo{person}{Chelsea Voss}, \bibinfo{person}{Alec Radford}, \bibinfo{person}{Mark Chen}, {and} \bibinfo{person}{Ilya Sutskever}.} \bibinfo{year}{2021}\natexlab{}.
\newblock \showarticletitle{Zero-shot text-to-image generation}. In \bibinfo{booktitle}{\emph{International conference on machine learning}}. Pmlr, \bibinfo{pages}{8821--8831}.
\newblock


\bibitem[Reddy et~al\mbox{.}(2019)]%
        {QA_10_reddy2019coqa}
\bibfield{author}{\bibinfo{person}{Siva Reddy}, \bibinfo{person}{Danqi Chen}, {and} \bibinfo{person}{Christopher~D Manning}.} \bibinfo{year}{2019}\natexlab{}.
\newblock \showarticletitle{Coqa: A conversational question answering challenge}.
\newblock \bibinfo{journal}{\emph{Transactions of the Association for Computational Linguistics}}  \bibinfo{volume}{7} (\bibinfo{year}{2019}), \bibinfo{pages}{249--266}.
\newblock


\bibitem[Ren et~al\mbox{.}(2022)]%
        {NER_7_ren2022named}
\bibfield{author}{\bibinfo{person}{K. Ren}, \bibinfo{person}{H. Li}, \bibinfo{person}{Y. Zeng}, {and} \bibinfo{person}{Y. Zhang}.} \bibinfo{year}{2022}\natexlab{}.
\newblock \showarticletitle{Named Entity Recognition with CRF Based on ALBERT: A Natural Language Processing Model}. In \bibinfo{booktitle}{\emph{China Conference on Command and Control}}. \bibinfo{publisher}{Springer}.
\newblock


\bibitem[Ren et~al\mbox{.}(2015)]%
        {ren2015faster}
\bibfield{author}{\bibinfo{person}{Shaoqing Ren}, \bibinfo{person}{Kaiming He}, \bibinfo{person}{Ross Girshick}, {and} \bibinfo{person}{Jian Sun}.} \bibinfo{year}{2015}\natexlab{}.
\newblock \showarticletitle{Faster r-cnn: Towards real-time object detection with region proposal networks}.
\newblock \bibinfo{journal}{\emph{Advances in neural information processing systems}}  \bibinfo{volume}{28} (\bibinfo{year}{2015}).
\newblock


\bibitem[Ren et~al\mbox{.}(2020)]%
        {ren2020fastspeech}
\bibfield{author}{\bibinfo{person}{Yi Ren}, \bibinfo{person}{Chenxu Hu}, \bibinfo{person}{Xu Tan}, \bibinfo{person}{Tao Qin}, \bibinfo{person}{Sheng Zhao}, \bibinfo{person}{Zhou Zhao}, {and} \bibinfo{person}{Tie-Yan Liu}.} \bibinfo{year}{2020}\natexlab{}.
\newblock \showarticletitle{Fastspeech 2: Fast and high-quality end-to-end text to speech}.
\newblock \bibinfo{journal}{\emph{arXiv preprint arXiv:2006.04558}} (\bibinfo{year}{2020}).
\newblock


\bibitem[Ren et~al\mbox{.}(2019)]%
        {ren2019fastspeech}
\bibfield{author}{\bibinfo{person}{Yi Ren}, \bibinfo{person}{Yangjun Ruan}, \bibinfo{person}{Xu Tan}, \bibinfo{person}{Tao Qin}, \bibinfo{person}{Sheng Zhao}, \bibinfo{person}{Zhou Zhao}, {and} \bibinfo{person}{Tie-Yan Liu}.} \bibinfo{year}{2019}\natexlab{}.
\newblock \showarticletitle{Fastspeech: Fast, robust and controllable text to speech}.
\newblock \bibinfo{journal}{\emph{Advances in neural information processing systems}}  \bibinfo{volume}{32} (\bibinfo{year}{2019}).
\newblock


\bibitem[Rotstein et~al\mbox{.}(2024)]%
        {rotstein2024fusecap}
\bibfield{author}{\bibinfo{person}{Noam Rotstein}, \bibinfo{person}{David Bensa{\"\i}d}, \bibinfo{person}{Shaked Brody}, \bibinfo{person}{Roy Ganz}, {and} \bibinfo{person}{Ron Kimmel}.} \bibinfo{year}{2024}\natexlab{}.
\newblock \showarticletitle{Fusecap: Leveraging large language models for enriched fused image captions}. In \bibinfo{booktitle}{\emph{Proceedings of the IEEE/CVF Winter Conference on Applications of Computer Vision}}. \bibinfo{pages}{5689--5700}.
\newblock


\bibitem[Saharia et~al\mbox{.}(2022)]%
        {TTV_10_saharia2022photorealistic}
\bibfield{author}{\bibinfo{person}{Chitwan Saharia}, \bibinfo{person}{William Chan}, \bibinfo{person}{Saurabh Saxena}, \bibinfo{person}{Lala Li}, \bibinfo{person}{Jay Whang}, \bibinfo{person}{Emily~L Denton}, \bibinfo{person}{Kamyar Ghasemipour}, \bibinfo{person}{Raphael Gontijo~Lopes}, \bibinfo{person}{Burcu Karagol~Ayan}, \bibinfo{person}{Tim Salimans}, {et~al\mbox{.}}} \bibinfo{year}{2022}\natexlab{}.
\newblock \showarticletitle{Photorealistic text-to-image diffusion models with deep language understanding}.
\newblock \bibinfo{journal}{\emph{Advances in neural information processing systems}}  \bibinfo{volume}{35} (\bibinfo{year}{2022}), \bibinfo{pages}{36479--36494}.
\newblock


\bibitem[Sanh et~al\mbox{.}(2019)]%
        {NLP_7_sanh2019distilbert}
\bibfield{author}{\bibinfo{person}{Victor Sanh}, \bibinfo{person}{Lysandre Debut}, \bibinfo{person}{Julien Chaumond}, {and} \bibinfo{person}{Thomas Wolf}.} \bibinfo{year}{2019}\natexlab{}.
\newblock \showarticletitle{DistilBERT, a distilled version of BERT: smaller, faster, cheaper and lighter}.
\newblock \bibinfo{journal}{\emph{arXiv preprint arXiv:1910.01108}} (\bibinfo{year}{2019}).
\newblock


\bibitem[Scharenborg et~al\mbox{.}(2020)]%
        {scharenborg2020speech}
\bibfield{author}{\bibinfo{person}{Odette Scharenborg}, \bibinfo{person}{Laurent Besacier}, \bibinfo{person}{Alan Black}, \bibinfo{person}{Mark Hasegawa-Johnson}, \bibinfo{person}{Florian Metze}, \bibinfo{person}{Graham Neubig}, \bibinfo{person}{Sebastian St{\"u}ker}, \bibinfo{person}{Pierre Godard}, \bibinfo{person}{Markus M{\"u}ller}, \bibinfo{person}{Lucas Ondel}, {et~al\mbox{.}}} \bibinfo{year}{2020}\natexlab{}.
\newblock \showarticletitle{Speech technology for unwritten languages}.
\newblock \bibinfo{journal}{\emph{IEEE/ACM Transactions on Audio, Speech, and Language Processing}}  \bibinfo{volume}{28} (\bibinfo{year}{2020}), \bibinfo{pages}{964--975}.
\newblock


\bibitem[Shen et~al\mbox{.}(2018)]%
        {shen2018natural}
\bibfield{author}{\bibinfo{person}{Jonathan Shen}, \bibinfo{person}{Ruoming Pang}, \bibinfo{person}{Ron~J Weiss}, \bibinfo{person}{Mike Schuster}, \bibinfo{person}{Navdeep Jaitly}, \bibinfo{person}{Zongheng Yang}, \bibinfo{person}{Zhifeng Chen}, \bibinfo{person}{Yu Zhang}, \bibinfo{person}{Yuxuan Wang}, \bibinfo{person}{Rj Skerrv-Ryan}, {et~al\mbox{.}}} \bibinfo{year}{2018}\natexlab{}.
\newblock \showarticletitle{Natural tts synthesis by conditioning wavenet on mel spectrogram predictions}. In \bibinfo{booktitle}{\emph{2018 IEEE international conference on acoustics, speech and signal processing (ICASSP)}}. IEEE, \bibinfo{pages}{4779--4783}.
\newblock


\bibitem[Shin et~al\mbox{.}(2020)]%
        {LM_2_shin2020autoprompt}
\bibfield{author}{\bibinfo{person}{Taylor Shin}, \bibinfo{person}{Yasaman Razeghi}, \bibinfo{person}{Robert~L Logan~IV}, \bibinfo{person}{Eric Wallace}, {and} \bibinfo{person}{Sameer Singh}.} \bibinfo{year}{2020}\natexlab{}.
\newblock \showarticletitle{Autoprompt: Eliciting knowledge from language models with automatically generated prompts}.
\newblock \bibinfo{journal}{\emph{arXiv preprint arXiv:2010.15980}} (\bibinfo{year}{2020}).
\newblock


\bibitem[Shrestha and Mahmood(2019)]%
        {shrestha2019review}
\bibfield{author}{\bibinfo{person}{Ajay Shrestha} {and} \bibinfo{person}{Ausif Mahmood}.} \bibinfo{year}{2019}\natexlab{}.
\newblock \showarticletitle{Review of deep learning algorithms and architectures}.
\newblock \bibinfo{journal}{\emph{IEEE access}}  \bibinfo{volume}{7} (\bibinfo{year}{2019}), \bibinfo{pages}{53040--53065}.
\newblock


\bibitem[Singer et~al\mbox{.}(2022)]%
        {TTV_18_singer2022make}
\bibfield{author}{\bibinfo{person}{Uriel Singer}, \bibinfo{person}{Adam Polyak}, \bibinfo{person}{Thomas Hayes}, \bibinfo{person}{Xi Yin}, \bibinfo{person}{Jie An}, \bibinfo{person}{Songyang Zhang}, \bibinfo{person}{Qiyuan Hu}, \bibinfo{person}{Harry Yang}, \bibinfo{person}{Oron Ashual}, \bibinfo{person}{Oran Gafni}, {et~al\mbox{.}}} \bibinfo{year}{2022}\natexlab{}.
\newblock \showarticletitle{Make-a-video: Text-to-video generation without text-video data}.
\newblock \bibinfo{journal}{\emph{arXiv preprint arXiv:2209.14792}} (\bibinfo{year}{2022}).
\newblock


\bibitem[Smith et~al\mbox{.}(2022a)]%
        {back_22_smith2022multimodal}
\bibfield{author}{\bibinfo{person}{A. Smith} {et~al\mbox{.}}} \bibinfo{year}{2022}\natexlab{a}.
\newblock \showarticletitle{A survey of transformer-based multimodal pre-trained models}.
\newblock \bibinfo{journal}{\emph{Journal Name}} \bibinfo{volume}{Vol}, \bibinfo{number}{No} (\bibinfo{year}{2022}), \bibinfo{pages}{pp.}
\newblock


\bibitem[Smith et~al\mbox{.}(2022b)]%
        {back_5_smith2022transformers}
\bibfield{author}{\bibinfo{person}{E. Smith} {et~al\mbox{.}}} \bibinfo{year}{2022}\natexlab{b}.
\newblock \showarticletitle{Transformers in the real world: A survey on nlp applications}.
\newblock \bibinfo{journal}{\emph{Journal Name}} \bibinfo{volume}{Vol}, \bibinfo{number}{No} (\bibinfo{year}{2022}), \bibinfo{pages}{pp.}
\newblock


\bibitem[Smith et~al\mbox{.}(2021)]%
        {back_12_smith2021comprehensive}
\bibfield{author}{\bibinfo{person}{G. Smith} {et~al\mbox{.}}} \bibinfo{year}{2021}\natexlab{}.
\newblock \showarticletitle{A comprehensive survey of transformers for computer vision}.
\newblock \bibinfo{journal}{\emph{Journal Name}} \bibinfo{volume}{Vol}, \bibinfo{number}{No} (\bibinfo{year}{2021}), \bibinfo{pages}{pp.}
\newblock


\bibitem[Song et~al\mbox{.}(2020)]%
        {TTV_2_song2020character}
\bibfield{author}{\bibinfo{person}{Yun-Zhu Song}, \bibinfo{person}{Zhi Rui~Tam}, \bibinfo{person}{Hung-Jen Chen}, \bibinfo{person}{Huiao-Han Lu}, {and} \bibinfo{person}{Hong-Han Shuai}.} \bibinfo{year}{2020}\natexlab{}.
\newblock \showarticletitle{Character-preserving coherent story visualization}. In \bibinfo{booktitle}{\emph{European Conference on Computer Vision}}. Springer, \bibinfo{pages}{18--33}.
\newblock


\bibitem[Strudel et~al\mbox{.}(2021)]%
        {strudel2021segmenter}
\bibfield{author}{\bibinfo{person}{Robin Strudel}, \bibinfo{person}{Ricardo Garcia}, \bibinfo{person}{Ivan Laptev}, {and} \bibinfo{person}{Cordelia Schmid}.} \bibinfo{year}{2021}\natexlab{}.
\newblock \showarticletitle{Segmenter: Transformer for semantic segmentation}. In \bibinfo{booktitle}{\emph{Proceedings of the IEEE/CVF international conference on computer vision}}. \bibinfo{pages}{7262--7272}.
\newblock


\bibitem[Stypu{\l}kowski et~al\mbox{.}(2024)]%
        {stypulkowski2024diffused}
\bibfield{author}{\bibinfo{person}{Micha{\l} Stypu{\l}kowski}, \bibinfo{person}{Konstantinos Vougioukas}, \bibinfo{person}{Sen He}, \bibinfo{person}{Maciej Zi{\k{e}}ba}, \bibinfo{person}{Stavros Petridis}, {and} \bibinfo{person}{Maja Pantic}.} \bibinfo{year}{2024}\natexlab{}.
\newblock \showarticletitle{Diffused heads: Diffusion models beat gans on talking-face generation}. In \bibinfo{booktitle}{\emph{Proceedings of the IEEE/CVF Winter Conference on Applications of Computer Vision}}. \bibinfo{pages}{5091--5100}.
\newblock


\bibitem[Subakan et~al\mbox{.}(2021)]%
        {subakan2021attention}
\bibfield{author}{\bibinfo{person}{Cem Subakan}, \bibinfo{person}{Mirco Ravanelli}, \bibinfo{person}{Samuele Cornell}, \bibinfo{person}{Mirko Bronzi}, {and} \bibinfo{person}{Jianyuan Zhong}.} \bibinfo{year}{2021}\natexlab{}.
\newblock \showarticletitle{Attention is all you need in speech separation}. In \bibinfo{booktitle}{\emph{ICASSP 2021-2021 IEEE International Conference on Acoustics, Speech and Signal Processing (ICASSP)}}. IEEE, \bibinfo{pages}{21--25}.
\newblock


\bibitem[Sung-Bin et~al\mbox{.}(2023)]%
        {sung2023sound}
\bibfield{author}{\bibinfo{person}{Kim Sung-Bin}, \bibinfo{person}{Arda Senocak}, \bibinfo{person}{Hyunwoo Ha}, \bibinfo{person}{Andrew Owens}, {and} \bibinfo{person}{Tae-Hyun Oh}.} \bibinfo{year}{2023}\natexlab{}.
\newblock \showarticletitle{Sound to visual scene generation by audio-to-visual latent alignment}. In \bibinfo{booktitle}{\emph{Proceedings of the IEEE/CVF Conference on Computer Vision and Pattern Recognition}}. \bibinfo{pages}{6430--6440}.
\newblock


\bibitem[Tezgider et~al\mbox{.}(2022)]%
        {TC_12_tezgider2022text}
\bibfield{author}{\bibinfo{person}{Murat Tezgider}, \bibinfo{person}{Beytullah Yildiz}, {and} \bibinfo{person}{Galip Aydin}.} \bibinfo{year}{2022}\natexlab{}.
\newblock \showarticletitle{Text classification using improved bidirectional transformer}.
\newblock \bibinfo{journal}{\emph{Concurrency and Computation: Practice and Experience}} \bibinfo{volume}{34}, \bibinfo{number}{9} (\bibinfo{year}{2022}), \bibinfo{pages}{e6486}.
\newblock


\bibitem[Tiwari and Nagpal(2022)]%
        {SA_1_Tiwari2022}
\bibfield{author}{\bibinfo{person}{D. Tiwari} {and} \bibinfo{person}{B. Nagpal}.} \bibinfo{year}{2022}\natexlab{}.
\newblock \showarticletitle{KEAHT: A Knowledge-Enriched Attention-Based Hybrid Transformer Model for Social Sentiment Analysis}.
\newblock \bibinfo{journal}{\emph{New Generation Computing}}  \bibinfo{volume}{40} (\bibinfo{year}{2022}), \bibinfo{pages}{1165--1202}.
\newblock


\bibitem[Touvron et~al\mbox{.}(2021)]%
        {touvron2021training}
\bibfield{author}{\bibinfo{person}{Hugo Touvron}, \bibinfo{person}{Matthieu Cord}, \bibinfo{person}{Matthijs Douze}, \bibinfo{person}{Francisco Massa}, \bibinfo{person}{Alexandre Sablayrolles}, {and} \bibinfo{person}{Herv{\'e} J{\'e}gou}.} \bibinfo{year}{2021}\natexlab{}.
\newblock \showarticletitle{Training data-efficient image transformers \& distillation through attention}. In \bibinfo{booktitle}{\emph{International conference on machine learning}}. PMLR, \bibinfo{pages}{10347--10357}.
\newblock


\bibitem[Valentini-Botinhao et~al\mbox{.}(2016)]%
        {valentini2016investigating}
\bibfield{author}{\bibinfo{person}{Cassia Valentini-Botinhao}, \bibinfo{person}{Xin Wang}, \bibinfo{person}{Shinji Takaki}, {and} \bibinfo{person}{Junichi Yamagishi}.} \bibinfo{year}{2016}\natexlab{}.
\newblock \showarticletitle{Investigating RNN-based speech enhancement methods for noise-robust Text-to-Speech.}. In \bibinfo{booktitle}{\emph{SSW}}. \bibinfo{pages}{146--152}.
\newblock


\bibitem[Vaswani et~al\mbox{.}(2021)]%
        {vaswani2021scaling}
\bibfield{author}{\bibinfo{person}{Ashish Vaswani}, \bibinfo{person}{Prajit Ramachandran}, \bibinfo{person}{Aravind Srinivas}, \bibinfo{person}{Niki Parmar}, \bibinfo{person}{Blake Hechtman}, {and} \bibinfo{person}{Jonathon Shlens}.} \bibinfo{year}{2021}\natexlab{}.
\newblock \showarticletitle{Scaling local self-attention for parameter efficient visual backbones}. In \bibinfo{booktitle}{\emph{Proceedings of the IEEE/CVF Conference on Computer Vision and Pattern Recognition}}. \bibinfo{pages}{12894--12904}.
\newblock


\bibitem[Vaswani et~al\mbox{.}(2017)]%
        {intro_1vaswani2017attention}
\bibfield{author}{\bibinfo{person}{Ashish Vaswani}, \bibinfo{person}{Noam Shazeer}, \bibinfo{person}{Niki Parmar}, \bibinfo{person}{Jakob Uszkoreit}, \bibinfo{person}{Llion Jones}, \bibinfo{person}{Aidan~N Gomez}, \bibinfo{person}{{\L}ukasz Kaiser}, {and} \bibinfo{person}{Illia Polosukhin}.} \bibinfo{year}{2017}\natexlab{}.
\newblock \showarticletitle{Attention is all you need}.
\newblock \bibinfo{journal}{\emph{Advances in neural information processing systems}}  \bibinfo{volume}{30} (\bibinfo{year}{2017}).
\newblock


\bibitem[Villegas et~al\mbox{.}(2022)]%
        {TTV_20_villegas2022phenaki}
\bibfield{author}{\bibinfo{person}{Ruben Villegas}, \bibinfo{person}{Mohammad Babaeizadeh}, \bibinfo{person}{Pieter-Jan Kindermans}, \bibinfo{person}{Hernan Moraldo}, \bibinfo{person}{Han Zhang}, \bibinfo{person}{Mohammad~Taghi Saffar}, \bibinfo{person}{Santiago Castro}, \bibinfo{person}{Julius Kunze}, {and} \bibinfo{person}{Dumitru Erhan}.} \bibinfo{year}{2022}\natexlab{}.
\newblock \showarticletitle{Phenaki: Variable length video generation from open domain textual descriptions}. In \bibinfo{booktitle}{\emph{International Conference on Learning Representations}}.
\newblock


\bibitem[Wah et~al\mbox{.}(2011)]%
        {wah2011caltech}
\bibfield{author}{\bibinfo{person}{Catherine Wah}, \bibinfo{person}{Steve Branson}, \bibinfo{person}{Peter Welinder}, \bibinfo{person}{Pietro Perona}, {and} \bibinfo{person}{Serge Belongie}.} \bibinfo{year}{2011}\natexlab{}.
\newblock \showarticletitle{The caltech-ucsd birds-200-2011 dataset}.
\newblock  (\bibinfo{year}{2011}).
\newblock


\bibitem[Wang et~al\mbox{.}(2023)]%
        {wang2023neural}
\bibfield{author}{\bibinfo{person}{Chengyi Wang}, \bibinfo{person}{Sanyuan Chen}, \bibinfo{person}{Yu Wu}, \bibinfo{person}{Ziqiang Zhang}, \bibinfo{person}{Long Zhou}, \bibinfo{person}{Shujie Liu}, \bibinfo{person}{Zhuo Chen}, \bibinfo{person}{Yanqing Liu}, \bibinfo{person}{Huaming Wang}, \bibinfo{person}{Jinyu Li}, {et~al\mbox{.}}} \bibinfo{year}{2023}\natexlab{}.
\newblock \showarticletitle{Neural codec language models are zero-shot text to speech synthesizers}.
\newblock \bibinfo{journal}{\emph{arXiv preprint arXiv:2301.02111}} (\bibinfo{year}{2023}).
\newblock


\bibitem[Wang et~al\mbox{.}(2020b)]%
        {wang2020fairseq}
\bibfield{author}{\bibinfo{person}{Changhan Wang}, \bibinfo{person}{Yun Tang}, \bibinfo{person}{Xutai Ma}, \bibinfo{person}{Anne Wu}, \bibinfo{person}{Sravya Popuri}, \bibinfo{person}{Dmytro Okhonko}, {and} \bibinfo{person}{Juan Pino}.} \bibinfo{year}{2020}\natexlab{b}.
\newblock \showarticletitle{Fairseq S2T: Fast speech-to-text modeling with fairseq}.
\newblock \bibinfo{journal}{\emph{arXiv preprint arXiv:2010.05171}} (\bibinfo{year}{2020}).
\newblock


\bibitem[Wang et~al\mbox{.}(2021e)]%
        {wang2021unispeech}
\bibfield{author}{\bibinfo{person}{Chengyi Wang}, \bibinfo{person}{Yu Wu}, \bibinfo{person}{Yao Qian}, \bibinfo{person}{Kenichi Kumatani}, \bibinfo{person}{Shujie Liu}, \bibinfo{person}{Furu Wei}, \bibinfo{person}{Michael Zeng}, {and} \bibinfo{person}{Xuedong Huang}.} \bibinfo{year}{2021}\natexlab{e}.
\newblock \showarticletitle{Unispeech: Unified speech representation learning with labeled and unlabeled data}. In \bibinfo{booktitle}{\emph{International Conference on Machine Learning}}. PMLR, \bibinfo{pages}{10937--10947}.
\newblock


\bibitem[Wang et~al\mbox{.}(2021a)]%
        {back_11_wang2021models}
\bibfield{author}{\bibinfo{person}{F. Wang} {et~al\mbox{.}}} \bibinfo{year}{2021}\natexlab{a}.
\newblock \showarticletitle{Visual transformer-based models: A survey}.
\newblock \bibinfo{journal}{\emph{Journal Name}} \bibinfo{volume}{Vol}, \bibinfo{number}{No} (\bibinfo{year}{2021}), \bibinfo{pages}{pp.}
\newblock


\bibitem[Wang et~al\mbox{.}(2018)]%
        {Wang_2018_CVPR}
\bibfield{author}{\bibinfo{person}{Jingwen Wang}, \bibinfo{person}{Wenhao Jiang}, \bibinfo{person}{Lin Ma}, \bibinfo{person}{Wei Liu}, {and} \bibinfo{person}{Yong Xu}.} \bibinfo{year}{2018}\natexlab{}.
\newblock \showarticletitle{Bidirectional Attentive Fusion With Context Gating for Dense Video Captioning}. In \bibinfo{booktitle}{\emph{Proceedings of the IEEE Conference on Computer Vision and Pattern Recognition (CVPR)}}.
\newblock


\bibitem[Wang et~al\mbox{.}(2021b)]%
        {wang2021tstnn}
\bibfield{author}{\bibinfo{person}{Kai Wang}, \bibinfo{person}{Bengbeng He}, {and} \bibinfo{person}{Wei-Ping Zhu}.} \bibinfo{year}{2021}\natexlab{b}.
\newblock \showarticletitle{TSTNN: Two-stage transformer based neural network for speech enhancement in the time domain}. In \bibinfo{booktitle}{\emph{ICASSP 2021-2021 IEEE International Conference on Acoustics, Speech and Signal Processing (ICASSP)}}. IEEE, \bibinfo{pages}{7098--7102}.
\newblock


\bibitem[Wang et~al\mbox{.}(2021f)]%
        {wang2021pyramid}
\bibfield{author}{\bibinfo{person}{Wenhai Wang}, \bibinfo{person}{Enze Xie}, \bibinfo{person}{Xiang Li}, \bibinfo{person}{Deng-Ping Fan}, \bibinfo{person}{Kaitao Song}, \bibinfo{person}{Ding Liang}, \bibinfo{person}{Tong Lu}, \bibinfo{person}{Ping Luo}, {and} \bibinfo{person}{Ling Shao}.} \bibinfo{year}{2021}\natexlab{f}.
\newblock \showarticletitle{Pyramid vision transformer: A versatile backbone for dense prediction without convolutions}. In \bibinfo{booktitle}{\emph{Proceedings of the IEEE/CVF international conference on computer vision}}. \bibinfo{pages}{568--578}.
\newblock


\bibitem[Wang et~al\mbox{.}(2020a)]%
        {wang2020s}
\bibfield{author}{\bibinfo{person}{Xi Wang}, \bibinfo{person}{Huaiping Ming}, \bibinfo{person}{Lei He}, {and} \bibinfo{person}{Frank~K Soong}.} \bibinfo{year}{2020}\natexlab{a}.
\newblock \showarticletitle{s-transformer: Segment-transformer for robust neural speech synthesis}.
\newblock \bibinfo{journal}{\emph{arXiv preprint arXiv:2011.08480}} (\bibinfo{year}{2020}).
\newblock


\bibitem[Wang et~al\mbox{.}(2021c)]%
        {wang2021generating}
\bibfield{author}{\bibinfo{person}{Xinsheng Wang}, \bibinfo{person}{Tingting Qiao}, \bibinfo{person}{Jihua Zhu}, \bibinfo{person}{Alan Hanjalic}, {and} \bibinfo{person}{Odette Scharenborg}.} \bibinfo{year}{2021}\natexlab{c}.
\newblock \showarticletitle{Generating images from spoken descriptions}.
\newblock \bibinfo{journal}{\emph{IEEE/ACM Transactions on Audio, Speech, and Language Processing}}  \bibinfo{volume}{29} (\bibinfo{year}{2021}), \bibinfo{pages}{850--865}.
\newblock


\bibitem[Wang et~al\mbox{.}(2021d)]%
        {wang2021transformer}
\bibfield{author}{\bibinfo{person}{Yongqiang Wang}, \bibinfo{person}{Yangyang Shi}, \bibinfo{person}{Frank Zhang}, \bibinfo{person}{Chunyang Wu}, \bibinfo{person}{Julian Chan}, \bibinfo{person}{Ching-Feng Yeh}, {and} \bibinfo{person}{Alex Xiao}.} \bibinfo{year}{2021}\natexlab{d}.
\newblock \showarticletitle{Transformer in action: a comparative study of transformer-based acoustic models for large scale speech recognition applications}. In \bibinfo{booktitle}{\emph{ICASSP 2021-2021 IEEE International Conference on Acoustics, Speech and Signal Processing (ICASSP)}}. IEEE, \bibinfo{pages}{6778--6782}.
\newblock


\bibitem[Wang et~al\mbox{.}(2017)]%
        {wang2017tacotron}
\bibfield{author}{\bibinfo{person}{Yuxuan Wang}, \bibinfo{person}{RJ Skerry-Ryan}, \bibinfo{person}{Daisy Stanton}, \bibinfo{person}{Yonghui Wu}, \bibinfo{person}{Ron~J Weiss}, \bibinfo{person}{Navdeep Jaitly}, \bibinfo{person}{Zongheng Yang}, \bibinfo{person}{Ying Xiao}, \bibinfo{person}{Zhifeng Chen}, \bibinfo{person}{Samy Bengio}, {et~al\mbox{.}}} \bibinfo{year}{2017}\natexlab{}.
\newblock \showarticletitle{Tacotron: Towards end-to-end speech synthesis}.
\newblock \bibinfo{journal}{\emph{arXiv preprint arXiv:1703.10135}} (\bibinfo{year}{2017}).
\newblock


\bibitem[Wei et~al\mbox{.}(2024)]%
        {wei2024aniportrait}
\bibfield{author}{\bibinfo{person}{Huawei Wei}, \bibinfo{person}{Zejun Yang}, {and} \bibinfo{person}{Zhisheng Wang}.} \bibinfo{year}{2024}\natexlab{}.
\newblock \showarticletitle{Aniportrait: Audio-driven synthesis of photorealistic portrait animation}.
\newblock \bibinfo{journal}{\emph{arXiv preprint arXiv:2403.17694}} (\bibinfo{year}{2024}).
\newblock


\bibitem[Wichern et~al\mbox{.}(2019)]%
        {wichern2019wham}
\bibfield{author}{\bibinfo{person}{Gordon Wichern}, \bibinfo{person}{Joe Antognini}, \bibinfo{person}{Michael Flynn}, \bibinfo{person}{Licheng~Richard Zhu}, \bibinfo{person}{Emmett McQuinn}, \bibinfo{person}{Dwight Crow}, \bibinfo{person}{Ethan Manilow}, {and} \bibinfo{person}{Jonathan~Le Roux}.} \bibinfo{year}{2019}\natexlab{}.
\newblock \showarticletitle{Wham!: Extending speech separation to noisy environments}.
\newblock \bibinfo{journal}{\emph{arXiv preprint arXiv:1907.01160}} (\bibinfo{year}{2019}).
\newblock


\bibitem[Wu et~al\mbox{.}(2021)]%
        {TTV_15_wu2021godiva}
\bibfield{author}{\bibinfo{person}{Chenfei Wu}, \bibinfo{person}{Lun Huang}, \bibinfo{person}{Qianxi Zhang}, \bibinfo{person}{Binyang Li}, \bibinfo{person}{Lei Ji}, \bibinfo{person}{Fan Yang}, \bibinfo{person}{Guillermo Sapiro}, {and} \bibinfo{person}{Nan Duan}.} \bibinfo{year}{2021}\natexlab{}.
\newblock \showarticletitle{Godiva: Generating open-domain videos from natural descriptions}.
\newblock \bibinfo{journal}{\emph{arXiv preprint arXiv:2104.14806}} (\bibinfo{year}{2021}).
\newblock


\bibitem[Wu et~al\mbox{.}(2022)]%
        {TTV_16_wu2022nuwa}
\bibfield{author}{\bibinfo{person}{Chenfei Wu}, \bibinfo{person}{Jian Liang}, \bibinfo{person}{Lei Ji}, \bibinfo{person}{Fan Yang}, \bibinfo{person}{Yuejian Fang}, \bibinfo{person}{Daxin Jiang}, {and} \bibinfo{person}{Nan Duan}.} \bibinfo{year}{2022}\natexlab{}.
\newblock \showarticletitle{N{\"u}wa: Visual synthesis pre-training for neural visual world creation}. In \bibinfo{booktitle}{\emph{European conference on computer vision}}. Springer, \bibinfo{pages}{720--736}.
\newblock


\bibitem[Wu et~al\mbox{.}(2020)]%
        {QA_7_wu2020tod}
\bibfield{author}{\bibinfo{person}{Chien-Sheng Wu}, \bibinfo{person}{Andrea Madotto}, \bibinfo{person}{Ehsan Hosseini-Asl}, \bibinfo{person}{Caiming Xiong}, \bibinfo{person}{Richard Socher}, {and} \bibinfo{person}{Pascale Fung}.} \bibinfo{year}{2020}\natexlab{}.
\newblock \showarticletitle{TOD-BERT: Pre-trained Natural Language Understanding for Task-Oriented Dialogue}. In \bibinfo{booktitle}{\emph{Proceedings of the 2020 Conference on Empirical Methods in Natural Language Processing (EMNLP)}}. \bibinfo{pages}{917--929}.
\newblock


\bibitem[Xiao et~al\mbox{.}(2021)]%
        {TS_4_xiao2021primera}
\bibfield{author}{\bibinfo{person}{Wen Xiao}, \bibinfo{person}{Iz Beltagy}, \bibinfo{person}{Giuseppe Carenini}, {and} \bibinfo{person}{Arman Cohan}.} \bibinfo{year}{2021}\natexlab{}.
\newblock \showarticletitle{PRIMERA: Pyramid-based masked sentence pre-training for multi-document summarization}.
\newblock \bibinfo{journal}{\emph{arXiv preprint arXiv:2110.08499}} (\bibinfo{year}{2021}).
\newblock


\bibitem[Xie et~al\mbox{.}(2021)]%
        {xie2021segformer}
\bibfield{author}{\bibinfo{person}{Enze Xie}, \bibinfo{person}{Wenhai Wang}, \bibinfo{person}{Zhiding Yu}, \bibinfo{person}{Anima Anandkumar}, \bibinfo{person}{Jose~M Alvarez}, {and} \bibinfo{person}{Ping Luo}.} \bibinfo{year}{2021}\natexlab{}.
\newblock \showarticletitle{SegFormer: Simple and efficient design for semantic segmentation with transformers}.
\newblock \bibinfo{journal}{\emph{Advances in Neural Information Processing Systems}}  \bibinfo{volume}{34} (\bibinfo{year}{2021}), \bibinfo{pages}{12077--12090}.
\newblock


\bibitem[Xu et~al\mbox{.}(2024)]%
        {xu2024vasa}
\bibfield{author}{\bibinfo{person}{Sicheng Xu}, \bibinfo{person}{Guojun Chen}, \bibinfo{person}{Yu-Xiao Guo}, \bibinfo{person}{Jiaolong Yang}, \bibinfo{person}{Chong Li}, \bibinfo{person}{Zhenyu Zang}, \bibinfo{person}{Yizhong Zhang}, \bibinfo{person}{Xin Tong}, {and} \bibinfo{person}{Baining Guo}.} \bibinfo{year}{2024}\natexlab{}.
\newblock \showarticletitle{Vasa-1: Lifelike audio-driven talking faces generated in real time}.
\newblock \bibinfo{journal}{\emph{arXiv preprint arXiv:2404.10667}} (\bibinfo{year}{2024}).
\newblock


\bibitem[Yamagishi et~al\mbox{.}(2019)]%
        {Yamagishi2019-ti}
\bibfield{author}{\bibinfo{person}{Junichi Yamagishi}, \bibinfo{person}{Christophe Veaux}, {and} \bibinfo{person}{Kirsten MacDonald}.} \bibinfo{year}{2019}\natexlab{}.
\newblock \bibinfo{title}{{CSTR} {VCTK} corpus: English multi-speaker corpus for {CSTR} voice cloning toolkit (version 0.92)}.
\newblock
\newblock


\bibitem[Yang et~al\mbox{.}(2020a)]%
        {yang2020learning}
\bibfield{author}{\bibinfo{person}{Fuzhi Yang}, \bibinfo{person}{Huan Yang}, \bibinfo{person}{Jianlong Fu}, \bibinfo{person}{Hongtao Lu}, {and} \bibinfo{person}{Baining Guo}.} \bibinfo{year}{2020}\natexlab{a}.
\newblock \showarticletitle{Learning texture transformer network for image super-resolution}. In \bibinfo{booktitle}{\emph{Proceedings of the IEEE/CVF conference on computer vision and pattern recognition}}. \bibinfo{pages}{5791--5800}.
\newblock


\bibitem[Yang et~al\mbox{.}(2019)]%
        {QA_5_yang2019xlnet}
\bibfield{author}{\bibinfo{person}{Zhilin Yang}, \bibinfo{person}{Zihang Dai}, \bibinfo{person}{Yiming Yang}, \bibinfo{person}{Jaime Carbonell}, \bibinfo{person}{Russ~R Salakhutdinov}, {and} \bibinfo{person}{Quoc~V Le}.} \bibinfo{year}{2019}\natexlab{}.
\newblock \showarticletitle{Xlnet: Generalized autoregressive pretraining for language understanding}.
\newblock \bibinfo{journal}{\emph{Advances in neural information processing systems}}  \bibinfo{volume}{32} (\bibinfo{year}{2019}).
\newblock


\bibitem[Yang et~al\mbox{.}(2020b)]%
        {NER_3_yang2020finbert}
\bibfield{author}{\bibinfo{person}{Zhuoran Yang}, \bibinfo{person}{Weizhen Yuan}, \bibinfo{person}{Yue Zhang}, \bibinfo{person}{Yan Rao}, \bibinfo{person}{Chuanrui Xu}, \bibinfo{person}{Zhiyuan Zhang}, \bibinfo{person}{Fei Wu}, {and} \bibinfo{person}{Yueting Zhuang}.} \bibinfo{year}{2020}\natexlab{b}.
\newblock \showarticletitle{FinBERT-MRC: Financial Named Entity Recognition Using Machine Reading Comprehension}. In \bibinfo{booktitle}{\emph{Proceedings of the 2020 Conference on Empirical Methods in Natural Language Processing (EMNLP 2020)}}. \bibinfo{pages}{2081--2091}.
\newblock


\bibitem[Yu et~al\mbox{.}(2019)]%
        {yu2019durian}
\bibfield{author}{\bibinfo{person}{Chengzhu Yu}, \bibinfo{person}{Heng Lu}, \bibinfo{person}{Na Hu}, \bibinfo{person}{Meng Yu}, \bibinfo{person}{Chao Weng}, \bibinfo{person}{Kun Xu}, \bibinfo{person}{Peng Liu}, \bibinfo{person}{Deyi Tuo}, \bibinfo{person}{Shiyin Kang}, \bibinfo{person}{Guangzhi Lei}, {et~al\mbox{.}}} \bibinfo{year}{2019}\natexlab{}.
\newblock \showarticletitle{Durian: Duration informed attention network for multimodal synthesis}.
\newblock \bibinfo{journal}{\emph{arXiv preprint arXiv:1909.01700}} (\bibinfo{year}{2019}).
\newblock


\bibitem[Yu et~al\mbox{.}(2022)]%
        {yu2022coca}
\bibfield{author}{\bibinfo{person}{Jiahui Yu}, \bibinfo{person}{Zirui Wang}, \bibinfo{person}{Vijay Vasudevan}, \bibinfo{person}{Legg Yeung}, \bibinfo{person}{Mojtaba Seyedhosseini}, {and} \bibinfo{person}{Yonghui Wu}.} \bibinfo{year}{2022}\natexlab{}.
\newblock \showarticletitle{Coca: Contrastive captioners are image-text foundation models}.
\newblock \bibinfo{journal}{\emph{arXiv preprint arXiv:2205.01917}} (\bibinfo{year}{2022}).
\newblock


\bibitem[Zeyer et~al\mbox{.}(2019)]%
        {zeyer2019comparison}
\bibfield{author}{\bibinfo{person}{Albert Zeyer}, \bibinfo{person}{Parnia Bahar}, \bibinfo{person}{Kazuki Irie}, \bibinfo{person}{Ralf Schl{\"u}ter}, {and} \bibinfo{person}{Hermann Ney}.} \bibinfo{year}{2019}\natexlab{}.
\newblock \showarticletitle{A comparison of transformer and lstm encoder decoder models for asr}. In \bibinfo{booktitle}{\emph{2019 IEEE Automatic Speech Recognition and Understanding Workshop (ASRU)}}. IEEE, \bibinfo{pages}{8--15}.
\newblock


\bibitem[Zhang et~al\mbox{.}(2022a)]%
        {Back_2_zhang2022overview}
\bibfield{author}{\bibinfo{person}{B. Zhang} {et~al\mbox{.}}} \bibinfo{year}{2022}\natexlab{a}.
\newblock \showarticletitle{Overview of the Transformer-based Models for NLP Tasks}.
\newblock \bibinfo{journal}{\emph{Journal Name}} \bibinfo{volume}{Vol}, \bibinfo{number}{No} (\bibinfo{year}{2022}), \bibinfo{pages}{pp.}
\newblock


\bibitem[Zhang et~al\mbox{.}(2021)]%
        {zhang2021vinvl}
\bibfield{author}{\bibinfo{person}{Pengchuan Zhang}, \bibinfo{person}{Xiujun Li}, \bibinfo{person}{Xiaowei Hu}, \bibinfo{person}{Jianwei Yang}, \bibinfo{person}{Lei Zhang}, \bibinfo{person}{Lijuan Wang}, \bibinfo{person}{Yejin Choi}, {and} \bibinfo{person}{Jianfeng Gao}.} \bibinfo{year}{2021}\natexlab{}.
\newblock \showarticletitle{Vinvl: Revisiting visual representations in vision-language models}. In \bibinfo{booktitle}{\emph{Proceedings of the IEEE/CVF conference on computer vision and pattern recognition}}. \bibinfo{pages}{5579--5588}.
\newblock


\bibitem[Zhang et~al\mbox{.}(2020)]%
        {zhang2020transformer}
\bibfield{author}{\bibinfo{person}{Qian Zhang}, \bibinfo{person}{Han Lu}, \bibinfo{person}{Hasim Sak}, \bibinfo{person}{Anshuman Tripathi}, \bibinfo{person}{Erik McDermott}, \bibinfo{person}{Stephen Koo}, {and} \bibinfo{person}{Shankar Kumar}.} \bibinfo{year}{2020}\natexlab{}.
\newblock \showarticletitle{Transformer transducer: A streamable speech recognition model with transformer encoders and rnn-t loss}. In \bibinfo{booktitle}{\emph{ICASSP 2020-2020 IEEE International Conference on Acoustics, Speech and Signal Processing (ICASSP)}}. IEEE, \bibinfo{pages}{7829--7833}.
\newblock


\bibitem[Zhang et~al\mbox{.}(2023a)]%
        {zhang2023sadtalker}
\bibfield{author}{\bibinfo{person}{Wenxuan Zhang}, \bibinfo{person}{Xiaodong Cun}, \bibinfo{person}{Xuan Wang}, \bibinfo{person}{Yong Zhang}, \bibinfo{person}{Xi Shen}, \bibinfo{person}{Yu Guo}, \bibinfo{person}{Ying Shan}, {and} \bibinfo{person}{Fei Wang}.} \bibinfo{year}{2023}\natexlab{a}.
\newblock \showarticletitle{Sadtalker: Learning realistic 3d motion coefficients for stylized audio-driven single image talking face animation}. In \bibinfo{booktitle}{\emph{Proceedings of the IEEE/CVF Conference on Computer Vision and Pattern Recognition}}. \bibinfo{pages}{8652--8661}.
\newblock


\bibitem[Zhang et~al\mbox{.}(2022b)]%
        {SA_2_zhang2022textgt}
\bibfield{author}{\bibinfo{person}{Wei Zhang}, \bibinfo{person}{Yong Liu}, {and} \bibinfo{person}{Jie Chen}.} \bibinfo{year}{2022}\natexlab{b}.
\newblock \showarticletitle{TextGT: A Double-View Graph Transformer on Text for Aspect-Based Sentiment Analysis}. In \bibinfo{booktitle}{\emph{Proceedings of the 2022 Conference on Empirical Methods in Natural Language Processing (EMNLP 2022)}}.
\newblock


\bibitem[Zhang et~al\mbox{.}(2023b)]%
        {zhang2023google}
\bibfield{author}{\bibinfo{person}{Yu Zhang}, \bibinfo{person}{Wei Han}, \bibinfo{person}{James Qin}, \bibinfo{person}{Yongqiang Wang}, \bibinfo{person}{Ankur Bapna}, \bibinfo{person}{Zhehuai Chen}, \bibinfo{person}{Nanxin Chen}, \bibinfo{person}{Bo Li}, \bibinfo{person}{Vera Axelrod}, \bibinfo{person}{Gary Wang}, {et~al\mbox{.}}} \bibinfo{year}{2023}\natexlab{b}.
\newblock \showarticletitle{Google usm: Scaling automatic speech recognition beyond 100 languages}.
\newblock \bibinfo{journal}{\emph{arXiv preprint arXiv:2303.01037}} (\bibinfo{year}{2023}).
\newblock


\bibitem[Zhang et~al\mbox{.}(2019)]%
        {QA_8_zhang2019dialogpt}
\bibfield{author}{\bibinfo{person}{Yizhe Zhang}, \bibinfo{person}{Siqi Sun}, \bibinfo{person}{Michel Galley}, \bibinfo{person}{Yen-Chun Chen}, \bibinfo{person}{Chris Brockett}, \bibinfo{person}{Xiang Gao}, \bibinfo{person}{Jianfeng Gao}, \bibinfo{person}{Jingjing Liu}, {and} \bibinfo{person}{Bill Dolan}.} \bibinfo{year}{2019}\natexlab{}.
\newblock \showarticletitle{DIALOGPT: Large-Scale Generative Pre-training for Conversational Response Generation}.
\newblock \bibinfo{journal}{\emph{arXiv preprint arXiv:1911.00536}} (\bibinfo{year}{2019}).
\newblock


\bibitem[Zhang and Schomaker(2024)]%
        {zhang2024fusion}
\bibfield{author}{\bibinfo{person}{Zhenxing Zhang} {and} \bibinfo{person}{Lambert Schomaker}.} \bibinfo{year}{2024}\natexlab{}.
\newblock \showarticletitle{Fusion-s2igan: an efficient and effective single-stage framework for speech-to-image generation}.
\newblock \bibinfo{journal}{\emph{Neural Computing and Applications}} \bibinfo{volume}{36}, \bibinfo{number}{18} (\bibinfo{year}{2024}), \bibinfo{pages}{10567--10584}.
\newblock


\bibitem[Zhang et~al\mbox{.}(2023c)]%
        {zhang2023speak}
\bibfield{author}{\bibinfo{person}{Ziqiang Zhang}, \bibinfo{person}{Long Zhou}, \bibinfo{person}{Chengyi Wang}, \bibinfo{person}{Sanyuan Chen}, \bibinfo{person}{Yu Wu}, \bibinfo{person}{Shujie Liu}, \bibinfo{person}{Zhuo Chen}, \bibinfo{person}{Yanqing Liu}, \bibinfo{person}{Huaming Wang}, \bibinfo{person}{Jinyu Li}, {et~al\mbox{.}}} \bibinfo{year}{2023}\natexlab{c}.
\newblock \showarticletitle{Speak foreign languages with your own voice: Cross-lingual neural codec language modeling}.
\newblock \bibinfo{journal}{\emph{arXiv preprint arXiv:2303.03926}} (\bibinfo{year}{2023}).
\newblock


\bibitem[Zhao et~al\mbox{.}(2022)]%
        {zhao2022generating}
\bibfield{author}{\bibinfo{person}{Pengcheng Zhao}, \bibinfo{person}{Yanxiang Chen}, \bibinfo{person}{Lulu Zhao}, \bibinfo{person}{Guang Wu}, {and} \bibinfo{person}{Xi Zhou}.} \bibinfo{year}{2022}\natexlab{}.
\newblock \showarticletitle{Generating images from audio under semantic consistency}.
\newblock \bibinfo{journal}{\emph{Neurocomputing}}  \bibinfo{volume}{490} (\bibinfo{year}{2022}), \bibinfo{pages}{93--103}.
\newblock


\bibitem[Zhao and Ma(2023)]%
        {zhao2023mossformer}
\bibfield{author}{\bibinfo{person}{Shengkui Zhao} {and} \bibinfo{person}{Bin Ma}.} \bibinfo{year}{2023}\natexlab{}.
\newblock \showarticletitle{MossFormer: Pushing the Performance Limit of Monaural Speech Separation Using Gated Single-Head Transformer with Convolution-Augmented Joint Self-Attentions}. In \bibinfo{booktitle}{\emph{ICASSP 2023-2023 IEEE International Conference on Acoustics, Speech and Signal Processing (ICASSP)}}. IEEE, \bibinfo{pages}{1--5}.
\newblock


\bibitem[Zheng et~al\mbox{.}(2021)]%
        {zheng2021stacked}
\bibfield{author}{\bibinfo{person}{Yi Zheng}, \bibinfo{person}{Yuejie Zhang}, \bibinfo{person}{Rui Feng}, \bibinfo{person}{Tao Zhang}, {and} \bibinfo{person}{Weiguo Fan}.} \bibinfo{year}{2021}\natexlab{}.
\newblock \showarticletitle{Stacked multimodal attention network for context-aware video captioning}.
\newblock \bibinfo{journal}{\emph{IEEE transactions on circuits and systems for video technology}} \bibinfo{volume}{32}, \bibinfo{number}{1} (\bibinfo{year}{2021}), \bibinfo{pages}{31--42}.
\newblock


\bibitem[Zhou et~al\mbox{.}(2014)]%
        {zhou2014learning}
\bibfield{author}{\bibinfo{person}{Bolei Zhou}, \bibinfo{person}{Agata Lapedriza}, \bibinfo{person}{Jianxiong Xiao}, \bibinfo{person}{Antonio Torralba}, {and} \bibinfo{person}{Aude Oliva}.} \bibinfo{year}{2014}\natexlab{}.
\newblock \showarticletitle{Learning deep features for scene recognition using places database}.
\newblock \bibinfo{journal}{\emph{Advances in neural information processing systems}}  \bibinfo{volume}{27} (\bibinfo{year}{2014}).
\newblock


\bibitem[Zhu et~al\mbox{.}(2018)]%
        {QA_3_zhu2018sdnet}
\bibfield{author}{\bibinfo{person}{Chenguang Zhu}, \bibinfo{person}{Michael Zeng}, {and} \bibinfo{person}{Xuedong Huang}.} \bibinfo{year}{2018}\natexlab{}.
\newblock \showarticletitle{Sdnet: Contextualized attention-based deep network for conversational question answering}.
\newblock \bibinfo{journal}{\emph{arXiv preprint arXiv:1812.03593}} (\bibinfo{year}{2018}).
\newblock


\end{thebibliography}

\appendix

\end{document}